\documentclass[11pt]{article}

\usepackage{pgf,pgfarrows,pgfautomata,pgfheaps,pgfnodes,pgfshade,multibox}
\usepackage{amsthm,latexsym,amssymb,amsfonts,array}
\usepackage[nosort]{cite}
\usepackage[centertags,intlimits]{amsmath}
\usepackage{graphicx}

\usepackage{longtable}

\makeatletter

\@addtoreset{equation}{section}
\makeatother
\newcommand{\be}{\begin{eqnarray}}
\newcommand{\ee}{\end{eqnarray}}
\newcommand{\nn}{\nonumber}

\newcommand{\lb}{\left[}
\newcommand{\rb}{\right]}
\newcommand{\cV}{{\cal V}}
\newcommand{\cQ}{{\cal Q}}
\newcommand{\cP}{{\cal P}}
\newcommand{\p}{\partial}
\newcommand{\cD}{D^{(0)}}
\newcommand{\eps}{\epsilon}

\newcommand{\ve}{\varepsilon}
\newcommand{\G}{\Gamma}
\newcommand{\al}{\alpha}

\newcommand{\beq}{\begin{equation}}
\newcommand{\eeq}{\end{equation}}

\newcommand{\f}{\frac}
\newcommand{\mc}{\mathcal}
\newcommand{\hs}{\hspace{0.1 cm}}

\newcommand{\mf}{\mathfrak}
\newcommand{\mbb}{\mathbb}
\newcommand{\noi}{\noindent}
\newcommand{\tten}{\tilde{10}}

\newcommand{\vet}{\varepsilon}
\newcommand{\bvet}{\bar\varepsilon}
\newcommand{\pt}{\psi}
\newcommand{\bpt}{\bar\psi}
\newcommand{\lt}{\lambda}
\newcommand{\blt}{\bar\lambda}

\newcommand{\venew}{\tilde{\varepsilon}}
\newcommand{\psinew}{\tilde{\psi}}
\newcommand{\lambdanew}{\tilde{\lambda}}

\newcommand{\om}{\omega}

\newcommand{\pa}{\partial}
\newcommand{\dg}{\delta_\Lambda}
\newcommand{\dS}{\delta_\Sigma}
\newcommand{\dcg}{\hat{\delta}_\Lambda}
\newcommand{\deps}{\delta_{\epsilon}}
\newcommand{\sP}{{\mathsf{P}}}
\newcommand{\sQ}{{\mathsf{Q}}}

\begin{document}

{\flushright {ULB-TH/08-36}\\
{LPTENS-08/59}\\}

\vskip .7 cm
\hfill
\vspace{13pt}
\begin{center}
{\Large {\bf \text{On the {\boldmath$ E_{10}$}/Massive Type IIA Supergravity Correspondence}}}

\vspace{1.1cm}

\rule[0.1in]{15cm}{0.5mm} 

\vspace{.8cm}

{\text{{\bf Marc Henneaux}}$\!$${}^{\spadesuit}$${}^{\diamondsuit}$\footnote{Also at \emph{Centro de Estudios Cient\'{\i}ficos
(CECS), Casilla 1469, Valdivia, Chile}}, \text{{\bf Ella Jamsin${}^{\spadesuit}$, Axel Kleinschmidt${}^{\spadesuit}$ and Daniel Persson}}${}^{\spadesuit}$${}^{\diamondsuit}$\footnote{Also at  \emph{Fundamental Physics, Chalmers University of Technology, SE-412 96, G\"oteborg, Sweden}}} \\

\vspace{10pt}
${}^{\spadesuit}${\em { Physique Th\'eorique et Math\'ematique,\\
Universit\'e Libre
de Bruxelles \& International Solvay Institutes,\\ 
ULB-Campus Plaine C.P. 231, B-1050 Bruxelles, Belgium}}\\

\vspace{8pt}
${}^{\diamondsuit}${\em {Laboratoire de Physique Th\'eorique, Ecole Normale Sup\'erieure,\\
24 rue Lhomond, F-75231 Paris Cedex 05, France}}
\vspace{4pt}

{\em { e-mail:}} \texttt{henneaux,ejamsin,axel.kleinschmidt,dpersson{ }@ulb.ac.be}
\end{center}

\vspace{10pt}
\begin{center}
\textbf{Abstract}\\[5mm]
\parbox{13cm}{\footnotesize
\noi In this paper we investigate in detail the correspondence between
$E_{10}$ and Romans' massive deformation of type IIA supergravity. We analyse
the dynamics of a non-linear sigma model for a spinning particle on the coset
space $E_{10}/K(E_{10})$ and show that it reproduces the dynamics of the
bosonic as well as the fermionic sector of the massive IIA theory, within the
standard truncation. The mass deformation parameter corresponds to a 
generator of $E_{10}$ outside the realm of the generators entering the usual
$D=11$ analysis, and is naturally included without any deformation of the
coset model for $E_{10}/K(E_{10})$. Our analysis thus provides a
dynamical unification of the massless and massive versions of type IIA
supergravity inside $E_{10}$. We discuss a number of additional and general
features of relevance in the analysis of any deformed supergravity in the
correspondence to Kac-Moody algebras, including recently studied deformations
where the trombone symmetry is gauged. } 
\end{center}

\noindent

\thispagestyle{empty}
\newpage
\setcounter{footnote}{0}

\tableofcontents

\section{Introduction}

Deformations of supergravity theories form an important part of the low energy
limit of M-theory. By a deformed supergravity we mean a theory that is different from standard Kaluza--Klein reductions of maximal $D=11$ supergravity (to which we restrict in this paper). Often, these theories cannot even be obtained by other types of reductions or compactifications from $D=11$ and are therefore to be interpreted as genuine new low energy actions for M-theory.
One of the pertinent features of deformed supergravity theories is
that they support domain wall solutions which are required to maintain
invariance under the duality symmetries expected from an underlying string
theory description. The most prominent example of this phenomenon is the
D8-brane of type IIA string theory that can be reached from lower-dimensional
branes by sequences of T-dualities~\cite{Polchinski:1995mt}. However, there is
no corresponding supergravity solution in undeformed type IIA supergravity in
$D=10$. Only when considering a (mass) deformation of type IIA supergravity
(constructed by Romans for different reasons~\cite{Romans:1985tz}) can one
accommodate the D8-brane~\cite{Polchinski:1995mt}. Similar results have been
obtained for many other deformed supergravity theories, most of which are
so-called gauged supergravities, see for
example~\cite{Roest:2004pk}.\footnote{Another common deformation is to add a
  cosmological constant but this is not always consistent with
  supersymmetry. Notably in $D=11$ the cosmological term is
  disallowed~\cite{Bautier:1997yp}.} 

Therefore any attempt at describing M-theory, at least at low energies, should
be able to reproduce these deformed supergravity theories. One possible
approach to M-theory is via Kac--Moody symmetries, notably $E_{10}$
\cite{Julia:1980gr,Julia:1982gx,Julia:1997cy,Damour:2002cu} and $E_{11}$ \cite{Julia:1980gr,Julia:1982gx,Julia:1997cy,West:2001as,Englert1,Englert:2003py}. Considering suitably truncated non-linear
realizations of these infinite-dimensional symmetries has led to perfect
agreement for the spectra of fields appearing in undeformed (super-)gravity
theories in various
dimensions, where the spectral analysis is carried out using a so-called level
decomposition~\cite{West:2001as,Damour:2002cu,West:2002jj,Nicolai:2003fw,Kleinschmidt:2003mf}.
For $E_{10}$ there exists a one-dimensional sigma model whose geodesic
equations of motion are conjectured to describe the dynamics of M-theory and
this has been verified in a low level approximation for undeformed maximal
$D=11$ supergravity~\cite{Damour:2002cu,Damour:2004zy}. A more
precise form of the conjecture states that one has to consider a constrained
geodesic, which mirrors the canonical constraints present in
supergravity~\cite{Damour:2007dt}. It has also been analysed how to add
fermions to the geodesic sigma model, with good agreement for both
kinematics and dynamics~\cite{deBuyl:2005zy,Damour:2005zs,deBuyl:2005mt,Kleinschmidt:2006tm,Damour:2006xu}.
Similarly, for $E_{11}$ a dynamical realization has been
discussed~\cite{West:2001as} based on non-linear realizations and an algebraic
construction of infinitesimal
diffeomorphisms~\cite{Borisov:1974bn,West:2000ga} and it was argued that this
construction agrees with undeformed supergravity.

In a recent development it was observed that the spectral analysis of $E_{10}$
and $E_{11}$ also yields all the fields required for deformed (maximal)
supergravity theories in various
dimensions~\cite{Riccioni:2007au,Bergshoeff:2007qi,Riccioni:2007ni}.\footnote{Related
  work for non-maximal supergravity can be found
  in~\cite{Gomis:2007gb,Bergshoeff:2007vb,Kleinschmidt:2008jj}.} 
These arise via forms of high rank, specifically $(D-1)$-forms for
deformations in $D$ dimensions.\footnote{In the case of gauged supergravity,
  there are (quadratic) constraints on these $(D-1)$-forms~\cite{Nicolai:2000sc,deWit:2005hv,deWit:2008ta} which appear in the tables of $E_{11}$ (only) as
$D$-forms~\cite{Riccioni:2007au,Bergshoeff:2007qi,Riccioni:2007ni}. For
$E_{10}$ the quadratic constraints appear to lie in a multiplet of
constraints~\cite{Bergshoeff:2008} together with the other canonical
constraints. } It is 
worthwhile to emphasize that the
Kac--Moody generators  associated to the usual gauge deformation parameters are
never part of the underlying $E_8$ or $E_9$ algebra but genuine $E_{10}$ (and
hence $E_{11}$) generators. This is in contradistinction to the fields of the
undeformed theory. 

This paper addresses the question to what extent the deformation parameters
also appear correctly in the dynamics of the geodesic model of $E_{10}$. This
will be done 
specifically in the case of the Romans' deformed IIA supergravity theory, where
the mass $m$ is the only deformation parameter. It corresponds to a generator
on level $\ell=4$ in parlance relevant for $D=11$ supergravity and hence
outside the realm of generators considered usually. The analysis here will be
carried out both for bosons and for fermions and we will show agreement
within the expected limitations of any known Kac--Moody/supergravity 
correspondence. Importantly, all terms associated with the mass deformation work perfectly.

The mass parameter was already identified for cosmological billiards~\cite{Damour:2002fz} and in the level decomposition
in~\cite{Kleinschmidt:2003mf} and studied in
\cite{Schnakenburg:2002xx,Kleinschmidt:2004dy}, with an emphasis on the D$8$ brane in~\cite{West:2004st,Englert:2007qb}. 
The present analysis completes
the picture by giving a complete and detailed account of the fermions and
bosons (as well as their supersymmetry) up to the level of the mass parameter
in an $\mf{sl}(9, \mbb{R}) \subset \mf{e}_{10}:= \text{Lie} \ E_{10}$ decomposition. Furthermore, our analysis
highlights several features relevant to deformed supergravity.
\begin{enumerate}
\item The $E_{10}$ coset model and fermionic representations we use are
  completely {\em  unaltered} with respect to the ones 
  successfully employed for the comparison with $D=11$ supergravity and
  undeformed IIA   and IIB supergravity~\cite{Kleinschmidt:2004dy}. In
  particular, there is no 
  deformation of the Kac--Moody structure underlying this algebraic approach
  to M-theory. This is different from proposals for the relation between
  $E_{11}$ and gauged supergravity~\cite{Riccioni:2007au,Riccioni:2007ni}.
\item In our approach the mass deformation can be included in the sigma model formulation without deforming the $E_{10}$ structure because the comparison always
  takes place in terms of {\em gauge-fixed} variables on both the supergravity
  and the Kac--Moody side. This allows to maintain the usual derivative and
  coset variables, as well as the gauge algebra structure. The gauge fixing
  re-enters in a consistent way via constraints. These issues are discussed in detail in Section \ref{sec:comments}.
  It is important to note that although this gauge fixing is chosen because it makes the $E_{10}$/supergravity correspondence particularly transparent, we of course expect that ultimately the physics should be independent of any particular gauge. 
\item The mass (or more generally all deformation
  parameters~\cite{Bergshoeff:2008}) arise as {\em integration constants} of
  the geodesic equations in a consistent truncation. This is in accordance with previous results in the context of massive IIA supergravity \cite{Bergshoeff:1996ui}. 
\item We analyse in addition the deformed supergravity theory obtained by gauging a global scaling symmetry of the equations of motion. This is known as trombone gauging~\cite{Howe:1997qt,Lavrinenko:1997qa} and has recently attracted attention in the $E_{10}$ and $E_{11}$ context~\cite{Diffon:2008sh}. Our findings suggest that the association of the deformation parameter with a certain mixed symmetry tensor appearing in the level decomposition is problematic.
\item As a by-product of our analysis we derive the commutation relations of
  $\mf{e}_{10}$ up to $\mf{sl}(10, \mbb{R})$ level $\ell=4$, confirming an
  earlier computer algebra analysis~\cite{Fischbacher:2005fy}. This result is then used to construct the $\ell=4$ action of $\mf{k}(\mf{e}_{10})$ on the Dirac spinor and vector-spinor representations of $ \mf{k}(\mf{e}_{10})$. These results are presented in Appendix~\ref{app:level4}. 
\end{enumerate}

This paper is structured as follows. We begin by presenting massive IIA
supergravity in our conventions in Section~\ref{sec:massiveIIA}, giving the complete Lagrangian including the fermionic sector. In this section we also provide the explicit form of the supersymmetry variations that leave the Lagrangian invariant. Then we proceed in Section~\ref{sec:E10} to introduce
the Kac-Moody algebra $\mf{e}_{10}$ in the $\mf{sl}(9, \mbb{R})$ level decomposition, relevant to the comparison with
Romans' theory. Here we also introduce the bosonic and fermionic parts of the $E_{10}$-invariant sigma model, and present their respective equations of motion. Section~\ref{sec:correspondence} is devoted to the comparison between the equations of motion of massive IIA supergravity and those of the geodesic sigma model for $E_{10}/K(E_{10})$. Here we establish the dictionary which yields a dynamical correspondence between the two models in the bosonic and fermionic sector. After the $E_{10}$/massive IIA correspondence is validated we discuss in more detail in Section~\ref{sec:comments} its implications, as well as some of the salient 
features of the analyis, including a derivation of the bosonic supersymmetry variations from the structure of $\mf{e}_{10}$. Finally, in Section~\ref{sec:trombone} we discuss the relation between our work and the recently proposed gauging of the so-called trombone symmetry. Since our analysis is quite technical and involves numerous lengthy calculations, we have for the benefit of the reader relegated most of the detailed calculations to appendices, where we for completeness also present an explicit analysis of $\mf{e}_{10}$ and $\mf{k}(\mf{e}_{10})$ at $\mf{sl}(10, \mbb{R})$ level four.

\section{Massive IIA supergravity}
\label{sec:massiveIIA}

Massive type IIA supergravity was first constructed by Romans \cite{Romans:1985tz} by deforming the standard type IIA supergravity through a St\"uckelberg mechanism, giving a mass to the two-form potential through the replacement $F_{\mu\nu}\rightarrow F_{\mu\nu}+mA_{\mu\nu}$, where $F_{\mu\nu}=2\partial_{[\mu}A_{\nu]}$. The potential $A_{\mu}$ can then be gauged away by a gauge transformation of $A_{\mu\nu}$. This process unfortunately obscures the massless limit to recover the standard IIA theory~\cite{Giani:1984wc,Campbell:1984zc,Huq:1983im} as some of the supersymmetry variations involve a coefficient $m^{-1}$. This is remedied by a field redefinition presented in \cite{Bergshoeff:1996ui,Lavrinenko:1999xi}, that we will make use of in this paper.

\begin{figure}
\begin{center}
\includegraphics{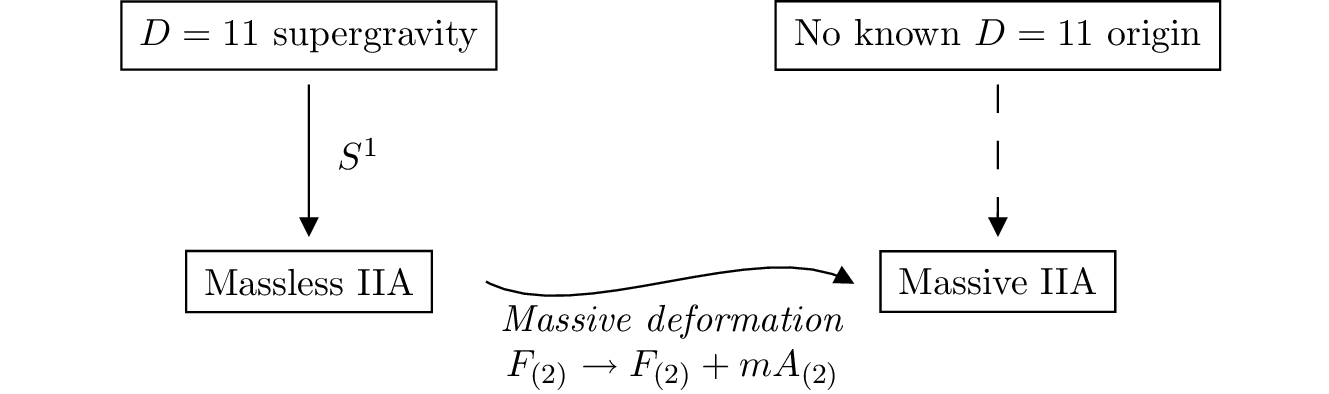}
\end{center}
\caption{\label{mIIA} \sl Massive IIA supergravity from $D=11$
    supergravity: Massive type IIA supergravity is obtained as a deformation of the standard type IIA supergravity, but unlike the latter, it does not possess any known eleven-dimensional origin. See Figure \ref{E10mIIA} for a pictorial description of how $D=11$ supergravity and massive IIA supergravity are unified inside $E_{10}$.}
\end{figure}

Moreover, a more democratic version of massive type IIA is given in \cite{Lavrinenko:1999xi,Bergshoeff:2001}, in which every form field comes with its dual. In particular, it involves a nine-form as a (potential) dual to the mass (seen as a field strength). This democratic formulation makes explicit the striking feature of massive IIA supergravity that it allows the existence of a D$8$-brane, that is known to exist in type IIA string theory. Indeed, this feature requires a nine-form potential that couples to the D$8$-brane, and accordingly does not appear in massless IIA supergravity. Although in this paper we will not use directly the democratic formulation, we will employ the field redefinitions used in \cite{Bergshoeff:1996ui} in order to clarify the massless limit.

\begin{figure}[t!]
\begin{center}
\includegraphics{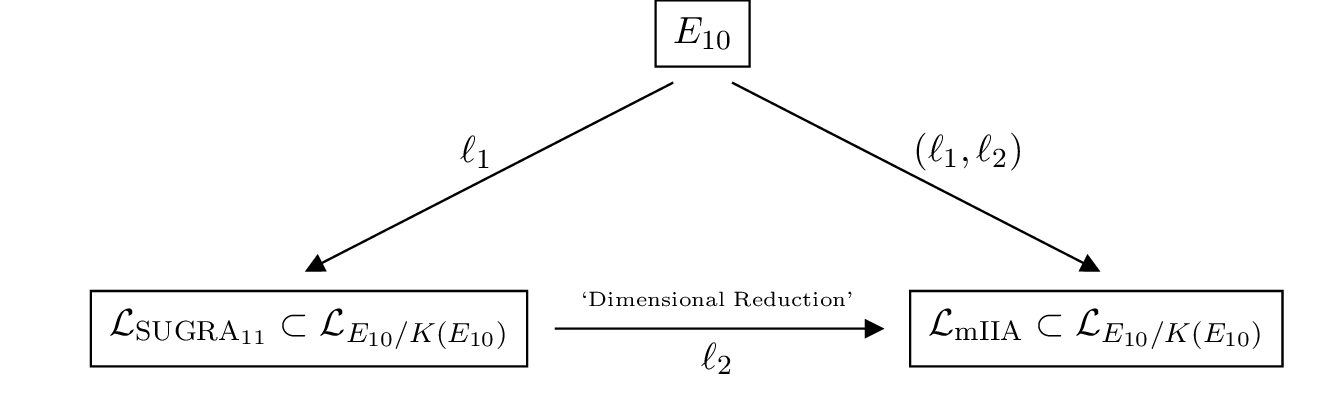}
\end{center}
\caption{\label{E10mIIA} \sl This picture describes the common $E_{10}$ origin
  of eleven-dimensional supergravity and massive type IIA supergravity. First,
  if one considers a level $\ell_1$ decomposition of $E_{10}$ with respect to
  $A_9$ (cf. Figure \ref{figure:E10}), one sees that the first levels
  ($\ell_1=0$ to $\ell_1=3$) of an $E_{10}/K(E_{10})$ sigma-model correspond
  to a truncated version of eleven-dimensional supergravity, with Lagrangian
  $\mc{L}_{\mathrm{SUGRA}_{11}}$ \cite{Damour:2002cu}. Taking this as a
  starting point, we can perform an additional level $\ell_2$ decomposition on
  the sigma model. On the lower $\ell_1$ levels ($\ell_1=0$ to $\ell_1=3$)
  this is equivalent to a dimensional reduction of eleven-dimensional
  supergravity, which gives massless IIA supergravity (cf. Figure
  \ref{mIIA}). However, if one includes one of the generators appearing at
  $\ell_1 = 4$, this leads to a theory that coincides with a truncated version
  of massive IIA supergravity, with Lagrangian
  $\mathcal{L}_{\mathrm{mIIA}}$. This procedure is equivalent to a multi-level
  $(\ell_1,\ell_2)$ decomposition of $E_{10}$ with respect to $A_8$. More
  details on this are given in Section \ref{sectionlevelE10}. By similar
  arguments, one could add IIB supergravity to this picture, in the sense that
  it has the same $E_{10}$ origin  in a level decomposition with respect to
  $A_8\times A_1$, i.e. with respect to node 8 in the Dynkin diagram in Figure
  \ref{figure:E10} \cite{Kleinschmidt:2004rg}.}  
\end{figure}

In addition, as a motivation for the present work, one also finds a nine-form in a certain decomposition of $E_{10}$, as will be developed in the next section, and this nine-form appears in the $E_{10}/K(E_{10})$ Lagrangian in the same way as the mass term in the massive IIA Lagrangian.
Massive IIA supergravity has in common with many other deformed maximal
supergravities that it does not possess any known higher dimensional origin,
as illustrated in Figure \ref{mIIA}. A consequence of the present work is to
show that, although they are not related by dimensional reduction, eleven
dimensional supergravity and massive IIA supergravity have the same $E_{10}$
origin as displayed in Figure \ref{E10mIIA}, see
also~\cite{Damour:2002fz,Kleinschmidt:2003mf,West:2004st}.\footnote{We note
  that the search for an eleven-dimensional origin of the Romans mass
  parameter, and hence the D$8$-brane, has led to studies of an M-theory
  M$9$-brane which is meant to exist in the presence of one Killing direction
  and then reduces to the IIA D$8$-brane, see
  e.g.~\cite{Bergshoeff:1997ak,Bergshoeff:1999bx}.} 

\subsection{Romans' theory}
In this section, and throughout the paper, the curved ten-dimensional space-time indices are denoted by $\mu$ and decompose into time $t$ and nine spatial indices $m$. The flat indices are similarly denoted by $\alpha=(0,a)$. Our space-time signature is $(-+\ldots +)$. More details on the conventions can be found in Appendix~\ref{Appendix:mIIAdetails}.

The complete supersymmetric Lagrangian (up to second order in fermions) in
Einstein frame is given by the sum of a bosonic and a fermionic part,
$\mathcal L=\mathcal L^{[B]}+\mathcal L^{[F]}$. The bosonic sector contains a
metric $G_{\mu\nu}$, a dilaton $\phi$, a one-form $A_{(1)}$ with field
strength $F_{(2)}$, a two-form $A_{(2)}$ with field strength $F_{(3)}$, a
three-form $A_{(3)}$ with field strength $F_{(4)}$, and a real mass parameter
$m$. The bosonic part of the Lagrangian reads {\allowdisplaybreaks
\begin{subequations}\label{lag}
\be
\mathcal L^{[B]} &=& \sqrt {-G} \Bigg[
  R-\f12|\p\phi|^2-\f14e^{3\phi/2}|F_{(2)}|^2-\f1{12}e^{-\phi}|F_{(3)}|^2
    -\f1{48}e^{\phi/2}|F_{(4)}|^2-\f12m^2e^{5\phi/2}\Bigg]\nn\\  
&+&\ve^{\mu_1\ldots\mu_{10}}\Bigg[\f1{144}
 \p_{\mu_1}A_{\mu_2\mu_3\mu_4}\p_{\mu_5}A_{\mu_6\mu_7\mu_8}A_{\mu_9\mu_{10}}  
 +\f{m}{288} \p_{\mu_1}A_{\mu_2\mu_3\mu_4} A_{\mu_5\mu_6} A_{\mu_7\mu_8}
    A_{\mu_9\mu_{10}}  \nn\\
&& \ \ \ \ \ \ \ \ \ \ \ \ + \f{m^2}{1280}
A_{\mu_1\mu_2}A_{\mu_3\mu_4}A_{\mu_5\mu_6}A_{\mu_7\mu_8}A_{\mu_9\mu_{10}}\Bigg]. 
\ee
Here, we already point out that the massive deformation induces a positive definite potential on the scalar sector. This is in marked contrast to what happens for gauged supergravity in lower dimensions where the scalar potential is generically indefinite~\cite{Nicolai:2000sc,deWit:2008ta}, causing also problems in the connection to $E_{10}$~\cite{Bergshoeff:2008}.

On the fermionic side, we have two gravitini, combined in a single $10\times32$ component
vector-spinor $\psi_\mu$, and two dilatini, combined in a single $32$
component Dirac-spinor $\lambda$, which decompose into two
fields of opposite chirality under $SO(1,9)$. For this sector the Lagrangian
takes the form 
\be
\mathcal L^{[F]} &=& i \sqrt {-G}\ \Bigg[ - 2\bpt_{\mu_1}\Gamma^{\mu_1\ldots\mu_3} D_{\mu_2}\pt_{\mu_{3}}
-\blt\G^{\mu} D_{\mu}\lt + \p_{\mu_1}\phi\blt\G^{\mu_2}\G^{\mu_1}\pt_{\mu_2} \nn\\
&+&\frac1{4}e^{3\phi/4}F_{\mu_1\mu_2}\Big(\bpt_{[\nu_1}\G^{\nu_1}\G^{\mu_1\mu_2}\G^{\nu_2}\G_{10}\pt_{\nu_2]} -\frac3{2}\blt\G^\nu\G^{\mu_1\mu_2}\G_{10}\pt_\nu+\frac58\blt\G^{\mu_1\mu_2}\G_{10}\lt\Big)\nn\\
&+&\frac1{12}e^{-\phi/2}F_{\mu_1\ldots\mu_3}\Big(\bpt_{[\nu_1}\G^{\nu_1}\G^{\mu_1\ldots\mu_3}\G^{\nu_2}\G_{10}\pt_{\nu_2]}-\blt\G^\nu\G^{\mu_1\ldots\mu_3}\G_{10}\pt_\nu\Big)\nn\\
&-&\frac1{48}e^{\phi/4}F_{\mu_1\ldots\mu_4}\Big(\bpt_{[\nu_1}\G^{\nu_1}\G^{\mu_1\ldots\mu_{4}}\G^{\nu_2}\pt_{\nu_2]}-\frac12\blt\G^\nu\G^{\mu_1\ldots\mu_{4}}\pt_\nu+\frac38\blt\G^{\mu_1\ldots\mu_{4}}\lt\Big) \nn\\
&+&\frac{21}{8}me^{5\phi/4}\blt\lt-\frac{1}{2}me^{5\phi/4}\bpt_{\mu_1}\G^{\mu_1\mu_2}\pt_{\mu_2}+\frac{5}{4}me^{5\phi/4}\blt\G^{\mu}\pt_{\mu}\Bigg].
\ee
\end{subequations}
The last line contains the explicit mass terms for the fermions. There are
also implicit mass deformations in the definitions of the field strength in
terms of the gauge potentials
\be
\label{curvs}
F_{\mu_1\mu_2} &=& 2\p_{[\mu_1} A_{\mu_2]} + m A_{\mu_1\mu_2}\,,\nn\\
F_{\mu_1\mu_2\mu_3} &=& 3\p_{[\mu_1} A_{\mu_2\mu_3]}\,,\nn\\
F_{\mu_1\mu_2\mu_3\mu_4} &=& 4\p_{[\mu_1} A_{\mu_2\mu_3\mu_4]} 
   + 4 A_{[\mu_1}F_{\mu_2\mu_3\mu_4]} 
   + 3 m A_{[\mu_1\mu_2} A_{\mu_3\mu_4]}\,.
\ee
In the Lagrangian, they are contracted without additional factors, for example
\beq |F_{(2)}|^2=F_{\mu_1\mu_2}F^{\mu_1\mu_2}.
\eeq 
In (\ref{curvs}) we see the
characteristic feature of a deformed theory that the usual tensor hierarchy of 
gauge fields is broken: there are forms coupling to potentials of higher
degree but only through terms proportional to the deformation parameter.
The tangent space field strengths are defined as usual by
conversion with the (inverse) vielbein $e_\al{}^\mu$, for example
$F_{\al_1\al_2}=e_{\al_1}{}^{\mu_1} e_{\al_2}{}^{\mu_2}F_{\mu_1\mu_2}$. Flat
indices are raised and lowered with the Minkowski metric and we will often
write contracted flat spatial indices on the same level. }

\subsection{Supersymmetry variations}

The supersymmetry variations leaving (\ref{lag}) invariant, up to total
derivatives and higher order fermion terms, are listed in this section. For the fermions, they read

\be
\delta_{\vet}\pt_\mu&=& D_\mu\vet-\frac1{32}me^{5\phi/4}\Gamma_\mu\vet-\frac1{64}e^{3\phi/4}F_{\nu\rho}({\Gamma_{\mu}}^{\nu\rho}-14\delta_{\mu}^{[\nu}\Gamma^{\rho]}\big)\G_{10}\vet\nn\\
&&+\frac1{96}e^{-\phi/2}F_{\nu\rho\sigma}\big({\Gamma_{\mu}}^{\nu\rho\sigma}-9\delta_{\mu}^{[\nu}\Gamma^{\rho\sigma]}\big)\Gamma_{10}\vet\nn\\
\label{deltapsi}
&&+\frac1{256}e^{\phi/4}F_{\nu\rho\sigma\gamma}\big({\Gamma_{\mu}}^{\nu\rho\sigma\gamma}-\f{20}{3}\delta_{\mu}^{[\nu}\Gamma^{\rho\sigma\gamma]}\big)\vet,
\ee
and 
\be
\delta_{\vet}\lt&=&\frac1{2}\p_\mu\phi\G^\mu\vet+\frac{5}{8}e^{5\phi/4}m\,\vet-\frac{3}{16}e^{3\phi/4}F_{\mu\nu}\G^{\mu\nu}\G_{10}\vet\nn\\
&&-\frac1{24}e^{-\phi/2}F_{\mu\nu\rho}\G^{\mu\nu\rho}\G_{10}\vet+\frac1{192}e^{\phi/4}F_{\mu\nu\rho\sigma}\G^{\mu\nu\rho\sigma}\vet,
\ee
while for the bosons we have
\beq
\delta_{\vet}{e_\mu}^\alpha= i \bvet\G^\alpha\pt_\mu,\ \ \ \ \ \delta_{\vet}\phi= i \blt\vet,
\eeq
and
\beq\label{susybos}
\delta_{\vet} A_{\mu}=\theta_{\mu},\ \ \ \ \ \delta_{\vet} A_{\mu\nu}=\theta_{\mu\nu},\ \ \ \ \ \delta_{\vet} A_{\mu\nu\rho}=\theta_{\mu\nu\rho}+6 A_{[\mu}\theta_{\nu\rho]},
\eeq
where
\be
\theta_\mu&:=&i
e^{-3\phi/4}\left(-\bpt_\mu-\f3{4}\blt\G_\mu\,\right)\G_{10}\vet,\nn\\ 
\theta_{\mu\nu}&:=&i e^{\phi/2}\left(\, 2\
  \bpt_{[\mu}\G_{\nu]}-\f1{2}\blt\G_{\mu\nu}\,\right)\G_{10}\vet,\nn \\ 
\theta_{\mu\nu\rho}&:=&i e^{-\phi/4}\left(\, 3\ \bpt_{[\mu}\G_{\nu\rho]}+\f1{4}\blt\G_{\mu\nu\rho}\,\right)\vet.
\ee
Note that the mass only enters in the supersymmetry variations of the fermions.
\section{On $E_{10}$ and the geodesic sigma model for $E_{10}/K(E_{10})$}
\label{sec:E10}

In this section we give some basic properties of the Kac--Moody
algebra $\mf{e}_{10}$ and explain how to construct a non-linear sigma model
for geodesic motion on the infinite-dimensional coset space
$E_{10}/K(E_{10})$. To this end we shall slice up the adjoint representation
of $\mf{e}_{10}$ in a multi-level decomposition, suitable to reveal the field
content of massive IIA supergravity~\cite{Schnakenburg:2002xx,Kleinschmidt:2003mf}. In order to incorporate also the
fermionic sector in the sigma model, we will analyse the relevant (unfaithful) 
Dirac-spinor and vector-spinor representations of $\mf{k}(\mf{e}_{10})$ up to
the desired level.\footnote{The vector-spinor representation $\Psi$ that we construct (following \cite{deBuyl:2005zy,Damour:2005zs,deBuyl:2005mt,Damour:2006xu}) is unfaithful in the sense that it is a finite-dimensional representation of the infinite-dimensional algebra $\mf{k}(\mf{e}_{10})$. However, we stress that in the spirit of the original `gradient conjecture' of \cite{Damour:2002cu} it would be more natural to regard $\Psi$ as the first component in a faithful (infinite-dimensional) representation $\hat{\Psi}:=(\Psi, \pa_a\Psi, \dots )$ of $\mf{k}(\mf{e}_{10})$, where the remaining components encode spatial gradients of the gravitino $\Psi$ \cite{Damour:2005zs,deBuyl:2005mt}. As the employed ${\bf 320}$ representation is a fully consistent representation of $K(E_{10})$, its transformations (that will be shown to be in good agreement with supergravity below) will never leave this $320$-dimensional (invariant) representation space. A reconciliation of this with the gradient conjecture is beyond the scope of this paper.} For a more detailed discussion of the general $\mf{e}_{10}$ methods employed here we
refer to the review papers \cite{Damour:2002et,Kleinschmidt:2006ad,Henneaux:2007ej}, for more
information about $\mf{k}(\mf{e}_{10})$ see
\cite{deBuyl:2005zy,Damour:2005zs,deBuyl:2005mt,Damour:2006xu},
and for a mathematical introduction to Kac-Moody algebras the canonical reference
is \cite{Kac}.  

\subsection{Generalities of the Kac-Moody algebra $\mf{e}_{10}$}
\label{sec:generalities}

\begin{figure}[t]
\centering
\includegraphics{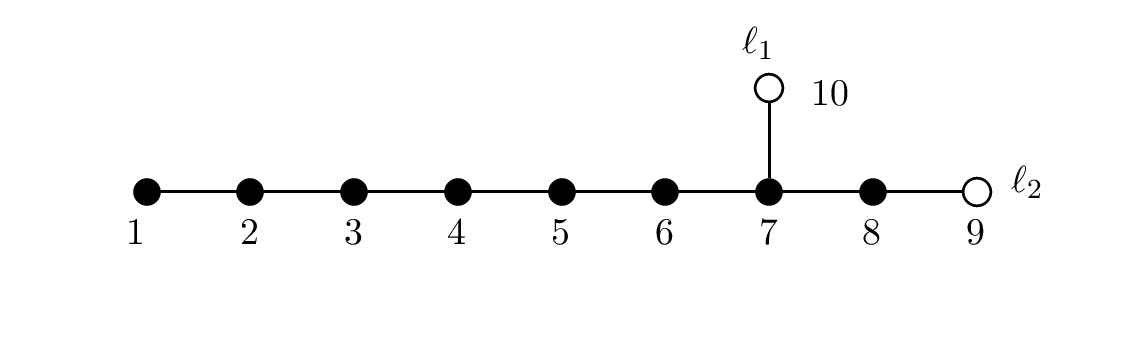}
\caption{\label{e10dynk}\sl The Dynkin diagram of $\mf{e}_{10}$ with the nodes associated with the level decomposition indicated in white.}
\label{figure:E10}
\end{figure}

Here we discuss the salient features of the hyperbolic Kac--Moody algebra $\mf{e}_{10}$, the group of which we shall denote by $E_{10}$. We will furthermore only be concerned with the split real form $\mf{e}_{10}:=\mf{e}_{10(10)}(\mbb{R})$ of the complex Lie algebra $\mf{e}_{10}(\mbb{C})$. The split real form is generated by ten triples $(e_i, f_i, h_i)$, $i=1,\dots ,10$, of Chevalley generators, each triple making up a distinguished subalgebra, 
\beq
\mf{sl}_i(2, \mbb{R})=\mbb{R}f_i \oplus \mbb{R}h_i \oplus \mbb{R}e_i \subset \mf{e}_{10}.
\eeq 
These subalgebras are intertwined inside $\mf{e}_{10}$ according to the stucture of the Dynkin diagram in Figure \ref{figure:E10}. The full structure of the algebra follows from multiple commutators of the form $[e_{i_1}, [e_{i_2}, \cdots [e_{i_{k-1}}, e_{i_k}]\cdots ]]$ (and similarly for the $f_i$'s) modulo the so-called Serre relations. We have the standard triangular decomposition
\beq
\mf{e}_{10}=\mf{n}_-\oplus \mf{h} \oplus \mf{n}_+,
\eeq
where $\mf{h}=\sum_{i}\mbb{R}h_i$ is the Cartan subalgebra and the nilpotent parts $\mf{n}_{\pm}$ are generated by the $e_i$'s and $f_i$'s, respectively. In other words, the subspace $\mf{n}_+$ contains the positive step operators, while $\mf{n}_-$ contains the negative step operators. 

The maximal compact subalgebra $\mf{k}(\mf{e}_{10})\subset \mf{e}_{10}$ is defined as the subalgebra which is pointwise fixed by the Chevalley involution $\om$,
\beq\label{ke10}
\mf{k}(\mf{e}_{10}):=\{ x\in \mf{e}_{10} \ |\  \om(x)=x\},
\eeq
where $\om$ is defined through its action on each triple $(e_i, f_i, h_i)$:
\beq
\om(e_i)=-f_i, \qquad \om(f_i)=-e_i, \qquad \om(h_i)=-h_i.
\eeq
By virtue of the existence of $\mf{k}(\mf{e}_{10})$ we have the standard Iwasawa  and Cartan decompositions (direct sums of vector spaces),
\be
\mf{e}_{10} &=& \mf{k}(\mf{e}_{10})\oplus \mf{h} \oplus \mf{n}_+,\quad \hs \text{(Iwasawa)}
\nn \\
\mf{e}_{10} &=& \mf{k}(\mf{e}_{10})\oplus \mf{p}, \qquad \qquad \text{(Cartan)},
\label{CartanIwasawa}
\ee
which will both be of importance in subsequent sections. The generators belonging to $\mf{p}$ are those which are anti-invariant under $\om$. Note that the subspace $\mf{p}$ does not close under the Lie bracket, but rather transforms in a representation of $\mf{k}$. On the other hand, the subspace $\mf{b}:=\mf{h}\oplus \mf{n}_+$ is a subalgebra of $\mf{e}_{10}$, known as a \emph{Borel subalgebra}. Using the Chevalley involution we project an arbitrary generator $x\in \mf{e}_{10}$ onto the different subspaces of  the Cartan decomposition according to
\be
\mc{Q}&:=& \f{1}{2}\big[x+\om(x)\big] \hs \in \hs   \mf{k}(\mf{e}_{10}), 
\nn \\
\mc{P}&:=& \f{1}{2}\big[x-\om(x)\big] \hs \in \hs \mf{p}.
\label{CartanProjection}
\ee

The Cartan matrix $A$ of $\mf{e}_{10}$, deduced from the Dynkin diagram in Figure \ref{figure:E10}, induces an indefinite and non-degenerate bilinear form $\left<\cdot |\cdot \right>$ on $\mf{h}$ as follows
\beq
\left<h_i|h_j\right>:=A_{ij}.
\eeq
By invariance, this bilinear form can be extended to all of $\mf{e}_{10}$, and in particular for the Chevalley generators we have
\beq
\left<e_i|f_j\right>=\delta_{ij}, \qquad \left<h_i|e_j\right>=0, \qquad \left<h_i|f_j\right>=0.
\eeq
This bilinear form will be used in subsequent sections to construct a manifestly $\mf{e}_{10}$-invariant Lagrangian.

\subsection{The IIA level decomposition of $\mf{e}_{10}$ and $\mf{k}(\mf{e}_{10})$}
\label{sectionlevelE10}
To elucidate the relation between the infinite-dimensional algebra $\mf{e}_{10}$ and the field content of massive type IIA supergravity, we shall perform a decomposition of the adjoint representation of $\mf{e}_{10}$ into representations of the finite-dimensional subalgebra $A_8\cong \mf{sl}(9, \mbb{R})$, defined by nodes $1, \dots, 8$ in the Dynkin diagram in Figure \ref{figure:E10} (see also~\cite{Kleinschmidt:2003mf}). Each generator of $\mf{e}_{10}$ is then represented as an $\mf{sl}(9, \mbb{R})$-tensor, say $X_{a_1\cdots a_k}\in \mf{e}_{10}$, where the indices are interpreted as the flat spatial indices of the ten-dimensional supergravity theory and transform in a given irreducible representation described by a set of Dynkin indices, or equivalently by a Young tableau. Since $\mf{e}_{10}$ is of rank 10 and $A_8$ is of rank 8, this decomposition is a multilevel $\ell:=(\ell_1, \ell_2)$-decomposition, with $\ell_1$ being associated with the exceptional node in the Dynkin diagram, while $\ell_2$ corresponds to the rightmost node. 

A useful alternative point of view on this decomposition is to first perform only the $\ell_1$ decomposition with respect to the horizontal $A_{9}\cong \mf{sl}(10, \mbb{R})$ subalgebra consisting of nodes $1$ through to $9$ in Figure~\ref{e10dynk}. As is well-known, at low $\ell_1$ levels, this decomposition gives rise to the field content of eleven-dimensional supergravity \cite{Damour:2002cu,Damour:2004zy}. We can then view the additional decomposition with respect to $\ell_2$ as a `dimensional reduction' from $D=11$ to $D=10$. It is this point of view which enables us to relate the mass term in massive type IIA supergravity to a generator of $\mf{e}_{10}$ at level $\ell_1=4$ in the standard `M-theory decomposition' of \cite{Damour:2002cu,Damour:2004zy,West:2004st}. This is intriguing since in $D=11$ the matching between supergravity and $\mf{e}_{10}$ has only been successful up to $\ell_1=3$. Thus, the mass term in $D=10$ provides a non-trivial check of $\mf{e}_{10}$ beyond its `$\mf{sl}(10, \mbb{R})$-covariantized $\mf{e}_8$' subset, i.e. the generators of $\mf{e}_8$ and their images under (the Weyl group of) $\mf{sl}(10, \mbb{R})$.

The relation between $\mf{e}_{10}$ and $\mf{e}_{11}$ and the mass deformation parameter of type IIA supergravity has been pointed out in various earlier references. In \cite{Schnakenburg:2002xx} the massive type IIA theory was reformulated as a non-linear realization where the mass term was associated with a certain nine-form generator of $\mf{e}_{11}$. Shortly after, in \cite{Damour:2002fz}, it was observed that the mass term in Romans' theory corresponds to a positive real root of $\mf{e}_{10}$. This was further elaborated upon in \cite{Kleinschmidt:2004dy} where a truncated version of massive type IIA supergravity in an $SO(9,9)$-covariant formulation was shown to be equivalent to the sigma model for $\mf{e}_{10}$ in a level decomposition with respect to a $D_9\cong \mf{so}(9,9)$-subalgebra. This analysis mainly focused on the bosonic sector, but a preliminary analysis of the fermionic sector was initiated by restricting to the zeroth level of the $\mf{e}_{10}$-decomposition. It was also pointed out in \cite{Kleinschmidt:2003mf,West:2004st,Henneaux:2007ej} that the mass term has a natural interpretation as a generator at level four in a decomposition of $\mf{e}_{10}$ with respect to $A_8\cong \mf{sl}(9, \mbb{R})$. From this point of view, the kinetic term in the sigma model associated with this generator naturally gives rise to the mass term of type IIA upon dimensional reduction. This is the viewpoint which we extend in the present work.

\subsubsection{Generators of $\mf{e}_{10}$}
The level decomposition of $\mf{e}_{10}$ under the $A_8\cong \mf{sl}(9,\mbb{R})$ subalgebra up to level $(\ell_1, \ell_2)=(4,1)$ is shown in Table~\ref{iiadec}. At level $(\ell_1, \ell_2)=(0,0)$ there is a copy of $\mf{gl}(9, \mbb{R})=\mf{sl}(9, \mbb{R})\oplus \mbb{R}$, as well as a scalar generator associated with the dilaton. The commutation relations at this level are ($a,b=1,\ldots, 9$)
\be
\lb K^a{}_b , K^c{}_d \rb &=& \delta^c_b K^a{}_d - \delta^a_d K^c{}_b,
\nn \\
\lb T, K^a{}_b \rb &=& 0,
\ee
and the bilinear form reads
\beq
\left<K^a{}_b| K^c{}_d\right> = \delta^a_d\delta^c_b- \delta^a_b\delta^c_d, \qquad \left<T|T\right>=\f{1}{2}, \qquad \left<T|{K^{a}}_b\right>=0.
\eeq
\begin{table}
\centering
\begin{tabular}{|c|c|c|}
\hline
$(\ell_1,\ell_2)$ & $\mf{sl}(9, \mbb{R})$ Dynkin labels & Generator of $\mf{e}_{10}$ \\
\hline
\hline
$(0,0)$ & $[1,0,0,0,0,0,0,1]\oplus[0,0,0,0,0,0,0,0]$ & $K^a{}_b$ \\
$(0,0)$ & $[0,0,0,0,0,0,0,0]$ & $T$ \\
$(0,1)$ & $[0,0,0,0,0,0,0,1]$ & $E^{a}$\\
$(1,0)$ & $[0,0,0,0,0,0,1,0]$ & $E^{a_1a_2}$\\
$(1,1)$ & $[0,0,0,0,0,1,0,0]$ & $E^{a_1a_2a_3}$\\
$(2,1)$ & $[0,0,0,1,0,0,0,0]$ & $E^{a_1\ldots a_5}$\\
$(2,2)$ & $[0,0,1,0,0,0,0,0]$ & $E^{a_1\ldots a_6}$\\
$(3,1)$ & $[0,1,0,0,0,0,0,0]$ & $E^{a_1\ldots a_7}$\\
$(3,2)$ & $[1,0,0,0,0,0,0,0]$ & $E^{a_1\ldots a_8}$\\
$(3,2)$ & $[0,1,0,0,0,0,0,1]$ & $E^{a_0|a_1\ldots a_7}$\\
$(4,1)$ & $[0,0,0,0,0,0,0,0]$ & $E^{a_1\ldots a_9}$\\
\hline
\end{tabular}
\caption{\label{iiadec}\sl IIA Level decomposition of $\mf{e}_{10}$ under $\mf{sl}(9, \mbb{R})$.}
\end{table}

All objects transform as $\mf{gl}(9, \mbb{R})$ tensors in the obvious way. The positive level generators are obtained through multiple commutators between the (fundamental) generators $E^{a}$ and $E^{ab}$ on levels $(0,1)$ and $(1,0)$, respectively. For example, the generator on level $(1,1)$ is obtained simply as the commutator
\beq
[E^{ab}, E^{c}]=E^{abc}.
\label{commutatorexample}
\eeq
All the remaining relevant commutators up to level $(4,1)$ are given in Appendix \ref{app:commutators}. No generators of $\mf{e}_{10}$ appear on mixed positive and negative levels, meaning that for any $X_{\ell}\in \mf{e}_{10}$ the levels $\ell_1$ and $\ell_2$ are either both non-positive or both non-negative. In terms of the multilevel $\ell=(\ell_1, \ell_2)$ we shall denote this by $\ell\leq 0$ or $\ell \geq 0$, respectively. This implies that the level decomposition induces a grading of $\mf{e}_{10}$ into an infinite set of finite-dimensional subspaces $\mf{g}_{\ell}$ with respect to the multilevel $\ell$. Let $E_{\ell}$ and $E_{\ell^{\prime}}$ be arbitrary root vectors in the subspaces $\mf{g}_{\ell}$ and $\mf{g}_{\ell^{\prime}}$ of $\mf{e}_{10}$. Then a generic commutator, generalizing (\ref{commutatorexample}), takes the form
\beq 
[E_{\ell}, E_{\ell^{\prime}}]=E_{\ell+\ell^{\prime}} \hs \in \hs \mf{g}_{\ell+\ell^{\prime}}.
\eeq
The negative level generators are simply obtained using the Chevalley involution, and for an arbitrary positive level generator $E_{\ell}\in \mf{g}_{\ell}$ we define the associated negative generator as
\beq
F_{\ell} :=-\om(E_{\ell}) \in \mf{g}_{-\ell}, \qquad   \ell\in \mbb{Z}^2_{\ge 0}.
\eeq
Because of the graded structure, commutators between generators in $\mf{g}_{\ell}$ and $\mf{g}_{-\ell}$ belong to the zeroth subspace $\mf{g}_0$. For example, for the fundamental generator $E^{ab}\in \mf{g}_{(0,1)}$ we have 
\beq
[E^{a_1a_2},F_{b_1b_2}]=-\frac12 \delta^{a_1a_2}_{b_1b_2} K 
    +4 \delta^{[a_1}_{[b_1} K^{a_2]}_{\,\,\,\,\, b_2]}   
    -2 \delta^{a_1a_2}_{b_1b_2} T.
\eeq
The explicit form of the remaining commutators can be found in Appendix
\ref{app:commutators}. 

\subsubsection{The `mass generator'}

Let us now discuss in more detail how the generator associated with the mass term appears in the level decomposition of $\mf{e}_{10}$. First consider the decomposition with respect to $\ell_1$ (see Appendix \ref{app:level4}). At level $\ell_1=4$ one finds a generator corresponding to an $\mf{sl}(10, \mbb{R})$-tensor with 12 indices of the form $E^{\dot{a}|\dot{b}|\dot{c}_1\cdots \dot{c}_{10}}$ (with $\dot{a}=(10, a),$ etc.), which has mixed Young symmetry, i.e. it is antisymmetric in the block of indices $\dot{c}_1, \dots, \dot{c}_{10}$ and symmetric in $\dot a, \dot b$. This generator has no physical interpretation in $D=11$ supergravity. However, consider now the further level decomposition with respect to $\ell_2$. This can be realized by doing a dimensional reduction on $E^{\dot{a}|\dot{b}|\dot{c}_1\cdots \dot{c}_{10}}$ along the direction `10'. We are interested in the generator obtained in this way by fixing the first three indices to $\dot{a}=\dot{b}=\dot{c}_1=10$, which yields \cite{West:2004st,Henneaux:2007ej}
\beq\label{massgen}
E^{a_1\cdots a_{9}}:= \f18E^{10|10|10 a_1\cdots a_{9}}.
\eeq
The resulting tensor corresponds to the $(\ell_1, \ell_2)=(4,1)$-generator $E^{a_1\cdots a_9}$ in Table \ref{iiadec}. This $\mf{sl}(9, \mbb{R})$-tensor is a nine-form, and is therefore associated with a supergravity potential $A_{a_1\cdots a_9}$ whose field-strength is a top-form in $D=10$. This is the generator associated with the mass term. From this analysis it is clear that the mass deformation of type IIA supergravity probes $\mf{e}_{10}$ beyond the realm of what has been successfully verified previously in the context of $D=11$ supergravity. This is especially interesting due to the fact that the $\ell_1=4$-generator $E^{\dot{a}|\dot{b}|\dot{c}_1\cdots \dot{c}_{10}}$ is a `genuine' $\mf{e}_{10}$ element, with no components contained in $\mf{e}_8$ nor in $\mf{e}_9$.

\subsubsection{Generators of $\mf{k}(\mf{e}_{10})$}

The level decomposition of $\mf{e}_{10}$ induces a decomposition of its maximal compact subalgebra $\mf{k}(\mf{e}_{10})$ which was defined in (\ref{ke10}). Using the Cartan decomposition, (\ref{CartanIwasawa}) and (\ref{CartanProjection}), we define compact and noncompact generators, $J_{\ell}\in \mf{k}(\mf{e}_{10})$ and $S_{\ell}\in \mf{p}$, as follows
\beq
J_{\ell}:= E_{\ell}-F_{\ell}, \qquad S_{\ell}:= E_{\ell}+F_{\ell}, \qquad \ell \in \mbb{Z}_{\ge 0}^2\backslash \{(0,0)\}.
\label{CompactAndNoncompactGenerators}
\eeq
Furthermore, at level $(0,0)$ we have the $\mf{so}(9)$ Lorentz generators
\beq\label{00gens}
M^{ab}:={K^a}_b-{K^b}_a.
\eeq
This decomposition of $\mf{k}(\mf{e}_{10})$ is not a gradation, but rather corresponds to a \emph{filtration} \cite{Damour:2006xu}, in the sense that an arbitrary commutator between two `positive level' generators exhibits a graded structure modulo lower-level generators only,
\beq
[J_{\ell}, J_{\ell^{\prime}}]=J_{|\ell-\ell^{\prime}|}+J_{\ell+\ell^{\prime}},
\eeq 
where it is understood that $J_{|\ell-\ell^{\prime}|}\neq 0$ if and only if $(\ell-\ell^{\prime})\geq 0$ or $(\ell-\ell^{\prime})\leq 0$. In other words, $J_{|\ell-\ell^{\prime}|}$ is non-zero only when the difference $(\ell-\ell^{\prime})$ involves no mixing between negative and positive levels. 

Using (\ref{CompactAndNoncompactGenerators}), together with the $\mf{e}_{10}$ commutators in Appendix \ref{app:commutators}, we may deduce the abstract $\mf{k}(\mf{e}_{10})$-relations at each `level'. Let us consider a few examples to illustrate the procedure. We begin by defining the $\mf{k}(\mf{e}_{10})$-generators associated with level $(0,0)$ and the fundamental generators at level $(0,1)$ and $(1,0)$:
\beq
 J^{a}:= E^{a}-F_a, \qquad J^{ab}:=E^{ab}-F_{ab}.
\eeq 
Since there are no generators of mixed level $(1,-1)$ the commutator $[E^{a}, F_b]$ vanishes, and the commutator between $J^{ab}$ and $J^{c}$ simply gives
\beq
[J^{ab}, J^{c}]=J^{abc},
\eeq
with $J^{abc}:=E^{abc}-F_{abc}$. Proceeding in the same way for $J^{a_1 a_2}$ and $J^{a_1a_2a_3}$, we obtain $J^{a_1\cdots a_5}$ modulo lower level terms, 
\beq
\lb J^{a_1a_2}, J^{a_3a_4a_5}\rb =J^{a_1\cdots a_5}-6 \delta_{a_1a_2}^{[a_3a_4}J^{a_5]}.
\eeq
Note that one may project onto $J^{a_1\cdots a_5}$ as follows,
\beq
J^{a_1\cdots a_5}= \lb J^{[a_1a_2}, J^{a_3a_4a_5]}\rb\,.
\eeq
Here, all indices are $\mf{so}(9)$ vector indices and are raised and lowered with the Euclidean $\delta_{ab}$ so that the position does not matter.
The other relevant $\mf{k}(\mf{e}_{10})$-commutators are listed in Appendix \ref{app:SpinorReps}. 

\subsection{Spinorial representations}

The fermionic degrees of freedom in the sigma model for $E_{10}/K(E_{10})$ transform in spinorial representations of $\mf{k}(\mf{e}_{10})$. The two relevant $\mf{k}(\mf{e}_{10})$ representations are finite-dimensional (unfaithful) and of dimensions $32$ and $320$, respectively.
In the decomposition of $\mf{k}(\mf{e}_{10})$ associated with eleven-dimensional supergravity these representations transform as a $32$-dimensional Dirac spinor representation $\epsilon$ of $\mf{so}(10)\subset \mf{k}(\mf{e}_{10})$ and a $320$-dimensional vector-spinor representation  $\Psi_{\dot{a}}, \hs \dot{a}=(10, a)$, of $\mf{so}(10)\subset \mf{k}(\mf{e}_{10})$, identified with the gravitino \cite{deBuyl:2005zy,Damour:2005zs,deBuyl:2005mt,Damour:2006xu}. Upon reduction to the IIA theory (through the additional level decomposition with respect to $\ell_2$) the gravitino decomposes into a 32-dimensional spinor $\Psi_{10}$ and a 288-dimensional vector spinor $\Psi_a$ of $\mf{so}(9)$. However, because they both descend from $\Psi_{\dot{a}}$, these two representations will mix under $\mf{k}(\mf{e}_{10})$~\cite{Kleinschmidt:2006tm}. No such complication arises for the supersymmetry parameter $\epsilon$, for which we keep the same notation in the IIA picture. The spinor $\Psi_{10}$ will be associated with the ten-dimensional dilatino, while the vector-spinor $\Psi_a$ is related to the gravitino. 

For the first three levels, the transformation properties of the spinor $\epsilon$ are
\be\label{vards}
M^{ab}\cdot \epsilon &=& \f{1}{2}\Gamma^{ab}\epsilon,
\nn \\
J^{a}\cdot \epsilon &=& \f{1}{2}\Gamma_{10}\Gamma^{a}\epsilon,
\nn \\
J^{ab}\cdot \epsilon &=& \f{1}{2}\Gamma_{10}\Gamma^{ab}\epsilon,
\ee
where $\Gamma^{a}$ and $\Gamma^{ab}$ are $\mf{so}(9)$ $\Gamma$-matrices (our $\Gamma$-matrix conventions are given in Appendix \ref{app:conventions}). The higher level transformations are now defined through the abstract $\mf{k}(\mf{e}_{10})$-relations; for example:
\be
J^{a_1a_2a_3}\cdot \epsilon &:=& \Big[J^{a_1a_2}, J^{a_3}\Big]\cdot \epsilon \qquad =\f{1}{2}\Gamma^{a_1a_2a_3}\epsilon,
\nn \\
J^{a_1\cdots a_5}\cdot \epsilon &:=& \Big[J^{[a_1a_2}, J^{a_3a_4a_5]}\Big] \cdot \epsilon = \f{1}{2}\Gamma_{10}\Gamma^{a_1\cdots a_5}\epsilon.
\ee
Similarly, we have for the spinor component $\Psi_{10}$ the following low-level transformations:
\be\label{varpsi10}
M^{a_1a_2}\cdot  \Psi_{10} &=&\f12\G^{a_1a_2}\Psi_{10}\nn\\
J^{a}\cdot \Psi_{10} &=& \f12 \G_{10}\G^a\Psi_{10}+\Psi^a,\nn\\
J^{a_1a_2}\cdot \Psi_{10} &=& \f16 \G_{10}\G^{a_1a_2}\Psi_{10}+\f43 \G^{[a_1}\Psi^{a_2]},
\ee
showing explicitly the mixing between $\Psi_{10}$ and $\Psi_{a}$ under $\mf{k}(\mf{e}_{10})$. Finally, we also have for $\Psi_a$:
\be\label{varpsia}
M^{a_1a_2}\cdot  \Psi_{b} &=& \f12\G^{a_1a_2}\Psi_{b}+2\delta_b^{[a_1}\Psi^{a_2]},\nn\\
J^{a}\cdot \Psi_{b}&=&\f12 \G_{10}\G^a\Psi_{b}-\delta^a_b\Psi_{10},\nn\\
J^{a_1a_2}\cdot \Psi_{b} &=& \f12\G_{10}\G^{a_1a_2}\Psi_b -\f43\G_{10}\delta^{[a_1}_b\Psi^{a_2]}+\f23\G_{10}\G_b^{\ [a_1}\Psi^{a_2]}\nn\\
&&+\f43\delta_b^{[a_1}\G^{a_2]}\Psi_{10}-\f13\G_b^{\ a_1a_2}\Psi_{10}.
\ee
More details on these $\mf{k}(\mf{e}_{10})$-representations can be found in Appendix \ref{app:SpinorReps}. We note that we could also have redefined the $SO(9)$ spinor $\Psi_a$ by a shift with $\G_a\G^{10}\Psi_{10}$ as one does in Kaluza--Klein reduction (cf. (\ref{kkferm})) but refrain from doing so here.

\subsection{The non-linear sigma model for $E_{10}/K(E_{10})$}
\label{sec:sigmamod}

A non-linear sigma model with rigid $E_{10}$-invariance and local $K(E_{10})$-invariance may now be constructed using the properties of $\mf{e}_{10}$ described in previous sections. By virtue of the Iwasawa decomposition we can always choose a coset representative in the partial `Borel gauge' by taking 
\beq
\cV(t) :=\mc{V}= \mc{V}_0  e^{\phi T} e^{A\star E}  \hs \in \hs E_{10}/K(E_{10}),
\eeq
where $\mc{V}_0=\text{exp}\ {{h^{a}}_b{K^{b}}_a}$ represents the $GL(9,
\mbb{R})$ inverse vielbein ${e_a}^m$, while $e^{A\star E}$ contains the
positive step operators of $\mf{e}_{10}$. We call this a partial Borel gauge
since $\mc{V}_0$ is not constrained to contain only positive step operators of
$\mf{gl}(9,\mbb{R})$. In other words, $\mc{V}_0$ is an arbitrary $GL(9,
\mbb{R})$-matrix and not a representative of the coset $GL(9,
\mbb{R})/SO(9)$. With some abuse of terminology, we shall sometimes refer to
the part of $E_{10}$ parametrized by $\mc{V}$ as the `Borel subgroup', and
denote it by $E_{10}^{+}$ \footnote{This differs from the `true' Borel subgroup $B:=
\text{exp}\ \mf{b}\subset E_{10}$ through the negative root generators in
$\mc{V}_0$ (see Section \ref{sec:generalities} for the definition of
$\mf{b}$). $E_{10}^{+}$ is only a parabolic subgroup of $E_{10}$.}. The positive
level part $ e^{A \star E}$ of $\mc{V}$ is defined as  
\beq\label{posexp}
 e^{A \star E}:=\text{exp}\big[A_{m}(t)E^m]\text{exp}\big[\tfrac{1}{2} A_{m_1m_2}(t) E^{m_1m_2}]\text{exp}\big[\tfrac{1}{3!} A_{m_1m_2m_3}(t) E^{m_1m_2m_3}\big]\cdots 
\eeq
with similar exponentials occurring for the higher levels. Note that in this expression the indices $m_1, m_2, \dots$ are $\mf{sl}(9, \mbb{R})$-indices, and hence correspond to curved spatial indices from a supergravity  point of view. 

The coset representative $\mc{V}$ transforms under global $g\in E_{10}$-transformations from the right and local $k\in K(E_{10})$-transformations from the left:
\beq
\mc{V} \longmapsto k\mc{V}g.
\label{transformationcoset}
\eeq
From $\mc{V}$ we construct the Lie algebra-valued Maurer--Cartan form 
\beq\label{mcform}
\p_t \cV \cV^{-1} = \cP +\cQ,
\eeq
where we also employed the Cartan decomposition, according to (\ref{CartanProjection}). The transformation property of $\mc{V}$ in (\ref{transformationcoset}) implies that the coset part $\mc{P}\in \mf{p}$ of the Maurer-Cartan form is globally $E_{10}$-invariant while transforms covariantly under $K(E_{10})$:
\beq
K(E_{10}) \hs : \hs \mc{P}\hs  \longmapsto\hs  k\mc{P}k^{-1}.
\eeq
On the other hand, $\mc{Q}\in \mf{k}(\mf{e}_{10})$ properly transforms as a connection,
\beq
K(E_{10}) \hs : \hs \mc{Q}\hs \longmapsto\hs  k\mc{Q}k^{-1} +\pa_t k k^{-1}.
\eeq

Using the bilinear form on $\mf{e}_{10}$, we can now construct a manifestly $E_{10}\times K(E_{10})_{\text{local}}$-invariant Lagrangian as follows \cite{Damour:2002cu,Damour:2004zy}
\beq\label{e10boslag}
\mc{L}^{[B]}_{E_{10}/K(E_{10})}=\f{1}{4}n(t) ^{-1}\left<\mc{P}\big|\mc{P}\right>,
\eeq
where the lapse function $n(t)$ ensures invariance under reparametrizations of the geodesic parameter $t$. We have also included a superscript $[B]$ to emphasize that this Lagrangian is only the bosonic part of a sigma model which also includes the fermionic degrees of freedom in the ${\bf 320}$ of $K(E_{10})$ to be introduced shortly. The equations of motion for $n(t)$ enforces a lightlike (Hamiltonian) constraint on the dynamics,
\beq 
\left<\mc{P}\big|\mc{P}\right>=0,
\eeq
while the equations of motion for $\mc{P}$ (in the gauge $n=1$) read
\beq
\mc{D}\mc{P}:= \pa_t \mc{P}-[\mc{Q}, \mc{P}]=0,
\label{bosoniceom}
\eeq
where we defined the $K(E_{10})$-covariant derivative $\mc{D}$. Equation (\ref{bosoniceom}) encodes the dynamics of the bosonic sector of the sigma model and is written out in detail in Appendix~\ref{Appendix:EOMBosonic}.

The fermionic degrees of freedom are included in the Lagrangian through the spinor representation $\Psi$ as follows \cite{Damour:2005zs,deBuyl:2005mt,Damour:2006xu}
\beq\label{e10fermlag}
\mc{L}^{[F]}_{E_{10}/K(E_{10})}=-\f{i}{2}\left<\Psi \big| \mc{D} \Psi\right>,
\eeq
where the bracket now denotes an invariant inner product on the representation space. The associated `Dirac equation' reads
\beq
\mc{D}\Psi:= \pa_t\Psi-\mc{Q}\cdot \Psi=0,
\label{fermioniceom}
\eeq
where it is understood that the connection $\mc{Q}$ acts on $\Psi$ in the vector-spinor representation constructed in the previous section. Equation (\ref{fermioniceom}) is written in terms of the full $320$-dimensional spinor $\Psi$, which encodes both the `gravitino' $\Psi_a$ and the `dilatino' $\Psi_{10}$. The separate equations of motion for these fields can be written out as follows
\be
\mc{D}\Psi_a&:=& \pa_t \Psi_{a\phantom{0}}-\mc{Q}\cdot \Psi_{a\phantom{0}}=0,
\nn \\
\mc{D}\Psi_{10}&:=& \pa_t\Psi_{10}-\mc{Q}\cdot \Psi_{10}=0.
\label{GravitinoDilatinoEOM}
\ee
Recall that the $\mf{k}(\mf{e}_{10})$-action on $\Psi_a$ contains $\Psi_{10}$-terms, and vice versa. This is important since the same mixing between the gravitino and the dilatino occurs in the corresponding supergravity equations of motion (see Appendix~\ref{app:sugraf}). The fermionic equations of motion in (\ref{GravitinoDilatinoEOM}) are written out in more detail in Appendix \ref{app:FermionicSigmaModelEOM}.

The bosonic equations of motion (\ref{bosoniceom}) were written for the gauge choice $n=1$. The lapse function $n$ has a superpartner $\Psi_t$, which is a Dirac spinor under $\mf{k}(\mf{e}_{10})$, and the associated supersymmetry transformations are
\be \label{SusyTransfSigmaModel}
\delta_{\epsilon} n &=& i\epsilon^T\Psi_t,\nn \\
\delta_{\epsilon} \Psi_t &=& \mc{D}\epsilon.
\ee
The fermionic equations of motion are then valid in the `supersymmetric gauge' $\Psi_t=0$. The associated constraint (analogous to the Hamiltonian constraint) that should be imposed is the supersymmetry constraint which states that the spin $\Psi$ is orthogonal to the velocity $\mc{P}$. The full form of this constraint is unknown due to the fact that it is not known how to supersymmetrize the $E_{10}$ model correctly. Nevertheless, low level expressions have been obtained in~\cite{Damour:2005zs,Damour:2006xu}, which the reader should also consult for further discussions of this point. Certain aspects of the supersymmetry constraint are also discussed in  Section \ref{constraints}.

Let us now analyse these properties of the geodesic sigma model in more detail by utilizing the level decomposition of $\mf{e}_{10}$ with respect to $\mf{sl}(9, \mbb{R})$. Expanding the coset element $\mc{P}$ up to level $(3,2)$ we obtain 
\be
\cP &=& \p_t\phi T + \frac12 p_{ab} (K^a{}_b+K^b{}_a) +  \sum_{\ell> 0}P_{\ell}\star S_{\ell}   \label{Pexpanded}
\ee
and the associated expansion for $\mc{Q}$ is given by
\beq
\mc{Q}=\frac12 q_{ab}M^{ab}+\sum_{\ell > 0}Q_{\ell}\star J_{\ell},
\eeq
where $\star$ schematically denotes the coupling between the dynamical fields $P_{\ell}$ and the associated generators $S_{\ell}$. The explicit form of this expansion will be given below. In  Borel gauge, the fields at non-zero levels are the same in $\mc{Q}$ and $\mc{P}$:
\beq
Q_{\ell}\equiv P_{\ell}, \qquad \ell \in \mathbb{Z}^2_{\ge 0}\backslash \{(0,0)\}\,.
\eeq
The explicit form of the higher level terms in the expansion of $\mc{P}$ reads
\be\label{cmform}
\sum_{\ell> 0}P_{\ell}\star S_{\ell}&:=&e^{3\phi/4} P_{a_1} S^{a_1} 
   + \frac12 e^{-\phi/2} P_{a_1a_2} S^{a_1a_2}
   + \frac1{3!} e^{\phi/4} P_{a_1a_2a_3} S^{a_1a_2a_3}\nn\\
&&   + \frac1{5!} e^{-\phi/4} P_{a_1\ldots a_5} S^{a_1\ldots a_5}
   + \frac1{6!} e^{\phi/2} P_{a_1\ldots a_6} S^{a_1\ldots a_6}
   + \frac1{7!} e^{-3\phi/4} P_{a_1\ldots a_7} S^{a_1\ldots a_7}\nn\\
&& + \frac1{8!} P_{a_1\ldots a_8}S^{a_1\ldots a_8} 
   +\frac1{8!}P_{a_0|a_1\ldots a_7}S^{a_0|a_1\ldots a_7} 
   + \frac1{9!} e^{-5\phi/4} P_{a_1\ldots a_9} S^{a_1\ldots a_9}+\ldots
   \nn \\
   \ee
with a similar expression for $\mc{Q}$ with the $S_{\ell}$'s replaced by the corresponding $J_{\ell}$'s.   
   
{}From a given parametrization of $\mc{V}$ as in (\ref{posexp}) one can work out explicit expressions for $\cP$ in terms of the `potentials' $A_{\ell}$ (which {\em a fortiori} carries flat indices as it transforms under $\mf{k}(\mf{e}_{10})$) which appear in the construction of $\mc{V}$. For example, one finds
\be 
P_a = \frac12 e_a{}^m\partial_t A_m =\frac12 e_a{}^m DA_m
\ee
in terms of the `covariant derivatives' of \cite{Damour:2002cu}. Here, we have written $\mc{V}_0=e_a{}^m$ as an inverse $GL(9, \mbb{R})$ vielbein. We do not require the exact expressions for the higher level components of $\mc{P}$ for establishing the correspondence and their expression will  be more complicated due to the appearance of additional terms, arising when expanding the Maurer-Cartan form $\pa_t\mc{V}\mc{V}^{-1}$ using the Baker--Campbell--Hausdorff formula $d e^X e^{-X} = dX + \frac12 [X,dX] +\ldots$. In Section \ref{sec:comments} when we discuss constraints and aspects of gauge-fixing the precise form will be more relevant and we will give more details there.

The geodesic equations $\p_t\cP =\lb\cQ,\cP\rb$ can be written conveniently by treating the $\cV_0$ contribution separately in a partially covariant derivative $\cD$, see~\cite{Damour:2004zy}. For example, for the fundamental generators at level $(0,1)$ and $(1,0)$ the equations of motion become (for $n=1$)
\be
\cD (e^{3\phi/2} P_a) &=&
    -e^{\phi/2} P_{ac_1c_2} P_{c_1c_2}   + \frac2{5!} e^\phi P_{ac_1\ldots c_5}P_{c_1\ldots c_5}     +\frac{12}{8!} P_{ac_1\ldots c_7}P_{c_1\ldots c_7} \nn\\
 & &      +\frac{1}{4\cdot 7!}P_{a|c_1\ldots c_7}P_{c_1\ldots c_7}
\label{eompa} \\
\cD (e^{-\phi} P_{a_1a_2}) &=& 
    2 e^{\phi/2}  P_{a_1a_2c}P_c +\frac13 e^{-\phi/2} P_{a_1a_2c_1c_2c_3}P_{c_1c_2c_3}\nn\\
 & &  +\frac2{5!}e^{-3\phi/2} P_{a_1a_2c_1\ldots c_5}P_{c_1\ldots c_5} 
     +\frac2{7!} e^{-5\phi/2}P_{a_1a_2c_1\ldots c_7}P_{c_1\ldots c_7} \nn\\
 & & +\frac1{6!}P_{a_1a_2c_1\ldots c_6}P_{c_1\ldots c_6} + \frac{1}{4\cdot 5!} P_{c_1|c_2\ldots c_6a_1a_2}P_{c_1\ldots c_6}.
\ee
In Section \ref{sec:correspondence} we will show that these equations are equivalent to the equations of motion for the electric fields $F_{ta}$ and $F_{tab}$, respectively. The equations for all the remaining levels are given in Appendix \ref{Appendix:EOMBosonic}. The  fermionic equations of motion can be found in~\ref{app:FermionicSigmaModelEOM} where the connection term $\cQ$ is evaluated in the vector-spinor representation. The supersymmetry variation (\ref{SusyTransfSigmaModel}) requires the evaluation of $\cQ$ in the Dirac-spinor representation, the resulting expression can be found in~\ref{app:susyvarcoset}.

\section{The correspondence}
\label{sec:correspondence}

In this section we make the $E_{10}/$massive IIA correspondence explicit by giving a dictionary between the dynamical variables of the $E_{10}/K(E_{10})$-sigma model and the fields of massive IIA supergravity. The comparison cannot be done at the level of the respective Lagrangians since the $E_{10}$ sigma model naturally incorporates kinetic terms for all fields as well as their duals, which is not the case for the IIA Lagrangian. The matching is rather done at the level of the equations of motion, where we will see that bosonic equations of motion and Bianchi identities on the supergravity side all become associated with geodesic equations of motion on the $E_{10}$ side. Similarly, the fermionic supergravity equations will be associated with the Dirac equation of the spinning coset particle.

\subsection{Bosonic equations of motion and truncation}
\label{EOMandTruncations}
To be able to compare the equations of motion on the supergravity side
(\ref{eoms}) and (\ref{lev0}) to the ones on the sigma model side (\ref{bosoniceom}), spelt out in Appendix \ref{Appendix:EOMBosonic},
we need to rewrite the former. As is customary in the correspondence between $E_{10}$ and supergravity we split the
indices $\al=(0,a)$ into temporal and spatial indices and also adopt
a pseudo-Gaussian gauge for the ten-dimensional vielbein:
\beq
\label{10v}
{e_{\mu}}^{\al}=\left(\begin{array}{cc}
N & 0 \\
0 & {e_m}^{a} \\
\end{array}\right)\,.
\eeq
In addition we demand that the spatial trace of the spin connection vanishes (see Appendix~\ref{app:conventions} for our conventions) 
\be\label{spintrace}
\omega_{a\,ab}=0 \quad\Rightarrow\quad \Omega_{ba\,a}=0
\ee 
and write the tracefree spin connection as $\tilde{\omega}_{a\,bc}$ as a reminder. We also choose temporal gauges for all supergravity gauge
potentials,
\be\label{tempgauge}
A_{t} = 0\,,\quad A_{tm} = 0\,,\quad A_{tm_1m_2} =0\,.
\ee

Moreover, we can only expect that a truncated version of the supergravity
equation corresponds to the coset model equations. 
This truncation was originally devised in the context of eleven-dimensional
supergravity, where it was strongly motivated by the billiard analysis of the
theory close to a spacelike singularity (the `BKL-limit')
\cite{Damour:2002cu,Damour:2004zy}. In this limit, spatial points decouple and
the dynamics becomes effectively time-dependent, ensuring that the truncation
is a valid one in this regime. In this paper, we analyse the same question in
the context of massive IIA supergravity, and an identical procedure requires
the truncation of a set of spatial gradients. These can be obtained from a
BKL-type analysis of massive IIA and their full list is presented in
Appendix~\ref{truncation}. Notice that except for the expression involving the
mass, all the spatial gradients to be truncated away can be obtained by
dimensional reduction of the eleven-dimensional truncation. 

Let us illustrate the implications of this truncation on the supergravity
equations in some detail for an explicit example. The following truncation of
massive IIA supergravity can be deduced from the billiard analysis: 
\beq
\label{trunc}
\p_a\left(N e^{3\phi/4} F_{b_1b_2}\right)=0.
\eeq
The effect of this truncation appears in the equation of motion for the
two-form field strength $F_{\alpha\beta}$ (see (\ref{eom1}) in Appendix
\ref{SugraBeom}). After splitting time and space indices, the space component
of this equation reads\footnote{$D^{(0)}$ here contains the contributions from
the time derivatives of the spatial vielbein $e_m{}^a$ and will be identified
with the corresponding $E_{10}$ coset derivative operator below.} 
\be
D^{(0)}(e^{3\phi/2}F_{tb})&=&-\frac12e^{\phi/2}F_{ta_1a_2b}F_{ta_1a_2}
+\frac1{3!}e^{\phi/2}N^{2}F_{a_1\ldots a_3b}F_{a_1\ldots a_3}\nn\\
&&+\frac34e^{3\phi/2}N^2\p_a\phi F_{ab}
-\frac12e^{3\phi/2}N^2\Omega_{a_1a_2b}F_{a_1a_2}\nn\\
&&+Ne^{3\phi/4} \p_a\left(N e^{3\phi/4} F_{ab}\right)\,.
\label{electricequation}
\ee
Using the truncation (\ref{trunc}), the last term on the right hand side
vanishes, and consequently the equation to compare with the $E_{10}$ equation
of motion (\ref{eompa}) is (\ref{electricequation}) without the
bottom line. Let us emphasize that the truncation (\ref{trunc}) also follows
from dimensional reduction of the associated truncation in the equation of
motion for the electric field $F_{0\dot{a}_1\dot{a}_2\dot{a}_3}$ in eleven
dimensions \cite{Damour:2002cu,Damour:2004zy}. Moreover, this example makes
clear the important comment that the truncation we impose is \emph{not}
equivalent to discarding all spatial gradients on the supergravity side, since
it is clear that spatial derivatives of the one-form potential $A_a$ are
implicitly contained in $F_{a_1a_2}$. 

From comparing in detail the supergravity equations (Appendix
\ref{Appendix:mIIAdetails}) with the truncations applied with the bosonic
equations of the coset model (Appendix \ref{Appendix:EOMBosonic}) one can
derive a correspondence between the components of the coset velocity $\cP$ and
the fields of supergravity. The dictionary is given in Table \ref{dicoeom}
where, in the last line,  $e=\text{det}\ {e_m}^{a}=\sqrt{\text{det}\
  g_{mn}}=\sqrt{g}$. It is perfectly consistent with identifying the form
equations of motion (\ref{eoms}) with the sigma model expressions
(\ref{boseom1}) to (\ref{boseom3}), the Bianchi identities (\ref{Bianchis})
with the equations (\ref{Bianchi1}) to (\ref{Bianchi3}) and the dilaton
equation of motion (\ref{dilsugra}) with (\ref{dilatoneom}). 

\begin{table}[t]
\centering
\begin{tabular}{|c|c|c|}
\hline
$(\ell_1,\ell_2)$&$E_{10}$ fields& Bosonic fields of supergravity\\
\hline
\hline
$(0,0)$&$p_{ab}$ & $-N\om_{a\,b0}$\\[.2cm]
$(0,0)$&$q_{a_1a_2}$ & $-N\om_{0\,a_1a_2}$\\[.2cm]
$(0,1)$&$P_a$ & $\f12 N F_{0a}$\\[.2cm]
$(1,0)$&$P_{a_1a_2}$ & $\f{1}{2}N F_{0a_1a_2}$\\[.2cm]
$(1,1)$&$P_{a_1a_2a_3}$ & $\f{1}{2}N F_{0a_1a_2a_3}$ \\[.2cm]
$(2,1)$&$P_{a_1\cdots a_5}$ & $\f{1}{2\cdot 4!}Ne^{\phi/2}{\epsilon_{a_1\cdots
    a_5}}^{b_1b_2b_3b_4}F_{b_1b_2b_3b_4}$ \\[.2cm]
$(2,2)$&$P_{a_1\cdots a_6}$ & $\f{1}{2\cdot 3!}N e^{-\phi}{\epsilon_{a_1\cdots
    a_6}}^{b_1b_2b_3}F_{b_1b_2b_3}$ \\[.2cm]
$(3,1)$&$P_{a_1\cdots a_7}$ & $-\f{1}{2\cdot 2!}Ne^{3\phi/2}{\epsilon_{a_1\cdots
    a_7}}^{b_1b_2}F_{b_1b_2}$ \\[.2cm]
$(3,2)$&$P_{a_1\cdots a_8}$ & $-\frac12N{\epsilon_{a_1\cdots a_8}}^{b}\p_{b}\phi$\\[.2cm]
$(3,2)$&$P_{a_0|a_1\cdots a_7}$ & $2N{\epsilon_{a_1\cdots a_7}}^{b_1b_2}\tilde{\Omega}_{b_1b_2\,a_0}$ \\[.2cm]
$(4,1)$&$P_{a_1\cdots a_9}$ & $\frac12 N e^{5\phi/2}\eps_{a_1\ldots a_9}m$\\[.2cm]
\hline
\hline
-&$n$&$Ne^{-1}$\\
\hline
\end{tabular}
\caption{\label{dicoeom}\sl Bosonic dictionary: This table shows the correspondence between $E_{10}$ fields and  bosonic fields of massive IIA supergravity that one obtains by considering the equations of motion and the Bianchi equations. }
\end{table}

The remaining equation, the Einstein equation (\ref{Einsteinsugra}), does not fit perfectly in this picture. More precisely, two terms do not match completely with (\ref{Einstein}). One is similar to the mismatch in
the $A_9$ decomposition relevant to $D=11$ supergravity and is a contribution
to the Ricci tensor $R_{ab}$ of the form $\Omega_{a\,cd}\Omega_{b\,dc}$ 
\cite{Damour:2004zy}. The other term is a mismatch of the coefficient in the
energy momentum tensor of the dilaton $T_{ab}\sim \p_a\phi \p_b\phi$; the
coefficient is off by a factor two. This can be traced back to $D=11$ where
both mismatches were part of the $D=11$ Ricci tensor. In this sense this is
not a new discrepancy but a known one. It is to be noted that all the terms
involved in the mismatch are related to contributions to the Lagrangian which
would give rise to walls corresponding to imaginary roots in the cosmological
billiards picture \cite{Damour:2002cu}. There is no mismatch in the equation
of motion of the dilaton $\phi$ since in the reduction the missing term in
$D=11$ does not contribute to this equation. 

Let us study the effect of the mass term more closely. In the bosonic
equations, the mass appears in five places: the dilaton equation
(\ref{dilsugra}), the Einstein equation (\ref{Einsteinsugra}), the equation of
motion for $F_{ta_1a_2}$ (\ref{eom2}), the Bianchi equation for $F_{a_1a_2}$
(\ref{SugraBianchi1}) and of course in its own equation of motion
$\p_tm=0$. It is remarkable that the contribution of the real root
corresponding to the mass deformation enters all equivalent sigma-model
equations correctly, even though it is beyond the realm of $E_8$ generators
and above height $30$. In particular, the $E_{10}$-invariant sigma model
produces the right potential for the scalar $\phi$. This is possible since the
supergravity potential is positive definite in agreement with positive
definiteness of the $E_{10}$-invariant Lagrangian (\ref{e10boslag}) away from
the Cartan subalgebra. By contrast, for gauged deformations in lower
dimensions where the supergravity potential is
indefinite and not reproduced fully by $E_{10}$~\cite{Bergshoeff:2008}. From
this point of view the massive Romans theory seems to be special since the $E_{10}$ model reproduces correctly all
the effects of the deformation. 

\subsection{The truncation revisited} 

The truncation we applied to supergravity, using billiard arguments, also
proves useful for ensuring the consistency of the dictionary. Indeed,
notice that all the sigma model variables depend on (sigma model) `time' only,
and hence we must demand that their spatial gradients vanish: 
\beq
 \pa_a\mc{P}(t)=0. 
 \label{GeneralTruncation}
 \eeq  
Applied to the Maurer--Cartan expansion, this equation translates to
constraints on the supergravity variables upon using the dictionary in
Table~\ref{dicoeom}. Of
course, (\ref{GeneralTruncation}) does not really make sense on the sigma
model side, where spatial gradients have no meaning, but must rather be
understood as a convenient way of encoding the relevant truncations on the
supergravity side. Nevertheless, the truncations encoded in
(\ref{GeneralTruncation}) correspond precisely to the truncations we have
just imposed.  Let us illustrate this for the example of the truncation on 
the magnetic field in (\ref{electricequation}).   
The level $(3,1)$ part of the expansion of the Maurer-Cartan form
contains the following term (see (\ref{cmform}))  
\beq 
\mc{P}(t)\big|_{(3,1)}=\f{1}{7!}e^{-3\phi/4}P_{a_1\cdots a_7}(t)S^{a_1\cdots a_7}.
\eeq 
The coefficient of the generator $S^{a_1\cdots a_7}\in \mf{e}_{10}$ is a dynamical quantity which is purely time-dependent. This implies that on the supergravity side we must make sure that the corresponding quantity is also purely time-dependent, i.e. we must demand 
\beq 
\pa_a(e^{-3\phi/4}P_{a_1\cdots a_7})=0,
\eeq  
which, after using the dictionary in Table \ref{dicoeom}, precisely yields the truncation in (\ref{trunc}). Similarly, one sees that requiring (\ref{GeneralTruncation}) for each term of the Maurer-Cartan expansion corresponds to the truncations of Appendix \ref{truncation}.

\subsection{Fermionic equations of motion} 
\label{feom} 
To make contact between the fermionic equations of Romans' theory, (equations of motion (\ref{eomlam}) and (\ref{eompsi}) and supersymmetry variation (\ref{deltapsi})), and the Kac-Moody side of the story, (equations of motion (\ref{eompsi10}) and (\ref{eomke10}) and supersymmetry variation (\ref{susyexpli})), we must make some further field redefinitions. We redefine the gravitino components, the dilatino and the supersymmetry parameter as
\be 
{} \psinew_0&\equiv&  g^{1/4}\Big(\pt_0-\Gamma_0\Gamma^{a}\pt_a\Big),
\nn \\ 
{} \psinew_a&\equiv&  g^{1/4}\Big(\pt_a-\f1{12} \Gamma_{a}\lt\Big),
\nn\\ 
{}\lambdanew&\equiv & \f{2}3 g^{1/4}\lt,
\nn \\ 
{}  \venew &\equiv& g^{-1/4}\vet.
\ee 
 
With these redefinitions, using the bosonic dictionary obtained in the
previous section and in the gauge 
\be \label{susygauge}
\psinew_0=0, 
\ee 
we can show that the
equations of motion of the fermions of massive IIA supergravity (\ref{eomlam})
and (\ref{eompsi}) are equivalent to the Kac-Moody fermionic spinor equations
(\ref{eompsi10}) and (\ref{eomke10}) if we assume a correspondence between the unfaithful
representation of $\mf{k}(\mf{e}_{10})$ and the redefined fermionic fields
displayed in the first half of Table \ref{dicofer}.  

\begin{table}[t]
\centering
\begin{tabular}{|c|c|}
\hline
$K(E_{10})$ representations&Fermionic fields of supergravity\\
\hline
\hline
$\Psi_a$ & $\psinew_a=g^{1/4}\Big(\pt_a-\f1{12} \Gamma_{a}\lt\Big)$\\[.2cm]
$\Gamma^{10}\Psi_{10}$ & $\lambdanew=\f{2}3 g^{1/4}\lt$\\[.2cm]
\hline
$\epsilon$ & $\venew=g^{-1/4}\vet$\\[.2cm]
$\Psi_t$ & $\psinew_t=n g^{1/4}\Big(\pt_0-\Gamma_0\Gamma^{a}\pt_a\Big)$\\[.2cm]
\hline
\end{tabular}
\caption{\label{dicofer}\sl Fermionic dictionary: This table presents the relation between the spinor and vector spinor unfaithful representations of $\mf{k}(\mf{e}_{10})$ and the fermionic fields of massive type IIA supergravity. The first half is obtained by requiring the equations of motion for the fermions to match with the  $\mf{k}(\mf{e}_{10})$ equation and the second half comes from the supersymmetry variation of $\psi_t$.}
\end{table}

To illustrate how the correspondence is proved, let us consider the mass terms of the equation of motion for the gravitino. On the supergravity side, one looks at the spatial components of the equation of motion (\ref{eompsi}). We only keep the terms involving the time derivative of $\pt_a$ and the mass:
\beq
\label{ex1}
\p_0\pt_a+\f1{16}e^{5\phi/4}m\G^0\G_{ab}\pt^b+\f5{16}e^{5\phi/4}m\G^0\pt_a+\f5{64}e^{5\phi/4}m\G^0\G_a\lt+\cdots=0.
\eeq
One the sigma model side, we must consider the terms at level $(4,1)$ of the explicit version of the equation of motion (\ref{eomke10}), that give
\beq
\label{ex2}
\pa_t \Psi_a-\f{1}{2\cdot9!}e^{-5\phi/4}P_{b_1\cdots b_9}\Gamma_{10}\Gamma^{b_1\cdots b_9}\Psi_a
 +\f{12}{9!}e^{-5\phi/4}P_{b_1\cdots b_9}\Gamma_{10}{\Gamma_{a}}^{b_1\cdots b_8}\Psi^{b_9}+\cdots=0.
\eeq
Using the bosonic and fermionic dictionaries given in Tables \ref{dicoeom} and \ref{dicofer}, it is now a purely algebraic exercise to see that (\ref{ex1}) and (\ref{ex2}) are equivalent.

\subsection{Supersymmetry variations of fermions}
Considering the supersymmetry variation of the gravitino provides us with a consistency check of the previously obtained bosonic and fermionic dictionaries. To this end it is natural to define 
\beq
\psinew_t \equiv n\psinew_0,
\eeq
where $t$ is considered as a `vector index' along the world line with respect to the einbein $n$.

Using the previous redefinitions yields the following supersymmetry transformation on the redefined gravitino 
\be
{} \delta_{\ve}\psinew_t &=& \p_t \ve +\f{1}{4}g^{-1}\p_t g \ve+\f{1}{4}N \om_{0ab}\Gamma^{ab}\ve+\f14 Ne^{5\phi/4}m\Gamma_0\ve 
\nn \\
{}& & + \f{1}{4\cdot 4!}e^{\phi/4}N\Gamma^{abcd}\Gamma_0F_{abcd}\ve-\f{1}{4\cdot 3!}e^{\phi/4}N\Gamma^{abc}F_{0abc}\ve
\nn \\
{}& & -\f{1}{8}e^{3\phi/4}N \Gamma^{ab}\Gamma_0\Gamma_{10}F_{ab}\ve+\f{1}{4}e^{3\phi/4}N\Gamma^{a}\Gamma_{10}F_{0a}\ve
\nn \\
{}& & -\f{1}{24}e^{-\phi/2}N\Gamma^{abc}\Gamma_0\Gamma_{10}F_{abc}\ve-\f{1}{8}e^{-\phi/2}N\Gamma^{ab}\Gamma_{10}F_{0ab}\ve
\nn \\
{}& & +\f{1}{4}N\om_{abc}\Gamma^{abc}\Gamma_0\ve -N\Gamma_0\Gamma^{a}\Big[\p_a \ve+\f{1}{4}g^{-1}\p_a g \ve\Big].
\label{redefinedgravitinovariation}
\ee

We can now identify the right hand side of (\ref{redefinedgravitinovariation}) with the $K(E_{10})$-covariant derivative acting on the $32$-dimensional spinor representation $\epsilon$ (\ref{SusyTransfSigmaModel}), given explicitly in (\ref{susyexpli}), using the bosonic and fermionic dictionaries already computed (Tables \ref{dicoeom} and \ref{dicofer}). We then obtain the second half of Table \ref{dicofer}, that is, the relations for the supersymmetry parameter and the time component of the gravitino.

In conclusion, we see that all the fermionic equations of motion as well as the supersymmetry variation of $\psi_t$ match with the $E_{10}/K(E_{10})$ fermionic theory. In particular, we notice that the mass enters these equations correctly everywhere.

\section{Gauge algebra and supersymmetry variations from $E_{10}$}
\label{sec:comments}

In this section we address and collect a few general points which
are of relevance for the analysis of hyperbolic Kac--Moody coset models in
relation to deformed or undeformed supergravity. In particular, we point out some crucial features related to gauge-fixing in the $E_{10}$ sigma model, and we discuss the associated constraints. We also show how the gauge algebra and supersymmetry variations of massive IIA supergravity may be obtained in a simple way from properties of $E_{10}$ and $K(E_{10})$. 

\subsection{The importance of being gauge-fixed}

As has been pointed out in~\cite{Kleinschmidt:2005bq}, in order to reveal the correspondence between the geodesic equation of the
sigma model and the equations of motion of supergravity it is essential that one fixes all
gauges on both sides. For $E_{10}$ this means that one should work in Borel
gauge, whereas the supergravity variables are subject to the pseudo-Gaussian
gauge (\ref{10v}) for the metric and also to temporal gauges for the $p$-form
gauge fields (\ref{tempgauge}) which are of particular importance in the context of the mass deformed
supergravity theory.\footnote{The same phenomenon occurs for gauged
  supergravity~\cite{Bergshoeff:2008}.} The dictionary as stated was a
correspondence at the level of the `velocities', i.e. first derivatives
of the basic variables. In the undeformed case, this correspondence also holds
for the basic variables so that the coordinates parametrising the
$E_{10}/K(E_{10})$ coset (in triangular gauge) can be identified with the
(gauge-fixed) variables of supergravity~\cite{Damour:2002cu}. For example, in
massless IIA one would write a coset element\footnote{There is some ambiguity
  in parametrising the positive step operators which is directly related to
  possible field redefinitions in supergravity.}
\be\label{Vexpl}
\cV=\cV_0 e^{\phi T} e^{A_{m}E^m} e^{\tfrac{1}{2} A_{m_1m_2} E^{m_1m_2}} 
  e^{\tfrac{1}{3!} A_{m_1m_2m_3} E^{m_1m_2m_3}} \cdots\,,
\ee
for which the Maurer--Cartan form (\ref{mcform}) expands to (using $\cV_0=e_a{}^m$ as
follows from Table~\ref{dicoeom})
\be
\partial_t\cV\,\cV^{-1} &\!\!=&\!\! -e_a{}^m\partial_t e_m{}^b K^a{}_b 
    + \partial_\phi T 
    + e^{3\phi/4} e_a{}^m  \partial_t A_{m}  E^{a}
     + \frac12 e^{-\phi/2} e_{a_1}{}^{m_1} e_{a_2}{}^{m_2} 
                   \partial_t A_{m_1m_2} E^{a_1a_2}\nn\\
&&\quad + \frac1{3!}    e^{\phi/4} e_{a_1}{}^{m_1} e_{a_2}{}^{m_2} e_{a_3}{}^{m_3}
      (\partial_t A_{m_1m_2m_3}-3A_{m_1}\partial_t  A_{m_2m_3})
   E^{a_1a_2a_3} +\ldots \,.
\ee
The fact that the coefficient of $E^{a_1a_2a_3}$ has several contributions is
due to the expansion when using the Baker--Campbell--Hausdorff formula. As
we are in Borel gauge this expansion is always guaranteed to
terminate. The level zero pieces merely `dress' the
  coefficients; $\cV_0$ is the inverse vielbein $e_a{}^m$ from supergravity
  and converts curved into flat indices while $e^{\phi T}$ introduces a level
  dependent dilaton prefactor. 
The precise form of the extra terms in the coefficients of the generators $E_{\ell}$ in the Maurer-Cartan form is prescribed by the $E_{10}$ structure and the
parametrization chosen. Clearly, writing out the corresponding field strength 
$F_{tm_1m_2m_3}$ (see~(\ref{curvs})) for vanishing mass gives in temporal gauge
\be
F_{tm_1m_2m_3} = \partial_t A_{m_1m_2m_3} 
   -3  A_{m_1} \partial_{t} A_{m_2m_3}\,,
   \label{E10FieldStrength}
\ee
which is exactly the coefficient of $E^{a_1a_2a_3}$ if the dilaton is
extracted. Therefore the structure constants of $E_{10}$ naturally encode the modified field strengths of supergravity, including Chern-Simons terms as well as extra terms involving lower-rank $p$-form potentials as in (\ref{E10FieldStrength}). Therefore we can, in more refined version of the dictionary in Table~\ref{dicoeom}, identify the supergravity fields with the fields appearing in the parametrization of $\mc{V}$.

Continuing to the massive case $m\ne 0$ this logic seems to run into problems
since there one has, for example,
\be\label{massfield}
F_{tn} = \p_t A_n + m A_{tn} \,.
\ee
The mass term was identified with the time derivative of the nine-form
potential appearing in the parametrisation of $E_{10}/K(E_{10})$ in Borel
gauge. Such a term cannot arise from the Maurer-Cartan form since in Borel gauge the
commutators in the expansion only increase the level, so that a coefficient of
a generator on a given level has only contributions from lower-lying
generators. Here, the gauge fixing (\ref{tempgauge}) becomes important because
it eliminates the second term in (\ref{massfield}) by simply setting $A_{tn}=0$. Hence, one can still
identify the coset coordinate $A_{m}$ with the supergravity gauge potential
$A_m$ even in the deformed case as long as one works in a completely
gauge-fixed framework.\footnote{This is in contrast to the analysis
  of~\cite{Schnakenburg:2002xx}, where an additional $(-1)$-form generator
  was introduced in the algebra in the form of a translation generator. With
  such a generator the commutators are not   only increasing in level; a
  similar logic was followed 
  in~\cite{Riccioni:2007au,Riccioni:2007ni}. Formal $(-1)$-forms also appear
  in~\cite{Lavrinenko:1999xi}.}

\subsection{Constraints  on the geodesic sigma model}
\label{constraints} 

Whenever gauges are fixed one has to remember to also impose the corresponding
constraints. In the context of $\mf{e}_{10}$ it has been observed in the
$\mf{sl}(10, \mbb{R})$ decomposition that the 
constraints can be imposed consistently on the
geodesic~\cite{Damour:2007dt}. However, the constraints transform only under a 
Borel subgroup $E_{10}^+\subset E_{10}$; the relevant linear representation
can be embedded into the highest weight representation $L(\Lambda_1)$ of 
$\mf{e}_{10}$ and its low level content in the $\mf{sl}(9, \mbb{R})$
decomposition used in this 
paper is displayed in table~\ref{l1dec}.\footnote{Except for the momentum  
  constraint there are `trailing' $\epsilon^{b_1\ldots b_9}$ factors which  
  have been suppressed in the table. They are not seen by $\mf{sl}(9,
  \mbb{R})$ but by $\mf{gl}(9, \mbb{R})$ and they correspond to volume
  factors, so that the corresponding  `tensors' transform as densities rather
  than true tensors. Their number can easily be recovered from demanding
  that the total number of indices on level $(\ell_1,\ell_2)$ be 
  $2\ell_1+\ell_2$. Furthermore, there is a factor $\exp[
  (-2\ell_1+3\ell_2)\phi/4]$  of the dilaton will multiply the product of $P$s
  in our conventions. We will exemplify this below.}  
$L(\Lambda_1)$ here indicates that 
the highest weight of this integrable $\mf{e}_{10}$ representation has its
weight given by the fundamental weight dual to the simple root of 
node~1. We note that the corresponding representation $L(\Lambda_1)$ of
$\mf{e}_{11}$ has been considered before both for the construction of
space-time~\cite{West:2003fc} and the construction of deformed
supergravities~\cite{Riccioni:2007au,Riccioni:2007ni}.  

In the context of $E_{10}$ one has to take into account the constraints that
arise from fixing the pseudo-Gaussian gauge (\ref{10v}) for the vielbein and
the temporal gauges (\ref{tempgauge}) for the vector potentials. The vanishing 
shift component in the vielbein gives rise to the spatial diffeomorphism
constraint, sometimes also called the momentum constraint. Fixing temporal 
gauge for a $p$-form potential one obtains an associated Gauss constraint
which carries $(p-1)$ antisymmetric indices. Furthermore, there are 
Bianchi-type constraints associated with all fields which arise from the split
between space and time. The details of the general construction can be found 
in~\cite{Damour:2007dt}, here we work out the details only for the case of the
constraint on the mass parameter. The other constraints follow from the
analysis of~\cite{Damour:2007dt} by dimensional reduction. 

\begin{table}[t] 
\centering 
\begin{tabular}{|c|c|c|}
\hline 
$(\ell_1,\ell_2)$ & $\mf{sl}(9, \mbb{R})$ Dynkin labels & Constraint\\ 
\hline 
\hline 
$(3,2)$ & $[1,0,0,0,0,0,0,0]$ & Momentum constraint \\ 
$(3,3)$ & $[0,0,0,0,0,0,0,0]$ & Gauss for $A_a$ \\ 
$(4,2)$ & $[0,0,0,0,0,0,0,1]$ & Gauss for $A_{a_1a_2}$\\
$(4,3)$ & $[0,0,0,0,0,0,1,0]$ & Gauss for $A_{a_1a_2a_3}$\\ 
$(5,3)$ & $[0,0,0,0,1,0,0,0]$ & Bianchi for $A_{a_1a_2a_3}$\\
$(5,4)$ & $[0,0,0,1,0,0,0,0]$ & Bianchi for $A_{a_1a_2}$\\ 
$(6,3)$ & $[0,0,1,0,0,0,0,0]$ & Bianchi for $A_a$\\
$(6,4)$ & $[0,1,0,0,0,0,0,0]$ & Bianchi for $\phi$\\ 
$(6,4)$ & $[0,0,1,0,0,0,0,1]\oplus [0,1,0,0,0,0,0,0]$ & Bianchi for Riemann\\
$(7,3)$ & $[1,0,0,0,0,0,0,0]$ & Bianchi for mass\\ 
\hline
\end{tabular} 
\caption{\label{l1dec}\sl IIA Level decomposition of $L(\Lambda_1)$
  representation of $\mf{e}_{10}$ under $\mf{sl}(9, \mbb{R})$.}
\end{table}

It turns out to be more convenient to 
think of the supergravity Bianchi constraint on the mass parameter $m$ as the
Gauss constraint for a nine-form potential with ten-form field strength. In
fact, all Bianchi constraints are secretly Gauss constraints of the dual
variables in temporal gauge. In this case it is of the form  
\be
\partial_{\mu}\left(\sqrt{-G} e^{-5\phi/2} F^{\mu\,m_1\ldots m_{8} 
    t}\right) = 0  \,, 
\ee 
where as usual one index was chosen to be $t$ in order to obtain the Gauss
constraint. This constraint also transforms manifestly as an eight-form, just
like the contraint listed on level $(7,3)$ in Table~\ref{l1dec}.

We now turn to the corresponding constraint on the $E_{10}$ side. In the 
general construction of the constraints~\cite{Damour:2007dt}, a constraint on
level $(\ell_1,\ell_2)$ is composed out of a product of two  
components $P$ of the Cartan form (\ref{cmform}) such that their levels add up
to $(\ell_1,\ell_2)$. For the mass term this leaves only the possibility to 
write\footnote{Here, we have written out the trailing $\epsilon$ (in the form
  of an extra set $[b_1\ldots b_9]$ of antisymmetric indices on the
  constraint) and the correct dilaton pre-factor for this level.}
\be \label{masscose10}
C^{b_1\ldots b_9|a_1\ldots a_8} =  e^{-5\phi/4} P^{b_1\ldots b_9} 
  P^{a_1\ldots  a_8} \,.
\ee 
This is indeed the correct corresponding constraint if set to
zero. Furthermore, this expression is weakly conserved along the geodesic 
motion, as immediately follows from the coset model equations of motion in
Appendix~\ref{Appendix:EOMBosonic}. 

After using the dictionary of Table~\ref{dicoeom}, the coset constraint
(\ref{masscose10}) turns into the expression
\be
C^{b_1\ldots b_9|a_1\ldots a_8} &=& -\frac14 N^2 m \,e^{5\phi/4}
\eps^{b_1\ldots   b_9}\eps^{a_1\ldots a_8c} \partial_c\phi\nn\\
&=& -\frac{N}4 \eps^{b_1\ldots   b_9}\eps^{a_1\ldots a_8c} 
\partial_c \left(e^{5\phi/4} m \right)
\ee
which indeed vanishes in supergravity as a consequence of the truncation
(\ref{truncsugra}). Therefore we find an extension of the set of constraints
of~\cite{Damour:2007dt} to include the mass generator. It would be interesting
to investigate in detail to what extent the Sugawara-type structure found
there continues to apply on this next level.

Besides the bosonic constraints one also has to ensure that the supersymmetry 
${\mathcal S}$ constraint arising from fixing (\ref{susygauge}) to zero is 
satisfied. This constraint was already studied in detail 
in~\cite{Damour:2006xu,Damour:2007dt}  where it was shown that it also gives 
rise to the bosonic constraints, in particular the Hamiltonian constraint, in 
a canonical analysis. This is in agreement with the general result 
of~\cite{Teitelboim:1977fs} that the supersymmetry constraint is the
`square-root' of the bosonic constraints.

\subsection{The gauge algebra}

We now consider the gauge algebra of massive IIA supergravity in $D=10$ and
its relation to global Kac--Moody transformations. It is instructive to start
the discussion in a covariant setting where the associated Kac--Moody
transformations  should belong to $E_{11}$~\cite{West:2000ga}. We then show
how the gauge-fixing in the case of $E_{10}$ reconciles the local nature of
gauge transformations with the global nature of $E_{10}$ rotations both in the
undeformed and in the deformed case.

From the definition of the
field strengths (\ref{curvs}) it follows that the tensor gauge transformations
of the massive supergravity gauge potentials are (see
also~\cite{Lavrinenko:1999xi,Bergshoeff:2007vb}) 
\be\label{gtrm}
\dg A_{\mu} &=& \partial_\mu \Lambda - m\Lambda_\mu\,,\nn\\
\dg A_{\mu_1\mu_2} &=& 2\partial_{[\mu_1} \Lambda_{\mu_2]}\,,\nn\\
\dg A_{\mu_1\mu_2\mu_3} &=& 3\partial_{[\mu_1} \Lambda_{\mu_2\mu_3]} 
   + 3 A_{[\mu_1\mu_2}\dg A_{\mu_3]} \,.
\ee
Each $p$-form has its own $(p-1)$-form gauge parameter and  the field
strengths in (\ref{curvs}) are invariant under these gauge transformations.  

In the massless case, the algebra of gauge transformations displays a
hierarchical or Borel-type structure in that every form only transforms under its own gauge
parameter or under gauge transformations of the lower rank $p$-forms. These
couplings are determined by Chern--Simons terms or the so-called transgression
terms (like $4A_{[\mu_1} F_{\mu_2\mu_3\mu_4]}$ in (\ref{curvs})). This
pattern persists also when higher rank $p$-forms are introduced in the
supersymmetry algebra~\cite{Bergshoeff:2006qw} and is reminiscent of a Borel
type structure on the positive step operators of $E_{11}$ (or also
$E_{10}$)~\cite{West:2000ga,Bergshoeff:2007vb}. In the massive case, however,
the hierarchy appears to be broken as now the vector field $A_\mu$ transforms
under the gauge parameter $\Lambda_\mu$ of the two-form, leading to the St\"uckelberg mechanism. 

In order to make the correspondence between $E_{10}$ or $E_{11}$ and the algebra of gauge
transformations more precise we rewrite (\ref{gtrm}) for any value of $m$ as 
\be\label{gstrm}
\dS A_{\mu} &=& \Sigma_{\mu}\,,\nn\\
\dS A_{\mu_1\mu_2} &=& \Sigma_{\mu_1\mu_2}\,,\nn\\
\dS A_{\mu_1\mu_2\mu_3} &=& \Sigma_{\mu_1\mu_2\mu_3} 
  + 3\Sigma_{[\mu_1}A_{\mu_2\mu_3]}\,, 
\ee
by absorbing the derivatives and the mass term into the definition of the
$\Sigma$.\footnote{Note that for $m\ne 0$, $\Sigma_\mu$ is no longer closed.}
It is straight-forward to show that for example  
\be\label{gcomm21}
\left[ \delta_{\Sigma_{\mu_1\mu_2}}, \delta_{\Sigma_{\mu_3}} \right] 
   = \delta_{\Sigma_{\mu_1\mu_2\mu_3}}
\ee
on all fields, where the parameter $\Sigma_{\mu_1\mu_2\mu_3}$ is expressed
through the constituent $\Lambda$ parameters. 

The commutator (\ref{gcomm21}) resembles the commutator (\ref{commutatorexample}) between the
level $(1,0)$ and $(0,1)$ generators. This is not purely coincidental as can
be seen from considering the action of a global $E_{10}$ transformation of the
coset element (\ref{Vexpl}) by a Borel element, i.e. $g\in E_{10}^+$ such that
$g$ is an exponential of positive step operators only. Evaluating the action
$\mc{V}\mapsto \mc{V} g$ requires no compensating $K(E_{10})$ transformation in this case
since Borel-valued $g$ preserve the Borel gauge. For example, for  
\be 
g=e^{\Sigma_m E^m} e^{\tfrac{1}{2} \Sigma_{m_1m_2} E^{m_1m_2}}
e^{\tfrac{1}{3!}\Sigma_{m_1m_2m_3} E^{m_1m_2m_3}} 
\ee
one can compute the resulting transformation of the coset fields to lowest
order in the $\Sigma$ parameters to be 
\be\label{e10trm}
A_{m} &\longmapsto& A_m + \Sigma_m\,,\nn\\
A_{m_1m_2} &\longmapsto& A_{m_1m_2} + \Sigma_{m_1m_2}\,,\nn\\
A_{m_1m_2m_3} &\longmapsto& A_{m_1m_2m_3} +\Sigma_{m_1m_2m_3}
  +3\Sigma_{[m_1}A_{m_2m_3]}\,. 
\ee
This is in exact agreement with the gauge transformations (\ref{gstrm}) for
the spatial components of the gauge fields. A similar reasoning can be carried
out for $E_{11}$~\cite{West:2000ga} and the formul\ae{}  are the same but with
covariant space-time indices.

However, this appears to pose a problem since the gauge transformations
(\ref{gstrm}) are local transformations and the $E_{10}$ (or
$E_{11}$) transformations (\ref{e10trm}) are global.
In~\cite{Bergshoeff:2007vb} the resolution of this 
problem was suggested to be to take the global limit of the $\Sigma$
parameters in supergravity.\footnote{This is the general way the global part of the higher-dimensional diffeomorphisms and gauge transformations gets inherited by the parabolic subgroup of the hidden symmetry group after dimensional reduction. It would be interesting to study the usual hidden symmetries of the massive theory, something that has not been done to the best of our knowledge.}
For undeformed supergravity, this is tantamount to
considering gauge parameters $\Lambda$ in (\ref{gtrm}) which are linearly
space-time dependent. As was already pointed out in~\cite{Bergshoeff:2007vb},
for the mass deformed model the appearance of a naked $\Lambda_\mu$ in the
variation of the vector field makes this limit inconsistent since then  
\be
\Sigma_\mu =\partial_\mu \Lambda -m \Lambda_\mu
\ee
is no longer constant. In the covariant $E_{11}$ formulation, one possibility
to resolve this inconsistency considered in~\cite{Bergshoeff:2007vb} is to
introduce so-called $(-1)$-forms (see
also~\cite{Lavrinenko:1999xi,Schnakenburg:2002xx}) which enlarge the gauge
algebra. The enlarged gauge algebra should be compared with the combined
algebra of $E_{11}$  and the momentum representation $l_1$ of
$E_{11}$~\cite{West:2003fc,Riccioni:2007au,Riccioni:2007ni}. The result
of~\cite{Bergshoeff:2007vb} was that these two algebras are different: the
additional fields required for closing the algebra once a $(-1)$-form is
present are not those of the $l_1$ representation. Consequently the problem was
not resolved in the covariant framework.

Here, we work in a gauge-fixed $E_{10}$ framework and therefore the question
has to be investigated anew. In the $E_{10}$ context, the global nature of a
transformation refers to its time-dependence. As all comparisons are assumed
to be carried out at a fixed spatial point all space-dependence is
`frozen'. Furthermore, only gauge-fixed quantities are compared. To preserve
the temporal gauge 
(\ref{tempgauge}) of supergravity one has to include compensating gauge
transformations. We will denote these compensated gauge transformations by
$\dcg$, for example 
\be 
\dcg A_{\mu} = \dg A_\mu + S_\mu
\ee
and $S_\mu$ denotes the compensator whose precise form we will not
require. The compensators are fixed by demanding that the temporal gauge is
preserved, i.e.   
\be
\dcg A_t = 0\,,\quad \dcg A_{tm} = 0\,,\quad \dcg A_{tm_1m_2} = 0 \,.
\ee
Since field strengths are gauge invariant, we obtain immediately
\be 
0 = \dcg F_{tm} = \partial_t \dcg A_m - \partial_m \dcg A_t + m \dcg A_{tm}
  = \partial_t \dcg A_m\,, 
\ee
so that $\dcg A_m$ is time-independent and hence can be associated with a
global $E_{10}$ transformation. In this way, the gauge-fixing resolves the
tension between the global nature of $E_{10}$ transformations and local gauge
transformations in supergravity. The local $K(E_{10})$ transformations have
been  fixed by adopting the (almost) Borel gauge so that there are no local
gauge transformations left in the $\sigma$-model (except for the level
$\ell=(0,0)$ Lorentz rotations), in agreement with the time-independence of
the gauge transformation in gauge-fixed supergravity.
This argument is valid both for the massless
and the massive case.

\subsection{Supersymmetry variation of the bosons from $E_{10}$}

We define a supersymmetry variation of the coset element $\cV$ via
\be
\deps \cV \cV^{-1} =   \sP + \sQ  \quad\in \mathfrak{e}_{10}\,.
\ee
{}From the explicit form of (\ref{Vexpl}) we obtain for the first few levels
\be
\sP &=& -\frac12 e_a{}^m\deps e_{mb} S^{ab} + \deps\phi T 
  +\frac12 e^{3\phi/4}e_a{}^m\deps A_m (E^a+F_a)\nn\\
&&  +\frac14 e^{-\phi/2}e_{a_1}{}^{m_1}e_{a_2}{}^{m_2}\deps A_{m_1m_2} (E^{a_1a_2}+F_{a_1a_2})\nn\\
&&
  +\frac1{12} e^{\phi/4}e_{a_1}{}^{m_1}e_{a_2}{}^{m_2}e_{a_3}{}^{m_3}\deps A_{m_1m_2m_3} (E^{a_1a_2a_3}+F_{a_1a_2a_3})
  +\ldots\,.
\ee

\begin{figure}
\begin{center}
\includegraphics{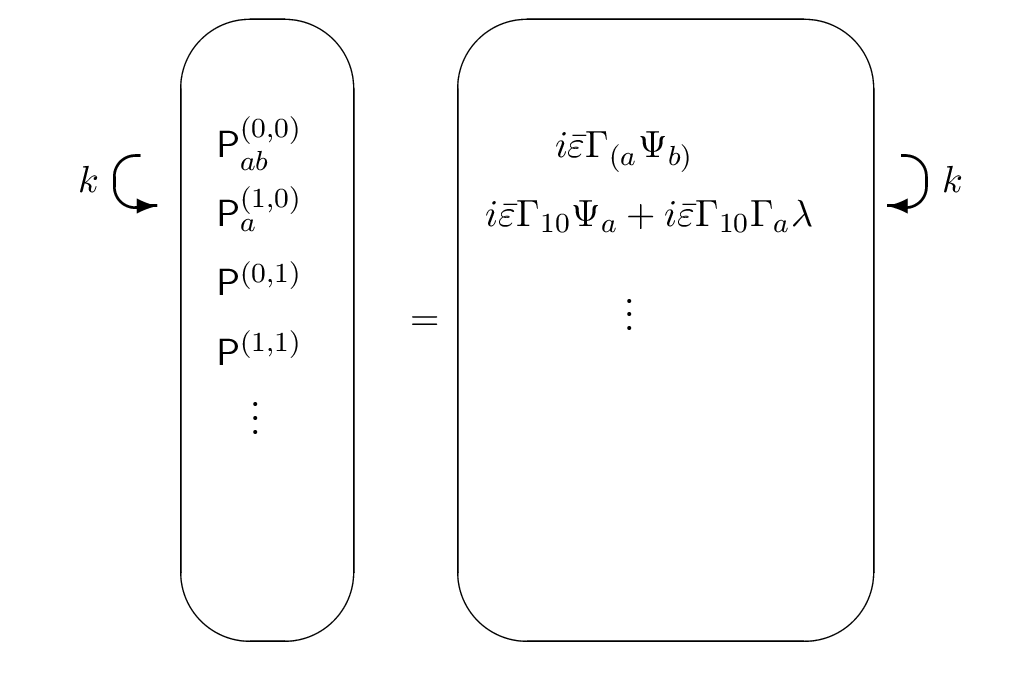}
\caption{\label{fig:susybos} \sl Illustration of the two ways of transforming
  the coset representation $\sP=\eps \odot \Psi$ with an element $k\in
  K(E_{10})$: Either using the abstract coset representation $\sP$ or the
  explicit expression in terms of $K(E_{10})$ fermion bilinears. The fermion bilinear expressions are only meant to be indicative for space reasons. The exact expressions are given in the text.} 
\end{center}
\end{figure}

The supersymmetry variations of the bosonic fields can now be computed from
the following observation~\cite{Damour:2006xu}. The Lie algebra element $\sP$
transforms under $K(E_{10})$ in the coset representation. So one can obtain
any higher level component from a $K(E_{10})$ rotation of level
zero.\footnote{In fact, one can obtain any element from the Cartan
  subalgebra. The converse is not true: One cannot conjugate an
  arbitrary element of the coset into the Cartan subalgebra, unlike in the
  finite-dimensional case. This derives from the general property of infinite-dimensional Kac-Moody algebras that any element of the algebra is no longer conjugate to a Cartan element \cite{KacPeterson}.} 
At the same time one can take the explicit expressions for the supersymmetry
variations of the level zero fields from supergravity in terms of fermions and
interpret the resulting expression as being constructed out of the tensor
product of the ${\bf 32}$ Dirac-spinor representation and the ${\bf 320}$
vector-spinor representation of $K(E_{10})$. Then one can compute the
$K(E_{10})$ action on an expression of the type $\sP = \epsilon\odot \Psi$ in
two different ways; either by transforming $\sP$ according to the coset
transformation, or by transforming the fermion bilinear according to the
unfaithful spinor representations, see Figure~\ref{fig:susybos}. Equating the
two results yields an 
expression for the supersymmetry variation of the higher level (bosonic) coset
fields. This procedure rests on the assumption that the expression bilinear in
the fermions transforms in exactly the same way as the coset variable. This is
known to be incorrect at higher levels~\cite{Damour:2006xu} (notably from the
level of the dual graviton onwards) and is related to the fact that there is a
certain disparity between the fermionic fields in unfaithful
finite-dimensional representations of $K(E_{10})$ whereas the bosons transform
in the infinite-dimensional coset representation. This point certainly
requires further investigation. 

Irrespective of the known obstacles at higher levels, we illustrate this
procedure now for the lowest lying bosonic fields. Under a $K(E_{10})$
rotation with generator  
\be
\delta k = \Lambda_a (E^a - F_a)
\ee
the coset element $\sP$ transforms infinitesimally as $\delta_{\delta k} \sP = [\delta k,\sP]$
leading to 
\be\label{susycm}
\delta_{\delta k} \sP &=&  e_{(a}{}^m\deps e_{mb)}\Lambda_b (E^a+F_a)
   -\frac34\deps \phi \Lambda_a (E^a+F_a)  
   +\frac32 e^{3\phi/4} e_c{}^m\deps A_m \Lambda_c T\nn\\  
&& -\frac1{16} e^{3\phi/4} e_c{}^m\deps A_m \Lambda_c \delta_{ab} S^{ab} 
    + \frac12 e^{3\phi/4} e_{a}{}^m\deps A_m\Lambda_b S^{ab} +\ldots\,.  
\ee 
This is the first way of transforming $\sP = \epsilon\odot \Psi$. Transforming
instead the fermion bilinear of level zero, using the formul\ae{}
(\ref{vards}), (\ref{varpsi10}) and (\ref{varpsia})
leads for the coefficient of the dilaton generator $T$ to  
\be
\delta_{\delta k} (i \bar\vet \lambda)   
=  i\Lambda_a\left(\frac{9}{8} \bar\vet\Gamma^a\Gamma_{10}\lambda 
  + \frac{3}{2}\bar\vet\Gamma_{10}\psi_a\right)\,,  
\ee
where also the dictionary of Table~\ref{dicoeom} was used. Comparing the two
expressions we deduce
\be 
\deps A_m = i e^{-3\phi/4} \left(\bar\vet \Gamma_{10}\psi_m 
    +\frac{3}{4} \bar\vet\Gamma_{m}\Gamma_{10}\lambda\right)
\ee  
in complete agreement with (\ref{susybos}). This procedure can be pushed to
higher  
level and yields the supersymmetry variations of all bosonic fields from the 
variation of the dilaton. From the level of the dual graviton onwards there 
appear the known problems~\cite{Damour:2006xu} related to the
non-supersymmetry  
of the bosonic-fermionic coset model of
section~\ref{sec:sigmamod}. Nevertheless this kind of calculation is a useful 
tool for studying and deriving supersymmetry transformation rules from 
Kac--Moody algebra. We also note that the same reasoning works in exactly the
same way for $E_{11}$ since the Maurer--Cartan form (\ref{susycm}) involves a 
supersymmetry variation and therefore no closure with the conformal group is
required.\footnote{This closure plays an important role in constructing field 
  strengths out of $\partial_\mu \cV \cV^{-1}$ for $E_{11}$ since the $\mu$ 
  index needs to be antisymmetrized with the indices on the
  generators~\cite{West:2000ga}. The conformal group enters in this discussion
due to a theorem by Ogievetsky~\cite{Ogievetsky:1973ik} giving an algebraic description of the
algebra of diffeomorphisms as the closure of affine diffeomorphisms with
conformal diffeomorphisms. The notion of conformal group is in fact misleading
here as there is no conformal invariance in the theory. All that is required
for the theorem is to have a suitable diffeomorphism generator which is
quadratic in the coordinates. After commutation it will yield
diffeomorphism generators with arbitrary polynomial dependence on the
coordinates.} 
We also checked that the same analysis in the case of $D=11$ supergravity
reproduces correctly the transformations of the three-form and the six-form  
including all contributions from the Chern--Simons term and transgression
terms.

\section{Trombone deformations and the dual graviton}
\label{sec:trombone}

In a recent interesting paper~\cite{Diffon:2008sh} it was remarked that it is
possible to systematically construct also deformed supergravity theories by using an appropriate embedding tensor if
one gauges a global scaling symmetry of the equations of motion. Under this scaling or {\em trombone symmetry}, all bosonic tensor
fields 
scale with a weight proportional to the number of their tensor indices: 
\be
G_{\mu\nu} \mapsto \Lambda^2 G_{\mu\nu}\,,\quad A_{\mu_1\ldots \mu_p}\mapsto
\Lambda^p A_{\mu_1\ldots \mu_p}\,. 
\ee
The fermions also scale appropriately such that the whole massless action
scales as\footnote{We note that for this to be true in Romans theory, the mass
  has to be replaced by a nine-form which scales according to the general
  rule.} 
\be
\mc{L}^{[B]}+\mc{L}^{[F]} \mapsto \Lambda^{D-2}
\left(\mc{L}^{[B]}+\mc{L}^{[F]}\right) 
\ee
in $D$ space-time dimensions. This is only a symmetry of the action
in $D=2$ space-time dimensions. Otherwise it is nevertheless an acceptable
global symmetry of the equations of motion. In the framework of gauging
subgroups of global symmetry groups one can also consider gauging this
trombone symmetry and obtain a deformed supergravity theory if one is willing
to give up the existence of an action from which the deformed equations of
motion are derived. This was already done in the $D=10$ type IIA case
in~\cite{Howe:1997qt,Lavrinenko:1997qa}. The analysis of~\cite{Diffon:2008sh}
sheds new light on 
this somewhat neglected class of deformed supergravities and the relation to
Kac--Moody symmetries by the following observation. In a given space-time
dimension $D\ge 3$, undeformed maximal supergravity has a global symmetry group
$E_{11-D}$ under which the vector fields transform in a certain
representation. The embedding tensor describing the gauging of the trombone
symmetry therefore has to transform in the representation dual to the vector
fields in order to construct the covariant derivative. By inspection of the
decomposition tables of~\cite{Bergshoeff:2007qi} 
one finds that exactly this representation occurs in the level decomposition
of $E_{10}$ or $E_{11}$ on the same level as the representations relevant for
the `usual' embedding tensor which describes gaugings of the global $E_{11-D}$
symmetry rather than of the trombone.  

The crucial difference is, however, that while from a space-time point of view
the usual embedding tensor is related to $(D-1)$-forms, the trombone embedding
tensor occurs in the tables of $E_{10}$ and $E_{11}$ as a {\em mixed symmetry
  tensor} according to the observation of~\cite{Diffon:2008sh}. The mixed
symmetry is a hook symmetry of type $(D-2,1)$ with also a total of $(D-1)$
boxes. While it is clear how to 
associate to a $(D-1)$-form a single constant deformation parameter $g$ (the
gauge coupling) by passing to a constant space-time filling field strength
(just like for the Romans mass), this is not so obvious for the $(D-2,1)$
mixed symmetry tensor conjectured to be of relevance for the trombone
deformation. One first observation is that a mixed $(D-2,1)$ tensor in $D$
dimensions carries no degrees of
freedom~\cite{Curtright,Bekaert:2002dt,Diffon:2008sh}. 

Here, we would like to point out that in $D=10$ this $(8,1)$ hook tensor
arises at level $(3,3)$ as can be seen from the extended level decomposition
in Table~\ref{longerdec} in Appendix \ref{app:commutators}. This implies that it is part of the
$D=11$ dual graviton (which is the only representation at
$\ell_1=3$ under $\mf{sl}(10, \mbb{R})$~\cite{Damour:2002cu}) 
and inspection of the associated root vector confirms 
that its lowest weight vector belongs to  $E_9$. This is again in contrast
with the other deformations which were all genuine $E_{10}$ generators not
visible in $D=11$ supergravity. The same is true when one continues to lower
dimensions where always the lowest weight vector is part of $E_9$. Acting with $GL(D-1,\mathbb{R})$ will of course also lead to root vectors which
are genuine $E_{10}$ elements. 

\subsection{Comparing the reduction of $p$-forms and their duals}

The fact that the hook arises from the $D=11$ dual graviton by dimensional
reduction is indicative of a general phenomenon. For forms of `extreme rank'
there is a difference in reducing the field or its dual. Before discussing
this for the trombone generator and the dual graviton, let us study this for
$p$-forms.  

In the antisymmetric $p$-form case `extreme rank' means $p=0$: The reduction
of an axionic scalar\footnote{`Axionic' here refers to the fact that the
  scalar enters the Lagrangian only with derivatives and so has a global shift
symmetry, see~(\ref{grs}).} 
from $D$ to $D-1$ dimensions gives a
scalar and nothing else. The dual of a scalar in $D$ dimensions is a
$(D-2)$-form which reduces to a $(D-3)$-form, which is dual to a scalar in
$D-1$ dimensions, and a $(D-2)$-form which carries no propagating degrees of
freedom in $D-1$ dimensions. From the point of view of counting degrees of
freedom this is satisfactory but it turns out that the $(D-2)$-form can have
an effect on the reduced theory. To see this consider the reduction of gravity
in $D$ dimensions coupled to a $(D-2)$-form 
\be\label{grd}
S^{(D)} = \int d^D x \sqrt{-\hat{G}} \left(\hat{R} - \frac1{2(D-1)!} \hat{F}_{(D-1)}^2\right)\,,
\ee
where we denote with hats the quantities in $D$ dimensions. This is classically equivalent to a scalar field $\chi$ coupled to gravity (for scalar fields hats are superfluous)
\be\label{grs}
S^{(D)} = \int d^D x \sqrt{-\hat{G}} \left(\hat{R} - \frac12 (\partial\chi)^2\right)\,,
\ee
where the two fields are dual via
\be\label{sdd}
\hat{F}_{\mu_1\ldots \mu_{D-1}} = \epsilon_{\mu_1\ldots \mu_{D-1}\nu} \partial^\nu\chi\,.
\ee
Returning to the action (\ref{grd}) in terms of the $(D-2)$-form and reducing one finds with the standard rules
\be\label{grdm1}
S^{(D-1)} &=& \int d^{D-1} x \sqrt{-G}\Big(R - \frac12(\partial\phi)^2 -\frac14 e^{\gamma\phi} F_{(2)}^2\nn\\ 
&& \quad\quad -\frac1{2(D-1)!} e^{\gamma\phi} F_{(D-1)}^2 -\frac1{2(D-2)!} F_{(D-2)}^2\Big)\,.
\ee
The first line contains the usual reduced metric, the dilaton $\phi$ and the
graviphoton field strength. Here, $\gamma=\pm\sqrt{2(D-2)/(D-3)}$, in
agreement with the exponent in (\ref{lag}). The definitions of the reduced
field strength $F_{(D-1)}$ involves a coupling to the graviphoton via a
transgression term whose
precise form will not be relevant as it does not influence the Bianchi
identity for $F_{(D-1)}$ which states that $F_{(D-1)}$ is a constant, which we
call $M$, times the $\epsilon$ tensor. Therefore, there will be an effective
(positive) cosmological term proportional to
$e^{\gamma\phi}M^2$, which also acts as a potential for the dilaton obtained
from reduction. Furthermore, it creates a linear interaction term of the type
$M*F_{(D-2)}$ in the equation of 
motion of the graviphoton $F_{(2)}$ (due to the transgression term) and also a
linear interaction term proportional to $M*F_{(2)}$ in the equation of motion
of $F_{(D-2)}$.  

These features are to be contrasted with what one would obtain from the
straight reduction of the scalar model (\ref{grs}). There no such effect arises: The scalar field is still free and there is no
cosmological constant. Hence, \emph{reducing a scalar $0$-form or its dual $(D-2)$-form gives
different results}. This can be understood also from the duality relation
(\ref{sdd}). Having a constant $F_{(D-1)}$ field strength means that $\chi$
has to be {\em linear} in the extra direction with slope $M$.\footnote{We
  ignore the transgression term in this discussion.} This shows that the
reduction of a scalar field in terms of its dual field is only an honest
dimensional reduction if $M=0$. For $M\ne 0$ one obtains a more general
reduction; in fact a Scherk--Schwarz
reduction~\cite{Scherk:1979zr,Lavrinenko:1997qa}.

\begin{figure}[t!]
\centering
\includegraphics{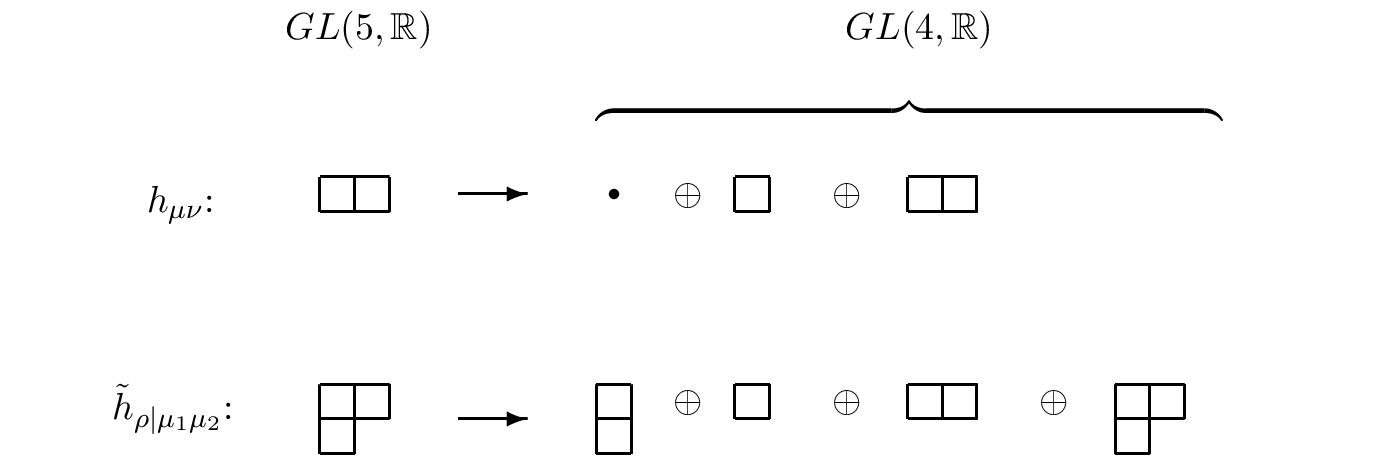}
\caption{\label{dgred}\sl Dimensional reduction of graviton $h_{\mu\nu}$ and
  dual graviton $\tilde{h}_{\rho|\mu_1\ldots \mu_{D-3}}$ from $D$ dimensions
  to $D-1$ dimensions in terms of irreducible representations of
  $GL(D,\mathbb{R})$ and $GL(D-1,\mathbb{R})$ for $D=5$. Objects in the same
  column are related by duality (in the linearised theory).} 
\end{figure}

\subsection{The trombone and the dual graviton}

After this digression let us return to the dual graviton and the trombone
deformations. It is known that linearized gravity in $D$ dimensions can be
rewritten in terms of a dual graviton $\tilde{h}_{\rho|\mu_1\ldots\mu_{D-3}}$
which transforms irreducibly in a $(D-3,1)$ hook representation of
$GL(D,\mathbb{R})$ and carries the same number of degrees of freedom as the
graviton $h_{\mu\nu}$~\cite{West:2001as,Olver,Hull:2001iu,DuboisViolette:2001jk,Bekaert:2002dt}.\footnote{The
  construction of an interacting theory solely in terms of the dual graviton
  is impossible under rather general assumptions~\cite{Bekaert:2002uh}. For a
  non-linear model containing both the graviton and its dual as well as an
  auxiliary field see~\cite{Boulanger:2008nd,Nurmagambetov:2008yg}.} 
Now
consider the dimensional reduction of the graviton and the dual graviton which
is summarized for the reduction from five to four dimensions in
Figure~\ref{dgred}. As is evident from the diagram, and also generally,
the reduction will not produce dual fields in a balanced way. The graviton in
$D$ dimensions gives a scalar, a Maxwell field and the graviton in $D-1$
dimensions. The dual graviton in $D$ dimensions gives a dual scalar, a dual
vector, a dual graviton in $D-1$ dimensions and one more field of mixed
symmetry type. This extra field has hook symmetry of type $(D-3,1)$ in $(D-1)$
dimensions and arises from the dual graviton in higher dimensions when none of
the indices lies in the extra direction. For the reduction from $D=11$ to
$D=10$ this is exactly the $(8,1)$ form that arises on level $(3,3)$ in the 
decomposition of $E_{10}$.\footnote{Going down in dimensions one finds parts
  of the $E_{11-D}$ multiplets of the conjectured trombone deformation tensors
  to arise in this way while others come also from different sources.}  

The extra field carries no degrees of freedom but it could change the dynamics 
if it is non-zero, in analogy with the scalar example above. The reason why
the dual scalar field in the $p$-form example influenced the reduced dynamics
was that one could associate in a gauge invariant manner a constant to its
field strength. In other words, the top de Rham cohomology is non-empty, and
one-dimensional. Repeating the same kind of analysis for the mixed symmetry
type tensor arising from the reduction of the dual graviton one finds instead
that there is no gauge invariant quantity one could associate with because the
corresponding cohomology is empty~\cite{Olver,DuboisViolette:2001jk,Bekaert:2002dt}. For this reason we
are led to the conclusion that the immediate association of the trombone
gauging with the $(D-2,1)$-forms appearing in the $E_{10}$ or $E_{11}$ tables
is problematic. One possible resolution of this puzzle could be to add a new
generator to the algebra which transforms as a $9$-form in eleven 
dimensions. This new generator would appear for the first time at height $30$
and change the multiplicity of the null root of $E_{10}$ (or $E_{11}$) from
eight to nine.In the eleven-dimensional language it would, after using the
dictionary~\cite{Damour:2002cu,Damour:2004zy}, correspond to  
the trace of the spin connection which is always set to zero when establishing
the correspondence. Naturally, this term would contribute to the 
$(D-1)$-forms in $D$ dimensions and would transform in the same $E_{11-D}$
representation as the $(D-2,1)$ hook which was conjectured for the trombone 
gauging. In this way one would also find the right object to treat trombone
gauging in the level decomposition. It is not clear, however, that this is the
right path to pursue since it is known that adding this generator creates
problems in the $E_{10}$ sigma model~\cite{Damour:2007dt}. One possible
habitat for such a new generator outside the $E_{10}$ algebra would be the 
vertex operator algebra based on the $E_{11}$ root lattice. This possibility
is investigated in some more depth in Appendix~\ref{app:trombone} where it is
argued that also the trombone gaugings would not be recovered correctly from
this modification of the algebra.

Irrespective of these speculations about new generators, one can analyse the
effect of the hook on level $(3,3)$ on the geodesic equations. This is carried
out in Appendix~\ref{app:trombone} and we verify that for $D=10$ the
interpretation of the mixed symmetry field as being associated to the trombone
is correct forasmuch as it modifies the equations of motion of the IIA
supergravity  fields in almost the same way as is to be expected from the
deformed theory of~\cite{Howe:1997qt,Lavrinenko:1997qa}. The final resolution
of this question could shed new light on the role of mixed symmetry fields
appearing in the level decomposition of $E_{10}$ (and $E_{11}$).

\vskip5mm 

\noindent {\bf Acknowledgements} \\ 
We are grateful to Xavier Bekaert, Eric A. Bergshoeff, Jarah Evslin, Ulf Gran,
Laurent Houart, Olaf 
Hohm, Bernard Julia, Carlo Maccaferri, Hermann Nicolai, Teake Nutma, Jakob Palmkvist,
Christoffer Petersson, Henning Samtleben and Alexander Wijns for interesting
discussions. We would also like to thank Ulf Gran for extensive aid in the use
of the computer program GAMMA. M.H. and D.P. are grateful to the `Laboratoire
de Physique Th\'eorique de l'Ecole Normale Sup\'erieure' for kind hospitality
while this work was being finalized. A.K. thanks the Eidgen\"ossische
Technische Hochschule Z\"urich for its generous hospitality during part of
this work. E.J. is a FRIA bursar of Fonds de la Recherche Scientifique--FNRS,
Belgium. A.K. is a Research Associate of the Fonds de la Recherche
Scientifique--FNRS, Belgium. Work supported in part by IISN-Belgium
(conventions 4.4511.06 and 4.4514.08), by the European Commission FP6 RTN
programme MRTN-CT-2004-005104 and by the Belgian Federal Science Policy Office
through the Interuniversity Attraction Pole P6/11. 

\newpage

\appendix 

\section{Details for massive IIA supergravity} 
\label{Appendix:mIIAdetails}

In this appendix we give all the relevant details of massive IIA
supergravity which are required to establish the correspondence with the $E_{10}$-sigma model. The complete Lagrangian and supersymmetry variations were already given in Section 2, and will not be repeated here. Here we complement this with information regarding our conventions, as well as explicit expressions for the bosonic and fermionic equations of motion, and the Bianchi identities. Moreover, we discuss in detail the truncations that we need to impose on the supergravity side to ensure the matching with the geodesic sigma model. Finally, we also discuss how our conventions for massless type IIA supergravity matches with a reduction from eleven-dimensional supergravity.

\subsection{Conventions} 
\label{app:conventions}

We use the signature $(-+\ldots +)$ for space-time. The indices $\mu=(t,m)$
are $(1+9)$-dimensional curved indices, whereas $\alpha=(0,a)$ are the
corresponding flat indices.  Partial derivatives with flat indices are defined
via conversion with the inverse vielbein: $\p_\alpha = e_\alpha{}^\mu
\p_\mu$. The Lorentz covariant derivative $D_\al$ acts on (co-)vectors via
$D_\al 
V_\beta= \p_\al V_\beta + \omega_{\al\,\beta}{}^\gamma V_\gamma$ in terms of
the Lorentz connection, which is defined in turn as
\be
\omega_{\al\,\beta\gamma} = \frac12\left(\Omega_{\al\beta\,\gamma} 
  + \Omega_{\gamma\al\,\beta} -\Omega_{\beta\gamma\,\al}\right)
\ee
in terms of the anholonomy of the orthonormal frame $e_\mu{}^\al$ defined by 
$\Omega_{\mu\nu}{}^\al= 2\p_{[\mu}e_{\nu]}{}^\al$.

Our fermions are Majorana--Weyl spinors of $SO(1,9)$ of real dimension
$16$. As the theory is type II non-chiral we can combine two spinors into a
$32$-dimensional Majorana representation on which the $\Gamma$-matrices of
$SO(1,10)$ act. These are the eleven real $32\times 32$ matrices $(\Gamma^0,
\Gamma^a, \Gamma^{10})$ which are symmetric except for $\Gamma^0$ which is
antisymmetric. We choose the representation such that $\Gamma^{10}$ is block
diagonal and projects on the two $16$ component spinors of opposite
chirality. $\Gamma^0$ is the charge conjugation matrix such that our
conventions are identical to those of~\cite{Kleinschmidt:2004dy}. 

A useful identity for our $\Gamma$-matrices is 
\be
\Gamma^{a_1\ldots a_k} = \frac{(-1)^{(k+1)(k+2)/2}}{(9-k)!}\epsilon^{a_1\ldots
  a_k b_1\ldots b_{9-k}} \Gamma_{b_1\ldots b_{9-k}} \Gamma^0\Gamma^{10}\,. 
\ee
The various $\epsilon$ tensors we use are such that 
\be
\eps^{0\,1\ldots 10} = +1\,,\quad 
\eps^{0\,1\ldots 9} = +1\,,\quad 
\eps^{1\,\ldots 9} = +1\,.
\ee

\subsection{Bianchi identities} 

For the comparison with $E_{10}$ it is useful to write all supergravity
equations in a non-coordinate orthonormal frame described by the vielbein
$e_\mu{}^\al$ and use only Lorentz covariant objects. In these flat 
indices the Bianchi identities following from (\ref{curvs}) are 
\begin{subequations}\label{Bianchis} 
\be
\label{SugraBianchi1} 3D_{[\al_1} F_{\al_2\al_3]} &=& m F_{\al_1\al_2\al_3} \,,\\ 
4D_{[\al_1} F_{\al_2\al_3\al_4]} &=& 0 \,,\\
5D_{[\al_1} F_{\al_2\al_3\al_4\al_5]} &=&  
  10 F_{[\al_1\al_2} F_{\al_3\al_4\al_5]} \,. 
\ee 
\end{subequations}

\subsection{Bosonic equations of motion}
\label{SugraBeom}

The form equations of motion can be rewritten in $10$-dimensional flat indices 
as 
\begin{subequations}\label{eoms} 
\be
\label{eom1}D_\alpha(e^{3\phi/2}F^{\alpha\beta}) 
   &=& -\frac1{3!}e^{\phi/2} F^{\alpha_1\alpha_2\alpha_3\beta}F_{\alpha_1\alpha_2\alpha_3} \,,\\
\label{eom2}D_\alpha(e^{-\phi}F^{\alpha\beta_1\beta_2})  
   &=& m e^{3\phi/2} F^{\beta_1\beta_2} +\frac1{2!}e^{\phi/2}
   F^{\beta_1\beta_2\alpha_1\alpha_2} F_{\alpha_1\alpha_2} \nn\\ 
    &&    - \frac1{1152}F_{\alpha_1\ldots \alpha_4} F_{\alpha_5\ldots \alpha_8}\eps^{\alpha_1\ldots \alpha_8\beta_1\beta_2}\,,\\ 
\label{eom3}D_\alpha(e^{\phi/2}F^{\alpha\beta_1\ldots \beta_3})
   &=& \frac1{144} F_{\alpha_1\ldots \alpha_4}F_{\alpha_5\ldots
     \alpha_7}\eps^{\alpha_1\ldots \alpha_7\beta_1\ldots \beta_3} ,
\ee
\end{subequations} 
while the dilaton and gravity equations are 
\begin{subequations}\label{lev0}
\be
D^\alpha \partial_\alpha \phi &=& \frac38e^{3\phi/2} |F_{(2)}|^2 
   - \frac1{12} e^{-\phi} |F_{(3)}|^2
   + \frac1{96} e^{\phi/2} |F_{(4)}|^2 
   + \frac54 m^2 e^{5\phi/2} \,,\label{dilsugra}\\
R_{\alpha\beta} &=& \frac12 \partial_\alpha\phi \partial_\beta\phi  
   +\frac{m^2}{16} \eta_{\alpha\beta}e^{5\phi/2} 
   +\frac12e^{3\phi/2}F_{\alpha\gamma}F_\beta{}^\gamma 
       -\frac1{32}\eta_{\alpha\beta} e^{3\phi/2}F_{\gamma_1\gamma_2}F^{\gamma_1\gamma_2} \nn\\
&&\quad  +\frac14 e^{-\phi}
F_{\alpha\gamma_1\gamma_2}F_\beta{}^{\gamma_1\gamma_2}   
       -\frac1{48} \eta_{\alpha\beta} e^{-\phi}
       F_{\gamma_1\gamma_2\gamma_3}F^{\gamma_1\gamma_2\gamma_3}\label{Einsteinsugra}\\ 
&&\quad  +\frac1{12} e^{\phi/2} F_{\alpha\gamma_1\gamma_2\gamma_3}F_\beta{}^{\gamma_1\gamma_2\gamma_3}  
       -\frac1{128} \eta_{\alpha\beta} e^{\phi/2}
       F_{\gamma_1\gamma_2\gamma_3\gamma_4}F^{\gamma_1\gamma_2\gamma_3\gamma_4}.\nn 
\ee
\end{subequations} 

\subsection{Fermionic equations of motion} 
\label{app:sugraf}

Besides the bosonic equations one also deduces the fermionic equations of 
motion from the Lagrangian (\ref{lag}) which we write out in flat indices. For
the dilatino this gives 
\be 
\label{eomlam} 
\G^\al D_\al \lt &-&
\f{5}{32}e^{3\phi/4}F_{\al_1\al_2}\G^{\al_1\al_2}\G_{10}\lt 
  +\f3{16}e^{3\phi/4}F_{\al_1\al_2}\G^{\beta}\G^{\al_1\al_2}\G_{10}\pt_\beta\nn\\ 
&+&\f1{24}e^{-\phi/2}F_{\al_1\cdots\al_3}\G^\beta\G^{\al_1\cdots\al_3}\G_{10}\pt_\beta\nn\\ 
&+&\f1{128}e^{\phi/4}F_{\al_1\cdots \al_4}\G^{\al_1\cdots \al_4}\lt
  - \f1{192} e^{\phi/4}F_{\al_1\cdots \al_4}\G^\beta\G^{\al_1\cdots \al_4}\pt_\beta\nn\\
&-&\f1{2}\p_\al\phi\G^\beta\G^\al\pt_\beta\nn\\ 
&-&\f{21}{16}me^{5\phi/4}\lt-\f5{8}me^{5\phi/4}\G^\al\pt_\al = 0, 
\ee
The  gravitino equation obtained directly from the variation of (\ref{lag}) is
of the form 
\beq
\label{psisimple}
\mathcal E^\alpha = \G^{\alpha\beta\gamma} D_\beta\pt_{\gamma}+R^{\alpha}=0.
\eeq
After multiplication with two gamma matrices it can be rewritten as
\beq
\label{psisimple+}
E_{\alpha}=\G^{\beta}(D_\alpha\pt_{\beta}-D_\beta\pt_{\al})+L_{\alpha}=0,
\eeq
where
\beq
L_{\alpha}=\f18(\G_{\alpha\beta}R^{\beta}-7R_{\alpha}),
\eeq
Although $\mathcal E^\alpha=0$ is equivalent to $E^\alpha=0$, the spatial
components $E^a$ and $\mathcal E^a$ are only equivalent when the
supersymmetry constraint $\mathcal E^0$ is also taken into account. It turns out
that the dynamical equation that corresponds directly to the $K(E_{10})$ Dirac
equation is $E^a=0$, not too surprisingly since in this form one obtains
directly a Dirac equation for $\psi_a$. The $SO(1,9)$ covariant equation
$E_{\alpha}=0$ reads explicitly as follows:  
\be\label{eompsi} 
\G^\beta\left( D_\al\pt_\beta -D_\beta\pt_\al\right)  
  &+&\f{21}{64} e^{3\phi/4} F_{\al\beta}\G^\beta \G_{10}\lt  
  -\f{3}{128} e^{3\phi/4} F_{\beta_1\beta_2}\G_{\al}{}^{\beta_1\beta_2}
  \G_{10}\lt\nn\\ 
&+& \f1{64}e^{3\phi/4} F_{\beta_1\beta_2} \G_{\al}{}^{\beta_1\beta_2\gamma}\G_{10}\pt^\gamma
  -\f1{32} e^{3\phi/4} F_{\beta_1\beta_2}
  \G_{\al}{}^{\beta_1}\G_{10}\pt^{\beta_2}\nn\\ 
&-&\f7{32} e^{3\phi/4} F_{\al\beta}\G^{\beta\gamma}\G_{10}\pt_\gamma
  -\f7{64} e^{3\phi/4} F_{\beta_1\beta_2}\G^{\beta_1\beta_2}\G_{10}\pt_\al
  \nn\\
&+& \f7{32} e^{3\phi/4} F_{\al\beta}\G_{10}\pt^\beta\nn\\ 
&+&\f1{96} e^{-\phi/2} F_{\beta_1\beta_2\beta_3}
\G_{\al}{}^{\beta_1\beta_2\beta_3}\G_{10}\lt 
 -\f3{32} e^{-\phi/2} F_{\al\beta_1\beta_2} \G^{\beta_1\beta_2}\G_{10}\lt\nn\\ 
&+& \f1{96} e^{-\phi/2} F_{\beta_1\beta_2\beta_3} \G_{\al}{}^{\beta_1\beta_2\beta_3\gamma}\G_{10}\pt^\gamma
  +\f1{32} e^{-\phi/2} F_{\beta_1\beta_2\beta_3}
  \G^{\beta_1\beta_2\beta_3}\G_{10}\pt_\al\nn\\ 
&-& \f1{32} e^{-\phi/2} F_{\beta_1\beta_2\beta_3}
\G_{\al}{}^{\beta_1\beta_2}\G_{10}\pt^{\beta_3} 
  -\f3{32} e^{-\phi/2} F_{\al\beta_1\beta_2} \G^{\beta_1\beta_2\gamma}\G_{10}\pt_\gamma\nn\\
&+& \f3{16} e^{-\phi/2} F_{\al\beta_1\beta_2}
\G^{\beta_1}\G_{10}\pt^{\beta_2}\nn\\ 
&-&\f1{512} e^{\phi/4} F_{\beta_1\ldots \beta_4} \G_{\al}{}^{\beta_1\ldots \beta_4}\lt
  +\f5{384} e^{\phi/4} F_{\al\beta_1\ldots \beta_3} \G^{\beta_1\ldots
    \beta_3}\lt\nn\\ 
&-& \f1{256} e^{\phi/4} F_{\beta_1\ldots \beta_4} \G_{\al}{}^{\beta_1\ldots\beta_4\gamma}\pt^\gamma
  +\f5{768} e^{\phi/4} F_{\beta_1\ldots \beta_4} \G^{\beta_1\ldots
    \beta_4}\pt_\al\nn\\ 
&+& \f1{64} e^{\phi/4} F_{\beta_1\ldots \beta_4} \G_{\al}{}^{\beta_1\ldots \beta_3}\pt^{\beta_4}
  +\f5{192} e^{\phi/4} F_{\al\beta_1\ldots \beta_3} \G^{\beta_1\ldots
    \beta_3\gamma}\pt_\gamma\nn\\ 
&-& \f5{64} e^{\phi/4} F_{\al\beta_1\ldots \beta_3}\G^{\beta_1\beta_2}\pt^{\beta_3} +\f12\p_\al \phi \lt\nn\\
&+& \f1{32}e^{5\phi/4}m \G_{\al\beta}\pt^\beta +\f9{32}e^{5\phi/4}m \pt_\al
+\f5{64}e^{5\phi/4} m \G_\al \lt=0 \,. 
\ee

\subsection{Truncation on the supergravity side}
\label{truncation} 

As explained in Section \ref{EOMandTruncations} of the main text, the correspondence between the dynamics of the $E_{10}$-invariant sigma model and the dynamics of type IIA supergravity only works if a certain truncation is applied. 

As dictated from the BKL analysis, the following truncations must be imposed on the equations of motion and Bianchi identities:
\begin{subequations}\label{truncbos}\be
\p_a (N \p_0 \phi) = \p_a ( N \omega_{0\,ab}) = \p_a (N\omega_{a\,b 0})  
= \p_a(N\p_b\phi) = \p_a (N\omega_{b\,cd}) =0 
\ee 
and 
\be \label{truncsugra}
\p_a\left(N e^{3\phi/4} F_{0b}\right)  
= \p_a\left(N e^{-\phi/2} F_{0b_1b_2}\right) 
= \p_a\left(N e^{\phi/4} F_{0b_1b_2b_3}\right)&=&0\,,\nn\\ 
\p_a\left(N e^{3\phi/4} F_{b_1b_2}\right) 
= \p_a\left(N e^{-\phi/2} F_{b_1b_2b_3}\right)
= \p_a\left(N e^{\phi/4} F_{b_1b_2b_3b_4}\right) &=&0\,,\nn\\
\p_a \left(N e^{5\phi/4}m\right) &=&0\,. 
\ee 
Furthermore, as already indicated in (\ref{spintrace}), the spatial trace of the spin connection has to be set to zero
\be
\omega_{b\,ba} = 0\,.
\ee
\end{subequations} 
Equations (\ref{truncbos}) exhaust all truncations of the bosonic variables. As explained in the text, with these truncations the bosonic geodesic equations agree with the supergravity up to one term in the Einstein equation coming from the contribution to the Ricci tensor $R_{ab}$ which is proportional to $\Omega_{ac\,d}\Omega_{bd\,c}$. We emphasize that there are no mismatches associated with the mass parameter $m$.
 
For the fermionic variables one also needs to apply appropriate truncations of spatial gradients. These turn out to be
\be\label{fermtrunc}
N^{-1} \p_a (N\lt) = N^{-1/2} \p_a ( N^{1/2} \pt_b ) = 0\,.
\ee
With this choice of truncation the Dirac equation of the coset match exactly
the fermionic equations of motion of supergravity if in addition the
supersymmetric gauge 
\be
\pt_0-\G_0\G^a\pt_a = 0 
\ee 
of (\ref{susygauge}) is adopted.

\subsection{Reduction from $D=11$} 
 Our conventions for massless type IIA supergravity are consistent with the
 reduction of eleven-dimensional supergravity through the reduction ansatz
 and field redefinitions given in this section. This construction of massless
 type IIA was first carried out
 in~\cite{Giani:1984wc,Campbell:1984zc,Huq:1983im}. In this section only, we
 will denote the $D=11$ gravitino by $\Psi_M$.
 
 The supersymmetry variation of the gravitino in eleven-dimensional
supergravity is  
\beq 
\delta_{\varepsilon^{(11)}}\Psi_M^{(11)}=D_M^{(11)}\varepsilon^{(11)}+\f{1}{288}\Big({\Gamma_M}^{N_1\cdots N_4}-8\delta_M^{[N_1}\Gamma^{N_2N_3N_4]}\Big)\varepsilon^{(11)}F^{(11)}_{N_1\cdots N_4}.
\eeq 
We reduce this expression along $x^{10}$ with the following ansatz for the eleven-dimensional vielbein:
\beq 
{E_M}^{A}=\left(\begin{array}{cc}
e^{-\f{1}{12}\phi}{e_\mu}^{\al} & e^{\f{2}{3}\phi}A_\mu\\ 
0 & e^{\f{2}{3}\phi}\\ 
\end{array}\right). 
\eeq
The four-form field strength is reduced as follows in curved indices
\be
{}F^{(10)}_{\mu\nu\rho}&\equiv & F^{(11)}_{\mu\nu\rho\tten}, 
\nn \\
{}F^{(10)}_{\mu\nu\rho\sigma}&\equiv &
F^{(11)}_{\mu\nu\rho\sigma}+4A_{[\mu}F^{(10)}_{\nu\rho\sigma]}, 
\ee
where $\tten$ denotes a curved index. The eleven-dimensional gravitino
$\Psi_M^{(11)}$ splits into the ten-dimensional gravitino $\pt_\mu$ and the
dilatino $\lt$, according to the following field redefinitions 
\be \label{kkferm}
{}\Psi^{(11)}_\mu &=& e^{-\f{1}{24}\phi}\big(\pt_\mu-\f{1}{12}\Gamma_\mu\lt\big)+\f{2}{3}e^{\f{1}{24}\phi}\Gamma_{\tten}A_\mu\lt,
\nn \\ 
{}\Gamma^{\tten}\Psi^{(11)}_{\tten}&=&\f{2}{3}e^{\f{1}{24}\phi}\lt. 
\ee 
We also rescale the supersymmetry parameter according to
\beq 
\vet\equiv e^{\f{1}{24}\phi}\varepsilon^{(11)}. 
\eeq
 
\section{Details on the IIA level decomposition of $\mf{e}_{10}$ and $\mf{k}(\mf{e}_{10})$}
\label{Appendix:LevelDecomp} 
 
In this appendix we give all the details of the level deomposition of $\mf{e}_{10}$ with respect to $\mf{sl}(9, \mbb{R})$, up to level $(\ell_1, \ell_2)=(4,1)$. In particular we give all the relevant $\mf{e}_{10}$ commutators which are needed to compute the explicit expressions of the bosonic and fermionic equations of motion in Appendix \ref{app:EOMe10}. Moreover, we give details on the spinor and and vector-spinor representations of $\mf{k}(\mf{e}_{10})$. Finally, we also extend the level decomposition up to level $(3,3)$, whose generator plays a role in the discussion in Section~\ref{sec:trombone} in relation to the trombone symmetry and whose role in $\mf{e}_{10}$ is studied in Appendix~\ref{app:trombone}.

\begin{table} 
\centering 
\begin{tabular}{|c|c|c|}
\hline
$(\ell_1,\ell_2)$&$\mf{sl}(9, \mbb{R})$ Dynkin labels&$\mf{e}_{10}$ root $\alpha$ of lowest weight \\
\hline\hline
$(0,0)$&$[1,0,0,0,0,0,0,1]\oplus [0,0,0,0,0,0,0,0]$&$(-1,-1,-1,-1,-1,-1,-1,-1,0,0)$\\
$(0,0)$&$[0,0,0,0,0,0,0,0]$&$(0,0,0,0,0,0,0,0,0,0)$\\
$(0,1)$&$[0,0,0,0,0,0,0,1]$&$(0,0,0,0,0,0,0,0,1,0)$\\
$(1,0)$&$[0,0,0,0,0,0,1,0]$&$(0,0,0,0,0,0,0,0,0,1)$\\
$(1,1)$&$[0,0,0,0,0,1,0,0]$&$(0,0,0,0,0,0,1,1,1,1)$\\
$(2,1)$&$[0,0,0,1,0,0,0,0]$&$(0,0,0,0,1,2,3,2,1,2)$\\
$(2,2)$&$[0,0,1,0,0,0,0,0]$&$(0,0,0,1,2,3,4,3,2,2)$\\
$(3,1)$&$[0,1,0,0,0,0,0,0]$&$(0,0,1,2,3,4,5,3,1,3)$\\
$(3,2)$&$[1,0,0,0,0,0,0,0]$&$(0,1,2,3,4,5,6,4,2,3)$\\
$(3,2)$&$[0,1,0,0,0,0,0,1]$&$(0,0,1,2,3,4,5,3,2,3)$\\
$(4,1)$&$[0,0,0,0,0,0,0,0]$&$(1,2,3,4,5,6,7,4,1,4)$\\
\hline
$(3,3)$&$[1,0,0,0,0,0,0,1]$&$(0,1,2,3,4,5,6,4,3,3)$\\
\hline
\end{tabular}
\caption{\label{longerdec}\sl $\mf{sl}(9, \mbb{R})$ level decomposition of
  $\mf{e}_{10}$ with root vectors. All shown levels are complete. The very
  last entry is that of a mixed symmetry generator not studied for the
  dictionary of Table~\ref{dicoeom}. Its possible relation to trombone
  gauging is discussed in Section~\ref{sec:trombone} and in
  Appendix~\ref{app:trombone}.} 
\end{table}

\subsection{Commutation relations for fields appearing in the dictionary}
\label{app:commutators}

At level $(\ell_1, \ell_2)=(0,0)$ there is a copy of $\mf{gl}(9, \mbb{R})=\mf{sl}(9, \mbb{R})\oplus \mbb{R}$, as well as a scalar generator associated with the dilaton. Their relations are ($a,b=1,\ldots, 9$)
\begin{align}
\lb K^a{}_b , K^c{}_d \rb &= \delta^c_b K^a{}_d - \delta^a_d K^c{}_b\,,\quad&
     \langle K^a{}_b| K^c{}_d \rangle &= \delta^a_d\delta^c_b 
       - \delta^a_b\delta^c_d \,,&\nn\\
\lb T, K^a{}_b \rb &= 0 \,,\quad &
    \langle T | T \rangle &=\frac12 \,,&\quad \langle T|K^a{}_b\rangle =0\,.
\end{align}
Here, $\langle\cdot|\cdot\rangle$ is the invariant bilinear form. We define also the trace $K=\sum_{a=1}^9 K^a{}_a$. For completeness
\be
K &=& 8 h_1 + 16 h_2 + 24 h_3 + 32 h_4 + 40 h_5 + 48 h_6 
      + 56 h_7 +37 h_8 + 18 h_9 + 27 h_{10}\,,\nn\\
T &=& \frac12 h_1 + h_2 +\frac32 h_3 + 2 h_4 + \frac52 h_5 + 3 h_6
      +\frac72 h_7 +\frac{25}{12}h_8 + \frac23 h_9 + \frac{23}{12}h_{10}\,.
\ee

All objects transform as $\mf{gl}(9, \mbb{R})$ tensors in the obvious way.
The $T$ commutator relations are
\begin{align}
\lb T, E^{a_1} \rb &= \frac34 E^{a_1} \,,& 
   \lb T, E^{a_1a_2} \rb &= -\frac12  E^{a_1a_2} \,, \nn\\
\lb T, E^{a_1a_2a_3} \rb &= \frac14 E^{a_1a_2a_3} \,,&
   \lb T, E^{a_1\ldots a_5} \rb &= -\frac14  E^{a_1\ldots a_5} \,, \nn\\
\lb T, E^{a_1\ldots a_6} \rb &= \frac12 E^{a_1\ldots a_6} \,,&
   \lb T, E^{a_1\ldots a_7} \rb &= -\frac34  E^{a_1\ldots a_7} \,, \\
\lb T, E^{a_1\ldots a_9} \rb &= -\frac54 E^{a_1\ldots a_9} \,,&\nn\\
\lb T, E^{a_0|a_1\ldots a_7} \rb &= 0 \,,&
   \lb T, E^{a_1\ldots a_8} \rb &= 0 \,.\nn 
\end{align}
The positive level generators are generated by the simple (fundamental)
generators on levels $(0,1)$ and $(1,0)$ by
\begin{align}\label{poslevs}
\lb E^{a_1}, E^{a_2} \rb &= 0 \,,&
  \lb E^{a_1a_2}, E^{a_3a_4} \rb &= 0 \,,&\nn\\
\lb E^{a_1a_2}, E^{a_3} \rb &= E^{a_1a_2a_3} \,,&
  \lb E^{a_1a_2}, E^{a_3\ldots a_5} \rb &= E^{a_1\ldots a_5} \,,& \nn\\
\lb E^{a_1a_2}, E^{a_3\ldots a_7} \rb &= E^{a_1\ldots a_7} \,,&
  \lb E^{a_1a_2}, E^{a_3\ldots a_9} \rb &= E^{a_1\ldots a_9} \,,&\\
\lb E^{a_1}, E^{a_2\ldots a_6} \rb &= E^{a_1\ldots a_6} \,,& 
  \lb E^{a_0} , E^{a_1\ldots a_7} \rb &= E^{a_0|a_1\ldots a_7} + \frac32
    E^{a_0a_1\ldots a_7} \,.\nn
\end{align}
These defining relations imply for example
\be
\lb
E^{a_1a_2a_3}, E^{a_4a_5a_6}\rb &=& -E^{a_1\ldots a_6} \,,\nn\\
\lb E^{a_1a_2}, E^{a_3\ldots a_8} \rb &=& 
    -2 E^{[a_1|a_2]a_3\ldots a_8} + E^{a_1\ldots a_8}\,,\nn\\
\lb E^{a_1a_2a_3}, E^{a_4\ldots a_8} \rb &=& 
    -3 E^{[a_1|a_2a_3]a_4\ldots a_8} -\frac12 E^{a_1\ldots a_8}\,,\nn\\
\lb E^{a_1\ldots a_5}, E^{a_6a_7a_8} \rb &=&
    5 E^{[a_1|a_2\ldots a_5]a_6a_7a_8} -\frac12 E^{a_1\ldots a_8}\,,\\
\lb E^{a_1\ldots a_6}, E^{a_7a_8} \rb &=&
    -6 E^{[a_1|a_2\ldots a_6]a_7a_8} - E^{a_1\ldots a_8}\,,\nn\\
\lb E^{a_1\ldots a_7}, E^{a_8} \rb &=&
    -7 E^{[a_1|a_2\ldots a_7]a_8} +\frac32 E^{a_1\ldots a_8}\,.\nn
\ee
The Young symmetry on the dual graviton implies $E^{a_0|a_1\ldots a_7} =
7E^{[a_1|a_2\ldots a_7]a_0}$. The two irreducible representations on $(3,2)$
are projected onto via
\be
E^{a_1\ldots a_8} &=& \frac23 \lb E^{[a_1}, E^{a_2\ldots a_8]} \rb \,,\quad
E^{a_0|a_1\ldots a_7} = \frac78\left(\lb E^{a_0}, E^{a_1\ldots a_7} \rb
       + \lb E^{[a_1}, E^{a_2\ldots a_7]a_0}\rb \right) \,,\nn\\
E^{a_0|a_1\ldots a_7} &=& -\frac78\left(\lb E^{a_0[a_1}, E^{a_2\ldots a_7]} \rb
       + \lb E^{[a_1a_2}, E^{a_3\ldots a_7]a_0}\rb \right) \,.
\ee 
The definitions (\ref{poslevs}) are such that the normalisations are
\be
\langle E^{a_1\ldots a_p} | F_{b_1\ldots b_p}\rangle
   &=& p!\,\delta^{a_1\ldots  a_p}_{b_1\ldots b_p} \,
              \quad\quad\quad\Rightarrow\quad 
   \langle E^{1\,\ldots\,p} |F_{1\,\ldots\,p} \rangle =1 \quad
   (p\ne8)\,,\nn\\ 
\langle E^{a_1\ldots a_8} | F_{b_1\ldots b_8}\rangle
  &=& \frac12\cdot 8! \delta^{a_1\ldots  a_8}_{b_1\ldots b_8} \,
              \quad\quad\Rightarrow\quad 
   \langle E^{1\,\ldots\,8} |F_{1\,\ldots\,8} \rangle =\frac12\,,\nn\\
\langle E^{a_0|a_1\ldots a_7} | F_{b_0|b_1\ldots b_7}\rangle
  &=& \frac78\cdot 7! \left(\delta^{a_0}_{b_0}
       \delta^{a_1\ldots a_7}_{b_1\ldots b_7} 
    + \delta^{[a_1}_{b_0} \delta^{a_2\ldots a_7]a_0}_{b_1\,\,\ldots\ldots\,\,
      b_7}\right)\,. 
\ee
The additional factor of $2$ in the normalisation of the $8$-form is chosen
such that all structure constants remain rational.

Defining the transposed generators via $F\equiv E^T$ as usual gives the
following commutations relation between the form generators and
their transposes:
\be\label{ef0}
\lb E^a, F_b \rb &=& -\frac18 \delta^a_b K + K^a{}_b 
   +\frac32 \delta^a_b T \,,\nn\\
\lb E^{a_1a_2}, F_{b_1b_2} \rb &=& -\frac12 \delta^{a_1a_2}_{b_1b_2} K 
    +4 \delta^{[a_1}_{[b_1} K^{a_2]}_{\,\,\,\,\, b_2]}   
    -2 \delta^{a_1a_2}_{b_1b_2} T \,,\nn\\
\lb E^{a_1a_2a_3}, F_{b_1b_2b_3} \rb &=& 
    -\frac38\cdot 3!\, \delta^{a_1a_2a_3}_{b_1b_2b_3} K 
    +3\cdot 3!\, \delta^{[a_1a_2}_{[b_1b_2} K^{a_3]}_{\,\,\,\,\, b_3]}   
    +3 \delta^{a_1a_2a_3}_{b_1b_2b_3} T \,,\nn\\
\lb E^{a_1\ldots a_5}, F_{b_1\ldots b_5} \rb &=& 
    -\frac58\cdot 5!\, \delta^{a_1\ldots a_5}_{b_1\ldots b_5} K 
    +5\cdot 5!\, \delta^{[a_1\ldots a_4}_{[b_1\ldots b_4} 
             K^{a_5]}_{\,\,\,\,\, b_5]}   
    -\frac12\cdot 5!\, \delta^{a_1\ldots a_5}_{b_1\ldots b_5} T \,,\nn\\
\lb E^{a_1\ldots a_6}, F_{b_1\ldots b_6} \rb &=& 
    -\frac34\cdot 6!\, \delta^{a_1\ldots a_6}_{b_1\ldots b_6} K 
    +6\cdot 6!\, \delta^{[a_1\ldots a_5}_{[b_1\ldots b_5} 
             K^{a_6]}_{\,\,\,\,\, b_6]}   
    +6!\, \delta^{a_1\ldots a_6}_{b_1\ldots b_6} T \,,\\
\lb E^{a_1\ldots a_7}, F_{b_1\ldots b_7} \rb &=& 
    -\frac78\cdot 7!\, \delta^{a_1\ldots a_7}_{b_1\ldots b_7} K 
    +7\cdot 7!\, \delta^{[a_1\ldots a_6}_{[b_1\ldots b_6} 
             K^{a_7]}_{\,\,\,\,\, b_7]}   
    -\frac32\cdot 7!\, \delta^{a_1\ldots a_7}_{b_1\ldots b_7} T \,,\nn\\
\lb E^{a_1\ldots a_9}, F_{b_1\ldots b_9} \rb &=& 
    -\frac98\cdot 9!\, \delta^{a_1\ldots a_9}_{b_1\ldots b_9} K 
    +9\cdot 9!\, \delta^{[a_1\ldots a_8}_{[b_1\ldots b_8} 
             K^{a_9]}_{\,\,\,\,\, b_9]}   
    -\frac52\cdot 9!\, \delta^{a_1\ldots a_9}_{b_1\ldots b_9} T \,,\nn\\
\lb E^{a_1\ldots a_8}, F_{b_1\ldots b_8} \rb &=&
   -\frac12\cdot 8!\, \delta^{a_1\ldots a_8}_{b_1\ldots b_8} K 
     + 4\cdot 8!\, \delta^{[a_1\ldots a_7}_{[b_1\ldots b_7} 
             K^{a_8]}_{\,\,\,\,\, b_8]} \,.\nn
\ee
The commutator of the dual graviton generator $E^{a_0|a_1\ldots a_7}$ can be
most conveniently written using a dummy tensor $X_{a_0|a_1\ldots a_7}$ as 
\be\label{efdg}
\lb  F_{b_0|b_1\ldots b_7} ,X_{a_0|a_1\ldots a_7}E^{a_0|a_1\ldots a_7}\rb = 
  7!\left(X_{b_0|b_1\ldots b_7} K - X_{c|b_1\ldots b_7} K^c{}_{b_0} 
   -7 X_{b_0|c[b_1\ldots b_6}K^c_{\,\,\,b_7]} \right).
\ee

The generators of different rank commute in the following non-trivial way:
\begin{align}\label{ef}
\lb E^a , F_{b_1b_2b_3} \rb &= 3 \delta^a_{[b_1} F_{b_2b_3]} \,,&
\lb E^{a_1a_2}, F_{b_1b_2b_3} \rb &= 
    -6 \delta^{a_1a_2}_{[b_1b_2} F_{b_3]}\,,&\nn\\
\lb E^{a_1a_2} ,F_{b_1\ldots b_5} \rb &=
   -20 \delta^{a_1a_2}_{[b_1b_2}  F_{b_3b_4b_5]} \,,&
\lb E^{a_1a_2a_3} ,F_{b_1\ldots b_5} \rb &=
   60 \delta^{a_1a_2a_3}_{[b_1b_2b_3}  F_{b_4b_5]} \,,& \nn\\
\lb E^a, F_{b_1\ldots b_6} \rb &=
   -6 \delta^a_{[b_1} F_{b_2\ldots b_6]} \,,&
\lb E^{a_1a_2a_3}, F_{b_1\ldots b_6} \rb &= 
   120 \delta^{a_1a_2a_3}_{[b_1b_2b_3} F_{b_4b_5b_6]} \,,&\nn\\
\lb E^{a_1\ldots a_5}, F_{b_1\ldots b_6} \rb &=
   -6!\, \delta^{a_1\ldots a_5}_{[b_1\ldots b_5} F_{b_6]} \,,&
\lb E^{a_1a_2}, F_{b_1\ldots b_7} \rb &=
   -7\cdot 6 \delta^{a_1a_2}_{[b_1b_2} F_{b_3\ldots b_7]} \,,&\\
\lb E^{a_1\ldots a_5}, F_{b_1\ldots b_7} \rb &=
   \frac12\cdot 7!\, \delta^{a_1\ldots a_5}_{[b_1\ldots
     b_5}F_{b_6b_7]}\,,&
\lb E^{a_1a_2}, F_{b_1\ldots b_9} \rb &=
   -9\cdot 8 \delta^{a_1a_2}_{[b_1b_2}F_{b_3\ldots b_9]}\,,&\nn\\
\lb E^{a_1\ldots a_7}, F_{b_1\ldots b_9} \rb &= 
   \frac12\cdot 9!\,\delta^{a_1\ldots a_7}_{[b_1\ldots b_7}F_{b_8b_9]}\,.&\nn
\end{align}
Anticipating the geodesic equation we know that (\ref{ef}) describes all the
couplings between the different forms occurring in the matter equations, so
that for example the $9$-form (=mass term) occurs only in the Bianchi identity
for $F_{(2)}$ and in the eom of $F_{(3)}$, consistent with (\ref{Bianchis})
and (\ref{eoms}). The dilaton and Einstein equation are described by the
couplings of equations (\ref{ef0}) and (\ref{efdg}).

The commutators with the dual dilaton are
\begin{align}
\lb E^a, F_{b_1\ldots b_8} \rb &=
   -6 \delta^a_{[b_1} F_{b_2\ldots b_8]} \,,&
    \lb E^{a_1a_2}, F_{b_1\ldots b_8} \rb &= 
    - 4\cdot 7 \delta^{a_1a_2}_{[b_1b_2}F_{b_3\ldots b_8]} \,,&\nn\\
\lb E^{a_1a_2a_3}, F_{b_1\ldots b_8} \rb &=
    2\cdot 7\cdot 6 \delta^{a_1a_2a_3}_{[b_1b_2b_3}F_{b_4\ldots b_8]} \,,&
  \lb E^{a_1\ldots a_5}, F_{b_1\ldots b_8} \rb &=
    2\cdot 7\cdot 5!\, \delta^{a_1\ldots a_5}_{[b_1\ldots b_5}F_{b_6b_7b_8]} \,,&\nn\\
\lb E^{a_1\ldots a_6}, F_{b_1\ldots b_8} \rb &=
    2\cdot 7!\, \delta^{a_1\ldots a_6}_{[b_1\ldots b_6}F_{b_7b_8]} \,,&
  \lb E^{a_1\ldots a_7}, F_{b_1\ldots b_8} \rb &=
    -6\cdot 7!\, \delta^{a_1\ldots a_7}_{[b_1\ldots b_7}F_{b_8]} \,,&
\end{align}
whereas for the dual graviton one finds
\be
\lb E^a, F_{b_0|b_1\ldots b_7} \rb &=& 
   -\frac78\left(\delta^a_{b_0} F_{b_1\ldots b_7} + \delta^a_{[b_1}F_{b_2\ldots b_7]b_0}\right) \,, \nn\\
\lb E^{a_1a_2}, F_{b_0|b_1\ldots b_7} \rb &=& 
   \frac{21}2\left(\delta^{a_1a_2}_{b_0[b_1} F_{b_2\ldots b_7]} + \delta^{a_1a_2}_{[b_1b_2}F_{b_3\ldots b_7]b_0}\right) \,, \nn\\
\lb E^{a_1a_2a_3}, F_{b_0|b_1\ldots b_7} \rb &=& 
  \f{45\cdot 7}{4}\left(\delta^{a_1a_2a_3}_{b_0[b_1b_2} F_{b_3\ldots b_7]} + \delta^{a_1a_2a_3}_{[b_1b_2b_3}F_{b_4\ldots b_7]b_0}\right) \,, \nn\\
\lb E^{a_1\ldots a_5}, F_{b_0|b_1\ldots b_7} \rb &=& 
  -\frac{5 \cdot 7!}{16}\left(\delta^{a_1\ldots a_5}_{b_0[b_1\ldots b_4} F_{b_5\ldots b_7]} + \delta^{a_1\ldots a_5}_{[b_1\ldots b_5}F_{b_6b_7]b_0}\right) \,, \nn\\
\lb E^{a_1\ldots a_6}, F_{b_0|b_1\ldots b_7} \rb &=& 
  \frac{3 \cdot 7!}{4}\left(\delta^{a_1\ldots a_6}_{b_0[b_1\ldots b_5} F_{b_6b_7]} + \delta^{a_1\ldots a_6}_{[b_1\ldots b_6}F_{b_7]b_0}\right) \,, \nn\\
  \lb E^{a_1\ldots a_7}, F_{b_0|b_1\ldots b_7} \rb &=& 
  \frac{7 \cdot 7!}{8}\left(\delta^{a_1\ldots a_7}_{b_0[b_1\ldots b_6} F_{b_7]} + \delta^{a_1\ldots a_7}_{b_1\ldots b_7}F_{b_0}\right) \,.
\ee

\subsection{Spinor representations of $\mf{k}(\mf{e}_{10})$ }
\label{app:SpinorReps}

For the $E_{10}$ model to incorporate all low energy limits of M-theory in a single model, the fermionic representations used in (\ref{e10fermlag}) should not depend on the particular supergravity one wishes to study. Rather the unfaithful ${\bf 320}$ and ${\bf 32}$ representations of $\mf{k}(\mf{e}_{10})$ should be decomposed under a suitable subalgebra. Here, this subalgebra is $\mf{so}(9)\subset \mf{gl}(9,{\mathbb{R}})$ and this appendix provides the details of the action of the $\mf{k}(\mf{e}_{10})$ generators in this basis. When doing the following calculations we found the computer package GAMMA \cite{GAMMA} useful.\footnote{We are grateful to Ulf Gran for generous help in modifying GAMMA to suit our needs.}

\subsubsection{Dirac spinor}
\label{app:DiracSpinor}

The result of writing the $K(E_{10})$ action on the Dirac spinor is (we recall the notation $M^{a_1a_2}$ for the level $(0,0)$ generator of $K(E_{10})$ from (\ref{00gens}))
\begin{align}
{}M^{a_1a_2}\cdot \epsilon &= \f{1}{2}\Gamma^{a_1a_2}\epsilon ,& 
  \qquad J_{(0,1)}^{a}\cdot \epsilon &= \f{1}{2}\Gamma_{10}\Gamma^{a}\epsilon,&\nn \\
{} J_{(1,0)}^{a_1a_2}\cdot \epsilon &= \f{1}{2}\Gamma_{10}\Gamma^{a_1a_2}\eps,&
  \qquad  J_{(1,1)}^{a_1a_2a_3}\cdot \eps &=\f{1}{2}\Gamma^{a_1a_2a_3}\eps,&\nn \\
{} J_{(2,1)}^{a_1\cdots a_5}\cdot \eps&=\f{1}{2}\Gamma_{10}\Gamma^{a_1\cdots a_5}\epsilon,&
  \qquad  J_{(2,2)}^{a_1\cdots a_6}\cdot \eps&= -\f{1}{2}\Gamma^{a_1\cdots a_6}\eps,&\nn \\
{} J_{(3,1)}^{a_1\cdots a_7}\cdot \eps&= \f{1}{2}\Gamma^{a_1\cdots a_7}\eps,&
  \qquad J_{(3,2)}^{a_0|a_1\cdots a_7}\cdot \eps&=\f{7}{2}\Gamma_{10}\delta_{a_0}^{[a_1}\Gamma^{a_2\cdots a_7]}\eps,&\nn \\
{} J_{(3,2)}^{a_1\cdots a_8}\cdot \eps&=0 ,&
  \qquad J_{(4,1)}^{a_1\cdots a_9}\cdot \eps&=\f{1}{2}\Gamma_{10}\Gamma^{a_1\cdots a_9}\eps,&
\end{align}
where the generators above levels $(0,1)$ and $(1,0)$ are defined through the lower levels as follows 
\be
{}J_{(1,1)}^{a_1a_2a_3}\cdot\epsilon &:=& \lb J_{(1,0)}^{[a_1a_2}, J_{(0,1)}^{a_3]}\rb\cdot\epsilon ,
\nn \\
{}J_{(2,1)}^{a_1\cdots a_5}\cdot\epsilon &:= & \lb J_{(1,0)}^{[a_1a_2}, J_{(1,1)}^{a_3a_4a_5]}\rb\cdot\epsilon ,
\nn\\
{} J_{(2,2)}^{a_1\cdots a_6}\cdot\epsilon &:= & \lb J_{(1,1)}^{[a_1}, J_{(1,1)}^{a_2\ldots a_6]}\rb\cdot\epsilon ,
\nn \\
J_{(3,1)}^{a_1\cdots a_7}\cdot\epsilon &:= & \lb J_{(1,0)}^{[a_1a_2}, J_{(2,1)}^{a_3\cdots a_7]}\rb\cdot\epsilon ,
\nn \\
J_{(3,2)}^{a_0|a_1\cdots a_7}\cdot\epsilon &:= &-\f{7}{8}\Big(\lb J_{(0,1)}^{a_0}, J_{(3,1)}^{a_1\cdots a_7}\rb+\lb J_{(0,1)}^{[a_1}, J_{(3,1)}^{a_2\cdots a_7]a_0}\rb\Big)\cdot\epsilon ,
\nn \\
J_{(3,2)}^{a_1\cdots a_8}\cdot\epsilon &:= & \lb J_{(1,0)}^{[a_1a_2}, J_{(2,2)}^{a_3\cdots a_8]}\rb\cdot\epsilon ,
\nn \\
J_{(4,1)}^{a_1\cdots a_9}\cdot\epsilon &:= &\lb J_{(1,0)}^{[a_1a_2}, J_{(3,1)}^{a_3\cdots a_9]}\rb \cdot\epsilon .
\ee
We stress that this kind of construction is guaranteed to yield a consistent unfaithful representation of all of $\mf{k}(\mf{e}_{10})$ given that a few simple consistency conditions between the lowest level fundamental generators are satisfied~\cite{Damour:2006xu}. That these conditions are satisfied here can be checked easily directly, but it also follows from the branching of the transformation rules given in~\cite{deBuyl:2005zy,Damour:2005zs}.

\subsubsection{Vector-spinor}
\label{app:VectorSpinor}

By reduction of the transformation rules of~\cite{Damour:2005zs,deBuyl:2005mt} one obtains that
the fundamental $\mf{k}(\mf{e}_{10})$-generators act on the vector spinor representation as follows on the $\Psi_{10}$ component 
\be
J_{(0,1)}^{a}\cdot \Psi_{10} &=& \f12 \G_{10}\G^a\Psi_{10}+\Psi^a\nn\\
J_{(1,0)}^{a_1a_2}\cdot \Psi_{10} &=& \f16 \G_{10}\G^{a_1a_2}\Psi_{10}+\f43 \G^{[a_1}\Psi^{a_2]}\nn.
\ee
On the $\Psi_a$ ($a=1,\ldots,9$) component they act as 
\be
J_{(0,1)}^{a}\cdot \Psi_{b}&=&\f12 \G_{10}\G^a\Psi_{b}-\delta^a_b\Psi_{10}\nn\\
J_{(1,0)}^{a_1a_2}\cdot \Psi_{b} &=& \f12\G_{10}\G^{a_1a_2}\Psi_b -\f43\G_{10}\delta^{[a_1}_b\Psi^{a_2]}+\f23\G_{10}\G_b^{\ [a_1}\Psi^{a_2]}\nn\\
&&+\f43\delta_b^{[a_1}\G^{a_2]}\Psi_{10}-\f13\G_b^{\ a_1a_2}\nn\Psi_{10}.
\ee
The other levels action is then computed to be on $\Psi_{10}$ as
\be
M^{a_1a_2}\cdot  \Psi_{10} &=&\phantom{-} \f12\G^{a_1a_2}\Psi_{10}\nn\\
J_{(1,1)}^{a_1\cdots a_3}\cdot \Psi_{10} &=& \phantom{-}  \f12\G^{a_1\cdots a_3}\Psi_{10}-\G_{10}\G^{[a_1a_2}\Psi^{a_3]}\nn\\
J_{(2,1)}^{a_1\cdots a_5}\cdot \Psi_{10} &=& - \f16 \G_{10}\G^{a_1\cdots a_5}\Psi_{10}-\f53\G^{[a_1\cdots a_4}\Psi^{a_5]}\nn\\
J_{(2,2)}^{a_1\cdots a_6}\cdot \Psi_{10} &=& -\f12\G^{a_1\cdots a_6}\Psi_{10}-4\G_{10}\G^{[a_1\cdots a_5}\Psi^{a_6]}\nn\\
J_{(3,1)}^{a_1\cdots a_7}\cdot \Psi_{10} &=& -\f32\G^{a_1\cdots a_7}\Psi_{10}+7\G_{10}\G^{[a_1\cdots a_6}\Psi^{a_7]}\nn\\
J_{(3,2)}^{a_1\cdots a_8}\cdot \Psi_{10} &=& - \f43 \G_{10}\G^{a_1\cdots a_8}\Psi_{10}-\f{32}3\G^{[a_1\cdots a_7}\Psi^{a_8]}\nn\\
J_{(3,2)}^{a_0|a_1\cdots a_7}\cdot \Psi_{10}&=& - \f{7}{2}\Gamma_{10}\delta_{a_0}^{[a_1}\Gamma^{a_2\cdots a_7]}\Psi_{10}\nn\\
J_{(4,1)}^{a_1\cdots a_9}\cdot \Psi_{10} &=& \phantom{-}  \f92 \G_{10}\G^{a_1\cdots a_9}\Psi_{10}-15\G^{[a_1\cdots a_8}\Psi^{a_9]}.
\label{onpsi10}
\ee

On $\Psi_a$ we obtain similarly
\be
M^{a_1a_2}\cdot  \Psi_{b} &=& \f12\G^{a_1a_2}\Psi_{b}+2\delta_b^{[a_1}\Psi^{a_2]}\nn\\
J_{(1,1)}^{a_1\cdots a_3}\cdot \Psi_{b} &=& \f12\G^{a_1\cdots a_3}\Psi_{b}+4\delta_b^{[a_1}\G^{a_2}\Psi^{a_3]}-\G_b{}^{[a_1a_2}\Psi^{a_3]}\nn\\
J_{(2,1)}^{a_1\cdots a_5}\cdot \Psi_{b} &=&  \f12 \G_{10}\G^{a_1\cdots a_5}\Psi_{b}+\f{20}{3}\Gamma_{10}\delta_b^{[a_1}\Gamma^{a_2\cdots a_4}\Psi^{a_5]}-\f{10}3\Gamma_{10}{\Gamma_b}^{[a_1\cdots a_4}\Psi^{a_5]}
\nn \\
& & +\f53\delta_b^{[a_1}\Gamma^{a_2\cdots a_5]}\Psi_{10}-\f{2}{3}{\Gamma_b}^{a_1\cdots a_5}\Psi_{10}\nn\\
J_{(2,2)}^{a_1\cdots a_6}\cdot \Psi_{b} &=& -\f12\G^{a_1\cdots a_6}\Psi_{b}+10\delta_b^{[a_1}\Gamma^{a_2\cdots a_5}\Psi^{a_6]}-4{\Gamma_b}^{[a_1\cdots a_5}\Psi^{a_6]}\nn\\
J_{(3,1)}^{a_1\cdots a_7}\cdot \Psi_{b} &=& \f12\G^{a_1\cdots a_7}\Psi_{b}-7{\Gamma_b}^{[a_1\cdots a_6}\Psi^{a_7]} -2\Gamma_{10}{\Gamma_b}^{a_1\cdots a_7}\Psi_{10}\nn\\
J_{(3,2)}^{a_1\cdots a_8}\cdot \Psi_{b} &=&-\f43\G_{10}\G_b^{\ [a_1\cdots a_7}\Psi^{a_8]}-\f{28}3\G_{10}\delta_b^{[a_1}\G^{a_2\cdots a_7}\Psi^{a_8]} \nn\\
&&-\f43\Gamma_b^{\ a_1\cdots a_8}\Psi_{10}+\f43\delta_b^{[a_1}\G^{a_2\cdots a_8]}\Psi_{10}\nn\\
J_{(3,2)}^{a_0|a_1\cdots a_7}\cdot \Psi_{b}&=&\f{7}{2}\Gamma_{10}\delta^{[a_1}_{a_0}\Gamma^{a_2\cdots a_7]}\Psi_{b}-\f{21}{2}\Gamma_{10}\delta_b^{[a_1}\Gamma_{a_0}^{\ \,a_2\cdots a_6}\Psi^{a_7]}-\f{49}{4}\Gamma_{10}\delta_b^{a_0}\Gamma^{[a_1\cdots a_6}\Psi^{a_7]}
\nn \\
 & & +42 \Gamma_{10}\delta_{a_0}^{[a_1}{\Gamma_b}^{a_2\cdots a_6}\Psi^{a_7]}-7\delta_{a_0}^{[a_1}{\Gamma_b}^{a_2\cdots a_7]}\Psi_{10}
 \nn \\
 & & +\f{7}{4}\Big(\Gamma_{10}{\Gamma_{b}}^{a_0[a_1\cdots a_6}\Psi^{a_7]}-\Gamma_{10}\delta_b^{[a_1}\Gamma^{a_2\cdots a_7]}\Psi^{a_0}
 \nn \\
 & &\phantom{7}+\Gamma_{10}{\Gamma_b}^{a_1\cdots a_7}\Psi^{a_0} +\delta_b^{[a_1}\Gamma^{a_2\cdots a_7] a_0}\Psi_{10} +\delta_b^{a_0}\Gamma^{a_1\cdots a_7}\Psi_{10}\Big)
  \nn\\
J_{(4,1)}^{a_1\cdots a_9}\cdot \Psi_{b} &=&\f{1}{2}\Gamma_{10}\Gamma^{a_1\cdots a_9}\Psi_b -12\Gamma_{10}{\Gamma_{b}}^{[a_1\cdots a_8}\Psi^{a_9]}
\nn \\
& & -24\Gamma_{10}\delta_b^{[a_1}\Gamma^{a_2\cdots a_8}\Psi^{a_9]}
-9\delta_b^{[a_1}\Gamma^{a_2\cdots a_9]}\Psi_{10}.
\label{onpsib}
\ee
We note that the mixed symmetry generator at level $(3,2)$ indeed satisfies 
\beq
J_{(3,2)}^{[a_0|a_1\cdots a_7]}\cdot \Psi_{b}=0
\eeq
as desired.

\section{Equations of motion of the $E_{10}/K(E_{10})$ coset model}
\label{app:EOMe10}

Using all the explicit commutators and representations of the Appendix \ref{Appendix:LevelDecomp}, we can now write out the bosonic and fermionic equations of motion in their full glory.

\subsection{Bosonic equations of motion}
\label{Appendix:EOMBosonic}

The level zero equations of motion read 
\begin{subequations}\label{levelzeroeom}
\be
\p^2_t\phi&=&e^{3\phi/2}P_aP_a-2 e^{-\phi}P_{a_1a_2}P_{a_1a_2}+\f{1}{3}e^{\phi/2}P_{a_1a_2a_3}P_{a_1a_2a_3}\nn\\
&&-\frac2{5!}e^{-\phi/2}P_{a_1\ldots a_5}P_{a_1\ldots a_5}+\frac4{6!}e^{\phi}P_{a_1\ldots a_6}P_{a_1\ldots a_6}\nn\\
&&-\frac6{7!}e^{-3\phi/2}P_{a_1\ldots a_7}P_{a_1\ldots a_7}-\frac{10}{9!}e^{-5\phi/2}P_{a_1\ldots a_9}P_{a_1\ldots a_9}\label{dilatoneom}\\
\cD p_{ab} &=& -\frac1{4}e^{3\phi/2}\delta_{ab}P_cP_c+2e^{3\phi/2}P_aP_b \nn \\
& & -\f{1}{4}e^{-\phi}\delta_{ab}P_{cd}P_{cd}+2e^{-\phi}P_{ca}P_{cb}
\nn \\
& & -\f1{8}e^{\phi/2}\delta_{ab}P_{c_1c_2c_3}P_{c_1c_2c_3}+e^{\phi/2}P_{ac_1c_2}P_{bc_1c_2}
\nn \\
& & -\f1{4\cdot 4!}e^{-\phi/2}\delta_{ab}P_{c_1\cdots c_5}P_{c_1\cdots c_5}+\f{2}{4!}e^{-\phi/2}P_{ac_1\cdots c_4}P_{bc_1\cdots c_4}
\nn \\
& & -\f{3}{2\cdot 6!}e^{\phi}\delta_{ab}P_{c_1\cdots c_6}P_{c_1\cdots c_6}+\f{2}{5!}e^{\phi}P_{ac_1\cdots c_5}P_{bc_1\cdots c_5}
\nn \\
& & -\f1{4\cdot 6!}e^{-3\phi/2}\delta_{ab}P_{c_1\cdots c_7}P_{c_1\cdots c_7}+\f{2}{6!}e^{-3\phi/2}P_{ac_1\cdots c_6}P_{bc_1\cdots c_6}
\nn \\
& & -\f1{8!}\delta_{ab}P_{c_1\cdots c_8}P_{c_1\cdots c_8}+\f1{7!}P_{ac_1\cdots c_7}P_{bc_1\cdots c_7}
\nn \\
& & +\f1{4\cdot 8!}\big(-\delta_{ab}P_{c_0|c_1\cdots c_7}P_{c_0|c_1\cdots c_7}+P_{a|c_1\cdots c_7}P_{b|c_1\cdots c_7}
\nn \\
& & +7P_{c_0|c_1\cdots c_6 a}P_{c_0|c_1\cdots c_6 b}\big)
\nn \\
& & -\f1{4\cdot 8!}e^{-5\phi/2}\delta_{ab}P_{c_1\cdots c_9}+\f{2}{8!}e^{-5\phi/2}P_{ac_1\cdots c_8}P_{bc_1\cdots c_8}\,.\label{Einstein}
\ee
\end{subequations}
{\allowdisplaybreaks For the higher level fields the equations of motion become
\begin{subequations}\label{higherleveom}
\be
\cD (e^{3\phi/2} P_a) &=&
    -e^{\phi/2} P_{ac_1c_2} P_{c_1c_2}   + \frac2{5!} e^\phi P_{ac_1\ldots c_5}P_{c_1\ldots c_5}     +\frac{12}{8!} P_{ac_1\ldots c_7}P_{c_1\ldots c_7} \nn\\
 &&       -\frac{7}{4\cdot 8!}(P_{c_1|ac_2\ldots c_7}P_{c_1\ldots c_7}+P_{a|c_1\ldots c_7}P_{c_1\ldots c_7})\,,\label{boseom1}\\
\cD (e^{-\phi} P_{a_1a_2}) &=& 
    2 e^{\phi/2}  P_{a_1a_2c}P_c +\frac13 e^{-\phi/2} P_{a_1a_2c_1c_2c_3}P_{c_1c_2c_3}\nn\\
 &&  +\frac2{5!}e^{-3\phi/2} P_{a_1a_2c_1\ldots c_5}P_{c_1\ldots c_5} 
     +\frac2{7!} e^{-5\phi/2}P_{a_1a_2c_1\ldots c_7}P_{c_1\ldots c_7} \nn\\
 && +\frac1{6!}P_{a_1a_2c_1\ldots c_6}P_{c_1\ldots c_6} \nn\\
 && + \frac{3}{8\cdot 6!} (P_{c_1|a_1a_2c_2\ldots c_6}P_{c_1\ldots c_6}+P_{a_1|a_2c_1\ldots c_6}P_{c_1\ldots c_6})\,, \label{boseom2} \\
\cD (e^{\phi/2}P_{a_1a_2a_3}) &=&
    -e^{-\phi/2} P_{a_1a_2a_3c_1c_2}P_{c_1c_2} 
        -\frac13 e^\phi P_{a_1a_2a_3c_1c_2c_3}P_{c_1c_2c_3}\nn\\
        && -\frac1{2\cdot 5!} P_{a_1a_2a_3c_1\ldots c_5}P_{c_1\ldots c_5}\nn\\
 && +\frac{1}{256}(P_{c_1|a_1a_2a_3c_2\ldots c_5}P_{c_1\ldots c_5}+P_{a_1|a_2a_3c_1\ldots c_5}P_{c_1\ldots c_5}) \,,\label{boseom3}\\
\cD (e^{-\phi/2} P_{a_1\ldots a_5}) &=&
    2 e^\phi P_{a_1\ldots a_5c}P_c-e^{-3\phi/2}P_{a_1\ldots a_4 c_1c_2}P_{c_1c_2}\nn\\
    && -\frac1{12}P_{a_1\ldots a_5c_1c_2c_3}P_{c_1c_2c_3}\nn\\
    && +\frac{1}{28}(P_{c_1|a_1\ldots a_5c_2c_3}P_{c_1\ldots c_3}+P_{a_1|a_2\ldots a_5c_1\ldots c_3}P_{c_1\ldots c_3}) \,,\label{Bianchi1}\\
\cD (e^\phi P_{a_1\ldots a_6}) &=&
   -\frac12 P_{a_1\ldots a_6c_1c_2} P_{c_1c_2}\nn\\
   &&-\frac{3}{16}(P_{c_1|a_1\ldots a_6c_2}P_{c_1c_2}+P_{a_1|a_2\ldots a_6c_1c_2}P_{c_1 c_2})\,,\label{Bianchi2} \\
\cD (e^{-3\phi/2} P_{a_1\ldots a_7}) &=&
  - e^{-5\phi/2}P_{a_1\ldots a_7c_1c_2} P_{c_1c_2}+\frac32P_{a_1\ldots a_7 c}P_{c} \nn\\
  &&-\frac7{32}(P_{c|a_1\ldots a_7}P_{c}+P_{a_1|a_2\ldots a_7c}P_{c})  \,,\\
\cD (e^{-5\phi/2} P_{a_1\ldots a_9}) &=& 0 \,,\label{Bianchi3}\\
\cD P_{a_1\ldots a_8} &=& 0 \,,\\
\cD P_{a_0|a_1\ldots a_7} &=& 0 \,.
\ee
\end{subequations}}

\subsection{Fermionic equations of motion}
\label{app:FermionicSigmaModelEOM}

The fermionic sector of the $E_{10}$-invariant Lagrangian involves a
Dirac-type kinetic term for the 320-dimensional vector-spinor representation
$\Psi$ of $\mf{k}(\mf{e}_{10})$ which was given in (\ref{e10fermlag}). The
resulting Dirac equation can be evaluated for both the $\Psi_a$ and the
$\Psi_{10}$ components as in (\ref{GravitinoDilatinoEOM})  using the
expressions for the $K(E_{10})$ action which were derived in
Appendix~\ref{app:SpinorReps}. 

The result for the $\Psi_{10}$ component up to level $(4,1)$ is {\allowdisplaybreaks
\be
\label{eompsi10}
0&=&\pa_t \Psi_{10}-\f{1}{4}q_{a_1a_2} \G^{a_1a_2}\Psi_{10}\nn\\
&-&\f12e^{3\phi/4}P_{a}\G_{10}\G^a\Psi_{10}-e^{3\phi/4}P_{a}\Psi^a\nn\\
&-&\f{1}{12}e^{-\phi/2}P_{a_1a_2}\G_{10}\G^{a_1a_2}\Psi_{10}
-\f23e^{-\phi/2}P_{a_1a_2} \G^{a_1}\Psi^{a_2}\nn\\
&-&\f{1}{12}e^{\phi/4}P_{a_1a_2a_3}\G^{a_1\cdots a_3}\Psi_{10}+\f16e^{\phi/4}P_{a_1a_2a_3}\G_{10}\G^{a_1a_2}\Psi^{a_3}\nn\\
&+& \f{1}{6!}e^{-\phi/4}P_{a_1\cdots a_5} \G_{10}\G^{a_1\cdots a_5}\Psi_{10}+\f1{3\cdot4!}e^{-\phi/4}P_{a_1\cdots a_5} \G^{a_1\cdots a_4}\Psi^{a_5}\nn\\
&+&\f{1}{2\cdot6!}e^{\phi/2}P_{a_1\cdots a_6}\G^{a_1\cdots a_6}\Psi_{10}+\f1{180}e^{\phi/2}P_{a_1\cdots a_6}\G_{10}\G^{a_1\cdots a_5}\Psi^{a_6}\nn\\
&+&\f{3}{2\cdot7!}e^{-3\phi/4}P_{a_1\cdots a_7}\G^{a_1\cdots a_7}\Psi_{10}
-\f1{6!}e^{-3\phi/4}P_{a_1\cdots a_7}\G_{10}\G^{a_1\cdots a_6}\Psi^{a_7}\nn\\
&+&\f{4}{3\cdot8!}P_{a_1\cdots a_8}\G_{10}\G^{a_1\cdots a_8}\Psi_{10}+\f{4}{3\cdot7!}P_{a_1\cdots a_8}\G^{a_1\cdots a_7}\Psi^{a_8} \nn \\
&+&\f{7}{2\cdot8!}P_{c|ca_1\cdots a_6}\Gamma_{10}\Gamma^{a_1\cdots a_6}\Psi_{10}\nn\\
&-&\f{1}{2\cdot8!}e^{-5\phi/4}P_{a_1\cdots a_9} \G_{10}\G^{a_1\cdots
  a_9}\Psi_{10}+\f{15}{9!}e^{-5\phi/4}P_{a_1\cdots a_9} \G^{a_1\cdots
  a_8}\Psi^{a_9} + \ldots\,,
\ee
and is related to the dilatino equation of motion in the body of the article.}

For the gravitino component $\Psi_a$ one finds similarly{\allowdisplaybreaks{
\be
\label{eomke10}
0 &=& \pa_t \Psi_a-\f{1}{4}q_{b_1b_2}\G^{b_1b_2}\Psi_{a}-q_{ab}\Psi_{b}
\nn\\
&-&\f12e^{3\phi/4}P_{b} \G_{10}\G^b\Psi_{a}+e^{3\phi/4}P_{a}\Psi_{10}
\nn \\
&-&\f{1}{4}e^{-\phi/2}P_{b_1b_2}\G_{10}\G^{b_1b_2}\Psi_a +\f23e^{-\phi/2}P_{ab}\G_{10}\Psi^{b}-\f13e^{-\phi/2}P_{b_1b_2}\G_{10}\G_a^{\ b_1}\Psi^{b_2}
\nn\\
&&-\f23e^{-\phi/2}P_{ab}\G^{b}\Psi_{10}
 +\f16e^{-\phi/2}P_{b_1b_1}\G_a^{\ b_1b_2}\Psi_{10}
\nn\\
&-&\f{1}{12}e^{\phi/4}P_{b_1b_2b_3}\G^{b_1\cdots b_3}\Psi_{a}
-\f23e^{\phi/4}P_{ab_1b_2}\G^{b_1}\Psi^{b_2}
+\f16e^{\phi/4}P_{b_1b_2b_3}\G_a{}^{b_1b_2}\Psi^{b_3}
\nn \\
&-& \f{1}{2\cdot5!}e^{-\phi/4}P_{b_1\cdots b_5} \G_{10}\G^{b_1\cdots b_5}\Psi_{a}
-\f{1}{18}e^{-\phi/4}P_{ab_1\cdots b_4} \Gamma_{10}\Gamma^{b_1\cdots
  b_3}\Psi^{b_4}\nn\\
&&+\f{1}{36}e^{-\phi/4}P_{b_1\cdots b_5} \Gamma_{10}{\Gamma_a}^{b_1\cdots
  b_4}\Psi^{b_5} 
 -\f1{3\cdot4!}e^{-\phi/4}P_{ab_1\cdots b_4}\Gamma^{b_1\cdots
   b_4}\Psi_{10}\nn\\ 
&&+\f{2}{3\cdot5!}e^{-\phi/4}P_{b_1\cdots b_5}{\Gamma_a}^{b_1\cdots b_5}\Psi_{10}\nn \\
&+&\f{1}{2\cdot6!}e^{\phi/2}P_{b_1\cdots b_6}\G^{b_1\cdots b_6}\Psi_{a}
-\f{1}{3\cdot 4!}e^{\phi/2}P_{ab_1\cdots b_5}\Gamma^{b_1\cdots b_4}\Psi^{b_5}
+\f4{6!}e^{\phi/2}P_{b_1\cdots b_6}{\Gamma_a}^{b_1\cdots b_5}\Psi^{b_6}
\nn \\
&-&\f{1}{2\cdot7!}e^{-3\phi/4}P_{b_1\cdots b_7}\G^{b_1\cdots b_7}\Psi_{a}
+\f1{6!}e^{-3\phi/4}P_{b_1\cdots b_7}{\Gamma_a}^{b_1\cdots b_6}\Psi^{b_7}\nn\\
&&+\f2{7!}e^{-3\phi/4}P_{b_1\cdots b_7}\Gamma_{10}{\Gamma_a}^{b_1\cdots b_7}\Psi_{10}
 \nn \\
&+&\f{1}{6\cdot7!}P_{b_1\cdots b_8}\G_{10}\G_a^{\ b_1\cdots b_7}\Psi^{b_8}
+\f1{6\cdot6!}P_{ab_1\cdots b_7}\G_{10}\G^{b_1\cdots b_6}\Psi^{b_7}
 \nn\\
&&+\f{1}{6\cdot7!}P_{b_1\cdots b_8}\Gamma_a^{\ b_1\cdots b_8}\Psi_{10}
-\f{1}{6\cdot7!}P_{ab_1\cdots b_7}\G^{b_1\cdots b_7}\Psi_{10}
\nn \\
&-&\f{7}{2\cdot8!}P_{c|cb_1\cdots b_6}\Gamma_{10}\Gamma^{b_1\cdots b_6}\Psi_{a}
+\f{21}{2\cdot8!}P_{b_0|ab_1\cdots b_6}\Gamma_{10}\Gamma^{b_0b_1\cdots
  b_5}\Psi^{b_6}\nn\\ 
&&+\f{49}{4\cdot8!}P_{a|b_1\cdots b_7}\Gamma_{10}\Gamma^{b_1\cdots b_6}\Psi^{b_7}
-\f{42}{8!} P_{c|cb_1\cdots b_6}\Gamma_{10}{\Gamma_a}^{b_1\cdots b_5}\Psi^{b_6}
 +\f7{8!}P_{c|cb_1\cdots b_6}{\Gamma_a}^{b_1\cdots b_6}\Psi_{10}
 \nn \\
 & & -\f{7}{4\cdot8!}P_{b_0|b_1\cdots
   c_7}\Gamma_{10}{\Gamma_{a}}^{b_0b_1\cdots b_6}\Psi^{b_7} 
+\f{7}{4\cdot 8!}P_{b_0|ab_1\cdots b_6}\Gamma_{10}\Gamma^{b_1\cdots
  b_6}\Psi^{b_0} 
 \nn \\
 & &-\f{7}{4\cdot 8!}P_{b_0|b_1\cdots b_7}\Gamma_{10}{\Gamma_a}^{b_1\cdots
   b_7}\Psi^{b_0} 
  -\f{7}{4\cdot 8!}P_{b_0|ab_1\cdots b_6}\Gamma^{b_0b_1\cdots
    b_6}\Psi_{10}\nn\\
&&   -\f{7}{4\cdot 8!}P_{a|b_1\cdots b_7}\Gamma^{b_1\cdots b_7}\Psi_{10}\nn
   \\ 
&-&\f{1}{2\cdot9!}e^{-5\phi/4}P_{b_1\cdots b_9}\Gamma_{10}\Gamma^{b_1\cdots b_9}\Psi_a
 +\f{12}{9!}e^{-5\phi/4}P_{b_1\cdots b_9}\Gamma_{10}{\Gamma_{a}}^{b_1\cdots b_8}\Psi^{b_9}
\nn \\
& & +\f{24}{9!}e^{-5\phi/4}P_{ab_1\cdots b_8}\Gamma_{10}\Gamma^{b_1\cdots b_7}\Psi^{b_8}
+\f1{8!}e^{-5\phi/4}P_{ab_1\cdots b_8}\Gamma^{b_1\cdots b_8}\Psi_{10}+
\ldots\,. 
\ee}}

\subsection{Supersymmetry variation}
\label{app:susyvarcoset}

In the same fashion, the supersymmetry variation given in
(\ref{SusyTransfSigmaModel})
can be written explicitly as 
\be
\label{susyexpli}
\delta\Psi_t &=& \pa_t \epsilon -\f{1}{4}q_{a_1a_2}\Gamma^{a_1a_2}\epsilon
- \f{1}{2}e^{3\phi/4}P_{a}\Gamma_{10}\Gamma^{a}\epsilon
-\f{1}{4}e^{-\phi/2}P_{a_1a_2}\Gamma_{10}\Gamma^{a_1a_2}\eps\nn\\
&-&\f{1}{2\cdot3!}e^{\phi/4}P_{a_1a_2a_3}\Gamma^{a_1a_2a_3}\eps
- \f{1}{2\cdot5!}e^{-\phi/4}P_{a_1\cdots a_5}\Gamma_{10}\Gamma^{a_1\cdots a_5}\epsilon
+\f{1}{2\cdot6!}e^{\phi/2}P_{a_1\cdots a_6}\Gamma^{a_1\cdots a_6}\eps\nn \\
& -&\f{1}{2\cdot7!}e^{-3\phi/4}P_{a_1\cdots a_7}\Gamma^{a_1\cdots a_7}\eps
-\f{7}{2\cdot8!}P_{c|ca_1\cdots a_6}\Gamma_{10}\Gamma^{a_1\cdots a_6}\eps\nn\\
&-&\f{1}{2\cdot9!}e^{-5\phi/4}P_{a_1\cdots a_9}\Gamma_{10}\Gamma^{a_1\cdots a_9}\eps+ \cdots
\ee

\section{The trombone generator}
\label{app:trombone}

The generator which was conjectured in~\cite{Diffon:2008sh} to be related to
deformations of supergravity involving a gauging of the on-shell trombone
symmetry appears in this case on level $(3,3)$ and has the index structure
$E^{a_0|a_1\ldots a_8}$ with vanishing totally antisymmetric part. As noted in
the text, it descends from the usual dual graviton in $D=11$. From
Table~\ref{longerdec} in Appendix \ref{app:commutators}  it can be seen that the conjectured trombone gauging
generator arises from a generator of $E_9$ since its 
first root entry is zero. More precisely, in current algebra language it is
the affine level one copy of the lowest root of $E_8$. 

\subsection{Appearance in the coset equations of motion}

We now work out the commutation relations of this level $(3,3)$ generator
without entering into all the details. It can be defined as arising from a
commutator of level $(0,1)$ and $(3,2)$ via
\be\label{trombdef}
\lb E^{a_0}, E^{a_1\ldots a_8} \rb = E^{a_0| a_1\ldots a_8} \,.
\ee
This definition implies that also $\lb E^{a_0|a_1\ldots a_7}, E^{a_8} \rb$ has
a contribution to the level $(3,3)$ generator. This coupling is related to a
Lorentz covariantization and we will not study it in detail here. The
generator $E^{a_0|a_1\ldots a_8}$ transforms in a hook representation of
$SL(9,{\mathbb R})$ and carries charge $3/4$ under the dilaton generator $T$
\be
E^{[a_0|a_1\ldots a_8]} = 0\,,\quad 
  \lb T, E^{a_0|a_1\ldots a_8} \rb = \frac34 E^{a_0|a_1\ldots a_8}\,.
\ee

{}From the commutation relations of Appendix~\ref{app:commutators}
and (\ref{trombdef}) 
one deduces that $E^{a_0|a_1\ldots a_8}$ has non-vanishing commutation
relations only with the negative level $F$-type generators as follows
\be\label{trombcom}
\lb E^{a_0|a_1\ldots a_8}, F_{b_1b_2b_3} \rb &=& 
  -7\cdot 12 \left( \delta_{b_1b_2b_3}^{a_0[a_1a_2} E^{a_3\ldots a_8]} 
    - \delta_{b_1b_2b_3}^{[a_1a_2a_3} E^{a_3\ldots a_8]a_0}\right)\,\nn\\
\lb E^{a_0|a_1\ldots a_8}, F_{b_1\ldots b_6} \rb &=& 
  2\cdot 7! \left( \delta_{b_1\ldots b_6}^{a_0[a_1\ldots a_5} E^{a_6a_7a_8]} 
    - \delta_{b_1\ldots b_6}^{[a_1\ldots a_6} E^{a_7a_8]a_0}\right) \,\nn\\
\lb E^{a_0|a_1\ldots a_8}, F_{b} \rb &=& 
    \delta_{b}^{a_0} E^{a_1\ldots a_8} 
    - \delta_{b}^{[a_1} E^{a_2\ldots a_8]a_0} \,,\nn\\
\lb  E^{a_0|a_1\ldots a_8}, F_{b_1\ldots b_8} \rb 
  &=& \frac12\cdot 8! \left( \delta^{a_1\ldots a_8}_{b_1\ldots b_8} E^{a_0} 
  - \delta^{a_0[a_1\ldots a_7}_{b_1\ldots b_8} E^{a_8]} \right)\,,
\ee
and of course with its own dual $F_{b_0|b_1\ldots b_8}$ and $F_{b_0|b_1\ldots
  b_7}$. We introduce the contribution of this
generator to the Maurer--Cartan form (\ref{cmform}) as
\be
\cP \to \cP + \frac1{9!} e^{3\phi/4} P_{a_0|a_1\ldots a_8} S^{a_0|a_1\ldots
  a_8} \,.
\ee

The commutators (\ref{trombcom}) imply that the geodesic equations
(\ref{levelzeroeom}) and (\ref{higherleveom}) get additional contributions in
the following way 
\be\label{treom1}
D^{(0)}(e^{3\phi/2} P_a) &=& \frac1{6\cdot 8!} e^{3\phi/2}
   P_{a|c_1 \ldots c_8} P_{c_1\ldots c_8} +\ldots\,,\nn\\ 
D^{(0)}(e^{\phi/2} P_{a_1a_2a_3}) &=& \frac1{2\cdot 6!} e^{3\phi/2}
   P_{c_1|c_2\ldots c_6a_1a_2a_3} P_{c_1\ldots c_6} +\ldots\,,\nn\\ 
D^{(0)}(e^{\phi} P_{a_1\ldots a_6}) &=& -\frac1{24} e^{3\phi/2}
   P_{c_1|c_2c_3a_1\ldots a_6} P_{c_1c_2c_3} +\ldots
\ee
and an additional contribution in the dual dilaton equation and dual graviton
equations. In addition, there are
contributions to the Einstein equation and scalar equations of the form
\be\label{treom2}
D^{(0)} p_{ab} &=& \frac1{4\cdot 9!} e^{3\phi/2}
  ( P_{a|c_1\ldots c_8} P_{b|c_1\ldots c_8}
  + 8 P_{c_0|c_1\ldots c_7 a}  P_{c_0|c_1\ldots c_7 b} \nn\\
  &&\phantom{ \frac1{4\cdot 9!} e^{3\phi/2}
  ( P_{a|c_1\ldots c_8} } - \delta_{ab} P_{c_0|c_1\ldots c_8} P_{c_0|c_1\ldots c_8})+\ldots\,,\nn\\ 
\partial_t^2\phi &= &  \frac{3}{24\cdot 9!} e^{3\phi/2}
    P_{c_0|c_1\ldots c_8} P_{c_0|c_1\ldots c_8}+\ldots\,.
\ee

\subsection{Comparison to the trombone gauged supergravity}

We now compare the new contributions in the geodesic equations (\ref{treom1}) and
(\ref{treom2}) to the equations of motion of the deformed supergravity where
the trombone symmetry has been gauged~\cite{Howe:1997qt,Lavrinenko:1997qa}. We
use the equations given in equation~(5.7) of~\cite{Lavrinenko:1997qa} where we note
that their dilaton has the opposite sign to ours. As is known, the equations
of motion after a trombone gauging contain bare vector potentials, denoted
by $\mathcal{A}_\mu$ in \cite{Lavrinenko:1997qa} and corresponding to our
vector potential $A_\mu$, together with the deformation
parameter. This deformation parameter is denoted by $m$ there but we call it $M$ in order to avoid confusion with the Romans mass parameter. For this Appendix only we turn off the
Romans deformation and consider instead the trombone deformation $M$. We do not know how
to describe the terms containing the bare $\mathcal{A}_\mu$; we ignore these
terms but instead focus on all contributions involving only the trombone
deformation parameter $M$. 

The relevant terms to be taken from (5.7) in \cite{Lavrinenko:1997qa} are then
(after changing the sign of the dilaton)
\be\label{trs1}
D_\al(e^{3\phi/2} F^{\al\beta}) &=& -12 m \partial^\beta \phi + \ldots\,,\nn\\
D_\al ( e^{\phi/2} F^{\al\beta_1\beta_2\beta_3}) &=& \phantom{-}6m e^{-\phi}
F^{\beta_1\beta_2\beta_3} + \ldots\,,
\ee
for the form equations of motion and modified Bianchi identites (cf. (5.9) in
\cite{Lavrinenko:1997qa}) 
\be\label{trs2}
4 D_{[\al_1} F_{\al_2\al_3\al_4]} = -3m F_{\al_1\al_2\al_3\al_4} + \ldots\,.
\ee
In the Einstein equation the correction is
\be\label{trs3}
R_{\al\beta} = 9 m^2 e^{-3\phi/2} \eta_{\al\beta}+\ldots\,,
\ee
while there is {\em no} new term in the dilaton equation of motion. 

Comparing the equations (\ref{trs1}) and (\ref{trs2}) to (\ref{treom1}) one
sees, upon use of the dictionary of Table~\ref{dicoeom} one sees that the new
terms in the $E_{10}$ equations (\ref{treom1}) appear precisely in the right
equations and with the correct dilaton prefactors in order to correspond to
(\ref{trs1}) and (\ref{trs2}) if one made a dictionary of the type
\be\label{trdic}
P_{a_0|a_1\ldots a_8} \sim e^{-3\phi/2} \epsilon_{a_0a_1\ldots a_8} m \,.
\ee
However, the hook symmetry of the $(3,3)$ generator forbids the use of the
epsilon. As discussed in Section~\ref{sec:trombone} and
in~\cite{Diffon:2008sh} there is no other way to associate a single parameter $m$ to
the hook tensor in an $SO(9)$ covariant way. We note that using the rough
correspondence (\ref{trdic}) in the Einstein equation of (\ref{treom2}) will
yield also a term with the correct dilaton prefactor compared to
(\ref{trs3}). In the dilaton equation, on the other hand, there is no new
contribution from the trombone gauging in supergravity, indicating that the
particular contraction in (\ref{treom2}) would have to vanish. This is,
however, not possible since it is a sum of 
squares.

Barring these difficulties we find it interesting that the hook generator
appears in the right equations and also yields the right dilaton
prefactor. Without a way of extracting a single deformation parameter $m$, we 
cannot confirm the association of the hook generator to the trombone
gaugings. 

One possible resolution of this puzzle was hinted at in
Section~\ref{sec:trombone} where it was suggested that in a vertex operator
algebra based on  the $E_{11}$ root lattice one would obtain a new generator
(associated with an imaginary simple root of the Borcherds algebra of physical
states) on level $(3,3)$ that transforms as a nine-form under $SL(9,{\mathbb
  R})$. 
In a way similar
to the discussion of the Romans mass one could associate a mass deformation
parameter with this new generator. However, it being a new {\em simple} root
it would not produce the same couplings to the lower level fields as in
(\ref{trombcom}) but commute with them. Therefore, this possibility also seems
ruled out as a resolution of the trombone gaugings.

\section{$\mf{e}_{10}$ and $\mf{k}(\mf{e}_{10})$ at $\mf{sl}(10,\mbb{R})$ level 4}
\label{app:level4}

The decomposition of $\mf{e}_{10}$ relevant for $D=11$ maximal supergravity is the one associated to the exceptional node $\ell_1$ in Figure \ref{figure:E10}. This is thus a decompositon with respect to the `horizontal' $\mf{sl}(10, \mbb{R})$ subalgebra. At levels $\ell_1=1, 2, 3$ one obtains respectively a three-form $E^{a_1a_2a_3}$, a six-form $E^{a_1\cdots a_6}$ and a nine-index tensor $E^{a|b_1\cdots b_8}$ with mixed Young symmetry \cite{Damour:2002cu}. These generators have a natural interpretation as $D=11$ supergravity fields following the dictionary derived in \cite{Damour:2002cu}. In this section only we use the convention that latin indices $a, b, c, \dots$ are $\mf{sl}(10, \mbb{R})$-indices and hence run from $1$ to $10$. 

Proceeding to level $\ell_1=4$ one obtains two distinct representations
corresponding to the tensors $E^{{a}|{b}|{c}_1\cdots {c}_{10}}$ and
$E^{{a}_1{a}_2{a}_3|{b}_1\cdots {b}_9}$, both with mixed Young
symmetry~\cite{Nicolai:2003fw,Fischbacher:2005fy}. Neither of these generators
has a physical interpretation in terms of $D=11$ supergravity degrees of
freedom. However, as we have seen in the main text of this paper, upon further
decomposion (`dimensional reduction') with respect to the node $\ell_2$ in
Figure \ref{figure:E10} the generator $E^{{a}|{b}|{c}_1\cdots {c}_{10}}$
descends to the `mass term generator' of massive type IIA supergravity in the
following way~\cite{West:2004st,Henneaux:2007ej} (see also (\ref{massgen}))
\beq
E^{\al_1\cdots \al_9}:=\f18E^{10|10|10\al_1\cdot \al_9},
\eeq
where $\al_i=1, \dots , 9$. Therefore, although no $D=11$ interpretation exists for the full level 4 generators, in this way one obtains a physical interpretation of part of the level 4 representation content in the context of the massive IIA theory. In the main text we have analysed $\mf{e}_{10}$ in a multilevel $\ell=(\ell_1, \ell_2)$ decomposition with respect to $\mf{sl}(9, \mbb{R})$, which directly reveals the spectrum of massive IIA supergravity. For completeness we shall in this appendix derive the commutation relations for $\mf{e}_{10}$ at level 4 also in the $D=11$ picture. Our results confirm a previous algorithmic computer algebra analysis in \cite{Fischbacher:2005fy}. In addition we shall use our $\ell_1=4$ results to extend the $\mf{k}(\mf{e}_{10})$-representations ${\bf 32}$ and ${\bf 320}$ up to `level 4'.  

\subsection{The $\mf{e}_{10}$ commutation relations at level four}

In the decomposition of the adjoint representation of $\mf{e}_{10}$ in terms of representations of $\mf{sl}(10, \mbb{R})$ we find two tensors at level 4, both with mixed Young tableaux symmetries:
\be
{}E^{{a}|{b}|{c}_1\cdots {c}_{10}} & = & E^{({a}|{b})|[{c}_1\cdots {c}_{10}]},
\nn \\
{} E^{{a}_1{a}_2{a}_3|{b}_1\cdots {b}_9}& = & E^{[{a}_1{a}_2{a}_3]|[{b}_1\cdots {b}_{9}]}.
\ee
Note that the first tensor can be simplified by pulling out the block of 10 antisymmetric indices,
\beq 
E^{a|b|{c}_1\cdots {c}_{10}}= \epsilon^{{c}_1\cdots {c}_{10}}E^{a|b}.
\eeq
In the following we shall use only $E^{{s}_1|{s}_2|{c}_1\cdots {c}_{10}}$ as it is more easily related to the mass of massive IIA supergravity. 
The second generator is also subject to the Young irreducibility constraint
\beq\label{39hook}
E^{{a}_1{a}_2[{a}_3|{b}_1\cdots {b}_9]}=0,
\eeq
which is crucial in simplifying expressions involving this generator.
To avoid an excess of indices in the text, we will often refer to the two level 4 generators schematically as $E^{1|1|10}$ and $E^{3|9}$, where the superscripts encode their respective index structures. Unless otherwise specified we shall throughout this section employ the convention that blocks of indices of  the same type, e.g. ${a}_1, \cdots, {a}_r$, are implicitly \emph{antisymmetrized}. For example,
\beq
E^{{a}_1{a}_2{a}_3|{a}_4\cdots {a}_9 {b}_1{b}_2{b}_3}=E^{[{a}_1{a}_2{a}_3|{a}_4\cdots {a}_9]{b}_1{b}_2{b}_3}.
\eeq
However, in cases where `0' is used as a subscript on an index, no antisymmetrization over this index  is assumed, i.e., $E^{{a}_0{a}_1{a}_2|{a}_3\cdots {a}_8{b}_1{b}_2{b}_3}=E^{{a}_0[{a}_1{a}_2|{a}_3\cdots {a}_8]{b}_1{b}_2{b}_3}$. In addition, we let blocks of indices of `$s$'-type, i.e., ${s}_1\cdots {s}_r$, have implicit \emph{symmetrization}.

We use slightly different conventions compared to \cite{Fischbacher:2005fy}, although we have checked that all our results are compatible with this reference. More specifically, our generators are related to the ones of \cite{Fischbacher:2005fy} as follows
\be
{} E^{{a}_1{a}_2{a}_3|{b}_1\cdots {b}_9}&=& -\f{210}{98}E_{\text{\cite{Fischbacher:2005fy}}}^{{a}_1{a}_2{a}_3|{b}_1\cdots {b}_9},
\nn \\
{} E^{{s}_1|{s}_2|a_1\cdots a_{10}} &=& -12 E_{\text{\cite{Fischbacher:2005fy}}}^{{s}_1|{s}_2|a_1\cdots a_{10}}.
\ee

The level $\ell_1=4$ generators can be obtained through commutators between generators at lower levels. 
Considering first the commutator between two level 2 generators, one finds that this yields only the generator $E^{3|9}$ at level 4, because no irreducible representation of $\mf{sl}(10, \mbb{R})$ which contains symmetric indices, as in $E^{1|1|10}$, can arise from the tensor product of two completely antisymmetric tensors. Thus, it is natural to define 
\beq
[E^{{a}_1\cdots {a}_6}, E^{{b}_1\cdots {b}_6}]:= E^{{a}_1{a}_2{a}_3|{a}_4{a}_5{a}_6{b}_1\cdots {b}_6}-E^{{b}_1{b}_2{b}_3|{b}_4{b}_5{b}_6{a}_1\cdots {a}_6},
\label{E2E2commutator}
\eeq
where the two terms on the right hand side are chosen in such a way that the antisymmetry of the commutator on the left hand side is taken into account. This relation can be inverted by using the irreducibility constraint (\ref{39hook}) to give 
\beq
E^{{a}_1{a}_2{a}_3|{b}_1\cdots {b}_9}= 42 [E^{{a}_1{a}_2{a}_3{b}_1{b}_2{b}_3}, E^{{b}_4\cdots {b}_9}].
\label{projection1}
\eeq
We note that the `hook'-symmetry of the left hand side of this equation is matched also on the right hand side, as can be seen by employing `overantisymmetrization' (also known a Schouten's identity). For example, we have
\beq
[E^{{a}_1[{a}_2{a}_3{b}_1{b}_2{b}_3}, E^{{b}_4\cdots {b}_9]}]=0\quad \Rightarrow \quad [E^{{a}_1{a}_2[{a}_3{b}_1{b}_2{b}_3}, E^{{b}_4\cdots {b}_9]}]=0.
\eeq

The second level $\ell_1=4$ generator is generated from the commutator
\be
{}[E^{{a}_1{a}_2{a}_3},E^{{b}_0|{b}_1\cdots {b}_8}] &:=& \big(E^{{a}_3|{b}_0|{b}_1\cdots {b}_8{a}_1{a}_2}+E^{{a}_3|{b}_8|{b}_1\cdots {b}_7{b}_0{a}_1{a}_2}\big)
\nn \\
{}& & +A\big(E^{{b}_0{b}_1{b}_2|{b}_3\cdots {b}_8{a}_1{a}_2{a}_3}-E^{{b}_1{b}_2{b}_3|{b}_4\cdots {b}_8{b}_0{a}_1{a}_2{a}_3}\big),
\label{E1E3commutator}
\ee 
where we have taken into account the mixed symmetry $E^{[{b}_0|{b}_1\cdots {b}_8]}=0$ of the level 3 tensor. We take this as a definition of the generator $E^{1|1|10}$, explaining the absence of a pre-factor in the first term on the right hand side of (\ref{E1E3commutator}). The coefficient $A$, however, is already fixed by (\ref{E2E2commutator}) and the Jacobi identity and will be determined shortly.

This definition of $E^{{s}_1|{s}_2|a_1\cdots a_{10}}$ can be also inverted by a suitable projection, and we obtain
\beq
E^{{s}_1|{s}_2|{a}_1\cdots {a}_{10}}=\f{10}{3}[E^{{a}_9{a}_{10}{s}_1}, E^{{s}_2|{a}_1\cdots {a}_8}],
\label{projection3}
\eeq
where we made use of the convention, introduced above, that indices of `$s$'-type are implicitly symmetric. To arrive at (\ref{projection3}) we also used the trick of overantisymmetrization on the indices ${a}_1\cdots {a}_{10}{s}_1$, which yields the following identity
\beq
E^{{s}_2|{a}_{10}|{a}_1\cdots {a}_9{s}_1}=\f{1}{10}E^{{s}_1|{s}_2|{a}_1\cdots {a}_{10}}.
\label{Schouten}
\eeq

The coefficient $A$ can now be fixed by requiring consistency, which in this case amounts to invoking the Jacobi identity. We shall need the following relations  (recall our conventions for implicit antisymmetrization on indices of the same kind)
\be
{}[E^{{a}_1{a}_2{a}_3}, E^{{a}_4{a}_5{a}_6}]&=&E^{{a}_1\cdots {a}_6}
\nn \\
{}[E^{{a}_1{a}_2{a}_3}, E^{{b}_1\cdots {b}_6}]&=&E^{{a}_1|{a}_2{a}_3{b}_1\cdots {b}_6}.
\ee
Using the Jacobi identity we may then write (\ref{projection1}) as
\beq
{}E^{{a}_1{a}_2{a}_3|{b}_1\cdots {b}_9} = -42[E^{{b}_1{b}_2{b}_3}, E^{{a}_1|{a}_2{a}_3{b}_4\cdots {b}_9}].
\label{projection2}
\eeq
Now, by inserting (\ref{E1E3commutator}) on the right hand side we find that this equation is only satisfied for the unique value 
\beq
A=-\f{28}{5}.
\eeq
Thus, equation (\ref{E1E3commutator}) becomes
\be
{}[E^{{a}_1{a}_2{a}_3},E^{{b}_0|{b}_1\cdots {b}_8}] &=& \big(E^{{a}_3|{b}_0|{b}_1\cdots {b}_8{a}_1{a}_2}+E^{{a}_3|{b}_8|{b}_1\cdots {b}_7{b}_0{a}_1{a}_2}\big)
\nn \\
{}& & +\f{28}{5}\big(E^{{b}_1{b}_2{b}_3|{b}_4\cdots {b}_8{b}_0{a}_1{a}_2{a}_3}-E^{{b}_0{b}_1{b}_2|{b}_3\cdots {b}_8{a}_1{a}_2{a}_3}\big).
\label{E1E3commutator2}
\ee

Finally, from our definitions (\ref{E2E2commutator}) and (\ref{E1E3commutator2}) and invariance of the bilinear form, we deduce that the level $4$ generators have the following normalisation:
\be
\left<E^{{a}_1{a}_2{a}_3|{b}_1\cdots {b}_9}|F_{{c}_1{c}_2{c}_3|{d}_1\cdots {d}_9}\right>&=&
(6\cdot7!)^2\left(45\delta^{a_1\cdots a_3}_{c_1[d_1d_2}\delta^{[b_1\cdots b_7}_{d_3\cdots d_9]}\delta^{b_8b_9]}_{c_2c_3}
-10\delta^{a_1\cdots a_3}_{[d_1\cdots d_3}\delta^{[b_1\cdots b_6}_{d_4\cdots d_9]} \delta^{b_7\cdots b_9]}_{c_1\cdots c_3}
\right)\nn\\
\left<E^{{s}_1|{s}_2|{a}_1\cdots {a}_{10}}|F_{{r}_1|{r}_2|{b}_1\cdots {b}_{10}}\right>&=&64\cdot 10!\,\delta^{(s_1}_{r_1}\delta^{s_2)}_{r_2}\,\delta^{{a}_1\cdots {a}_{10}}_{{b}_1\cdots {b}_{10}}\,.
\ee

\subsection{The Dirac-spinor representation $\epsilon$ of $\mf{k}(\mf{e}_{10})$}

We proceed to define the $\mf{k}(\mf{e}_{10})$-generators at `level 4' as
\be
{}J_{(4)}^{s_1|s_2|a_1\cdots a_{10}}&= & E^{s_1|s_2|a_1\cdots a_{10}}-F_{s_1|s_2|a_1\cdots a_{10}},
\nn \\
{}J_{(4)}^{{a}_1{a}_2{a}_3|{b}_1\cdots {b}_9}&=&E^{{a}_1{a}_2{a}_3|{b}_1\cdots {b}_9}-F_{{a}_1{a}_2{a}_3|{b}_1\cdots {b}_9},
\ee
with the $\mf{k}(\mf{e}_{10})$-generators at levels 0, 1, 2 and 3 defined similarly. As for the $\mf{e}_{10}$-generators at level 4 we will sometimes denote these generators by $J_{(4)}^{1|1|10}$ and $J_{(4)}^{3|9}$. We want to find how they act on the object $\epsilon$, which is a 32-dimensional spinor representation of $\mf{k}(\mf{e}_{10})$. Subsequently, we shall generalize this to the 320-dimensional vector-spinor representation $\Psi$. 

For the levels 0, 1, 2 and 3 generators we have \cite{deBuyl:2005zy,Damour:2005zs,deBuyl:2005mt,Damour:2006xu}
\be
{} M^{{a}_1{a}_2}\cdot \epsilon &=& \f{1}{2}\Gamma^{{a}_1{a}_2}\epsilon,
\nn \\
{}J_{(1)}^{{a}_1{a}_2{a}_3}\cdot \epsilon & =& \f{1}{2}\Gamma^{{a}_1{a}_2{a}_3}\epsilon,
\nn \\
{}J_{(2)}^{{a}_1\cdots {a}_6}\cdot \epsilon &=& \f{1}{2}\Gamma^{{a}_1\cdots {a}_6}\epsilon,
\nn \\
{}J_{(3)}^{{a}_0|{a}_1\cdots {a}_8}\cdot \epsilon&=& 12\delta^{a_0{a}_1}\Gamma^{{a}_2\cdots {a}_8}\epsilon.
\label{spinorrepresentationuptolevel3}
\ee

The extension of this representation to level 4 can be obtained by considering the commutators $\lb J_{(1)}, J_{(3)}\rb $ and $\lb J_{(2)}, J_{(2)}\rb$. Schematically, they have the following structures
\be
\lb J_{(1)}, J_{(3)}\rb &=& J_{(2)}+J_{(4)}^{1|1|10}+J_{(4)}^{3|9},
\nn \\
\lb J_{(2)}, J_{(2)}\rb &=& M+J_{(4)}^{3|9},
\ee
where the $J_{(4)}^{1|1|10}$ generator is absent from the second commutator because of (\ref{E2E2commutator}).

We can then derive $J_{(4)}^{{a}_1{a}_2{a}_3|{b}_1\cdots {b}_9}$ by using the projection (\ref{projection1}), which yields
\be 
{}J_{(4)}^{{a}_1{a}_2{a}_3|{b}_1\cdots {b}_9}\cdot \epsilon &=& 42\lb J_{(2)}^{{a}_1{a}_2{a}_3{b}_1{b}_2{b}_3}, J_{(2)}^{{b}_4\cdots {b}_9}\rb \cdot \epsilon
\nn \\
{} & =& - 42\cdot 60\delta^{{a}_1{a}_2{a}_3}_{{b}_1{b}_2{b}_3}\Gamma_{{b}_4\cdots {b}_9}\epsilon+42\cdot 9\delta^{{a}_1}_{{b}_1}{\Gamma^{{a}_2{a}_3}}_{{b}_2\cdots {b}_9}\epsilon.
\ee
We now proceed to the other level 4 generator, $J_{(4)}^{1|1|10}$. This arises from the commutator 
\beq
\lb J_{(1)}^{{a}_1{a}_2{a}_3}, J_{(3)}^{{b}_0|{b}_1\cdots {b}_8}\rb\cdot \epsilon.
\label{J1J3commutatorSpinor}
\eeq 
Using the projection in (\ref{projection3}) yields
\beq
J_{(4)}^{{s}_1|{s}_2|{a}_1\cdots {a}_{10}}\cdot \epsilon =\f{10}{3}\lb J_{(1)}^{{a}_9{a}_{10}{s}_1}, J_{(3)}^{{s}_2|{a}_1\cdots {a}_8}\rb\cdot \epsilon= 10\cdot 4 \delta^{{s}_1}_{{a}_1}{\Gamma^{{s}_2}}_{{a}_2\cdots {a}_{10}}\epsilon.
\label{J11Spinor}
\eeq
We may now use equation (\ref{Schouten}) to rewrite (\ref{J11Spinor}) as
\beq
J_{(4)}^{{s}_1|{s}_2|{a}_1\cdots {a}_{10}}\cdot \epsilon = 4\,\delta^{{s}_1s_2}\Gamma^{{a}_1\cdots {a}_{10}}\epsilon.
\eeq

\subsection{The vector-spinor representation $\Psi$ of $\mf{k}(\mf{e}_{10})$}

Let us now move on to the 320-dimensional vector-spinor representation $\Psi$ of $\mf{k}(\mf{e}_{10})$. In $\mf{so}(10)$ language it is written as $\Psi_a$. For the levels 0, 1, 2 and 3 generators we have \cite{Damour:2005zs,deBuyl:2005mt,Damour:2006xu}
\be
{} M^{{a}_1{a}_2}\cdot \Psi_{{b}} &=& \f{1}{2}\Gamma^{{a}_1{a}_2}\Psi_{ b} +2\delta^{{a}_1}_{ b}\Psi^{{a}_2},
\nn \\
{}J_{(1)}^{{a}_1{a}_2{a}_3}\cdot \Psi_{ b} & =& \f{1}{2}\Gamma^{{a}_1{a}_2{a}_3}\Psi_{ b}+4\delta^{{a}_1}_{ b}\Gamma^{{a}_2}\Psi^{{a}_3}-{\Gamma_{ b}}^{{a}_1{a}_2}\Psi^{{a}_3},
\nn \\
{}J_{(2)}^{{a}_1\cdots {a}_6}\cdot \Psi_{ b} &=& \f{1}{2}\Gamma^{{a}_1\cdots {a}_6}\Psi_{ b}-10\delta_{ b}^{{a}_1}\Gamma^{{a}_2\cdots {a}_5}\Psi^{{a}_6}+4{\Gamma_{ b}}^{{a}_1\cdots {a}_5}\Psi^{{a}_6},
\nn \\
{}J_{(3)}^{{a}_0|{a}_1\cdots {a}_8}\cdot \Psi_{ b}&=& \f{16}{3}\big({\Gamma_{ b}}^{{a}_1\cdots {a}_8}\Psi^{{a}_0}-{\Gamma_{ b}}^{{a}_0{a}_1\cdots {a}_7}\Psi^{{a}_8}\big)+12\delta^{{a}_1}_{{a}_0}\Gamma^{{a}_2\cdots {a}_8}\Psi_{ b}
\nn \\
{}& & -168 \delta_{{a}_0}^{{a}_1}{\Gamma_{ b}}^{{a}_2\cdots {a}_7} \Psi^{{a}_8}-\f{16}{3}\big(\delta^{{a}_1}_{ b}\Gamma^{{a}_2\cdots {a}_8}\Psi^{{a}_0}
\nn \\
{}& & -8\delta^{{a}_0}_{ b}\Gamma^{{a}_1\cdots {a}_7}\Psi^{{a}_8}-7\delta^{{a}_1}_{ b}{\Gamma_{{a}_0}}^{{a}_2\cdots {a}_7}\Psi^{{a}_8}\big).
\label{vectorspinorrepresentationuptolevel3}
\ee
We proceed in the same way as before, and derive the $J_{(4)}^{3|9}$-generator from the commutator $\lb J_{(2)},J_{(2)}\rb$. Using (\ref{vectorspinorrepresentationuptolevel3}) we must compute 
\beq
\lb J_{(2)}^{{a}_1\cdots {a}_6}, J_{(2)}^{{b}_1\cdots {b}_6}\rb\cdot \Psi_{ c}.
\eeq
Performing this calculation and projecting onto $J_{(4)}^{{a}_1{a}_2{a}_3|{b}_1\cdots {b}_9}$ using (\ref{projection1}) then yields\footnote{In doing this calculation the computer package GAMMA \cite{GAMMA} proved to be very useful.} 
\be
{} J_{(4)}^{{a}_1{a}_2{a}_3|{b}_1\cdots {b}_9}\cdot \Psi_{ c} &=& 42  \lb J_{(2)}^{{a}_1{a}_2{a}_3{b}_1{b}_2{b}_3}, J_{(2)}^{{b}_4\cdots {b}_9}\rb \cdot \Psi_{ c}
\nn \\
{} &=& - 50400 \delta^{{a}_1{a}_2{a}_3 c}_{{b}_1{b}_2{b}_3{b}_4}\Gamma_{{b}_5 \cdots {b}_8}\Psi_{{b}_9}- 5040\delta^{{a}_1{a}_2  c}_{{b}_1{b}_2{b}_3}\Gamma_{{b}_4\cdots {b}_9}\Psi^{{a}_3}
\nn \\
& & + 30240\delta^{{a}_1{a}_2 c}_{{b}_1{b}_2{b}_3}{\Gamma^{{a}_3}}_{{b}_3\cdots {b}_8}\Psi_{{b}_9}+ 25200\delta^{{a}_1{a}_2{a}_3}_{ c\ \,{b}_1\,{b}_2}\Gamma_{{b}_3\cdots {b}_8}\Psi_{{b}_9}
\nn \\
{} & & +  40320\delta^{{a}_1{a}_2{a}_3}_{{b}_1{b}_2{b}_3}\Gamma_{{b}_4\cdots {b}_8 c}\Psi_{{b}_9}+ 5040\delta^{c\ {a}_1}_{{b}_1{b}_2}{\Gamma^{{a}_2}}_{{b}_3\cdots {b}_9}\Psi^{{a}_3}
\nn \\
{} & & - 2520\delta^{{a}_1{a}_2}_{ c\ \,{b}_1}\Gamma_{{b}_2s {b}_9}\Psi^{{a}_3}+ 5040\delta^{{a}_1{a}_2}_{{b}_1{b}_2}\Gamma_{{b}_3\cdots {b}_9 c}\Psi^{{a}_3}
\nn \\
{} & & - 25200\delta^{ c{a}_1}_{{b}_1{b}_2}{\Gamma^{{a}_2{a}_3}}_{{b}_3\cdots {b}_8}\Psi_{{b}_9}- 5040\delta^{{a}_1{a}_2}_{ c{b}_1}{\Gamma^{{a}_3}}_{{b}_2\cdots {b}_8}\Psi_{{b}_9}
\nn\\
{} & & + 5040 \delta^{{a}_1{a}_2}_{{b}_1{b}_2}{\Gamma^{{a}_3}}_{{b}_3s {b}_8  c}\Psi_{{b}_9}+ 672\delta^{{a}_1}_{ c}{\Gamma^{{a}_2}}_{{b}_1\cdots {b}_9}\Psi^{{a}_3}
\nn \\
{} &  & - 4032\delta^{{a}_1}_{{b}_1}{\Gamma^{{a}_2}}_{{b}_2\cdots {b}_9  c}\Psi^{{a}_3}- 2688\delta^{ c}_{{b}_1}{\Gamma^{{a}_1{a}_2{a}_3}}_{{b}_2\cdots {b}_8}\Psi^{{a}_3}
\nn \\
{}& & - 2016 \delta^{{a}_1}_{ c}{\Gamma^{{a}_2{a}_3}}_{{b}_1\cdots {b}_8}\Psi_{{b}_9}- 16128\delta^{{a}_1}_{{b}_1}{\Gamma^{{a}_2{a}_3}}_{{b}_2\cdots {b}_8 c}\Psi_{{b}_9}
\nn \\
{} & &- 2520\delta^{{a}_1{a}_2{a}_3}_{{b}_1{b}_2{b}_3}\Gamma_{{b}_4\cdots {b}_9}\Psi_{ c}+ 378\delta^{{a}_1}_{{b}_1}{\Gamma^{{a}_2{a}_3}}_{{b}_2\cdots {b}_9}\Psi_{ c}.
\ee
We compute the action of $J_{(4)}^{{s}_1|{s}_2|{a}_1\cdots {a}_{10}}$ by projection, as in (\ref{J11Spinor}), with the result
\be
{} J_{(4)}^{{s}_1|{s}_2|{a}_1\cdots {a}_{10}}\cdot  \Psi_{ b} &=&\f{10}3 \lb J_{(1)}^{{a}_9{a}_{10}{s}_1}, J_{(3)}^{{s}_2|{a}_1\cdots {a}_8}\rb \cdot  \Psi_{ b}
\nn \\
{} &=&- \f{2960}{3} \delta^{{s}_1}_{ b}\delta^{{s}_2}_{{a}_1}\Gamma_{{a}_2\cdots {a}_9}\Psi_{{a}_{10}}-\f{640}{3}\delta^{{s}_1}_{{s}_2}\delta^{ b}_{{a}_1}\Gamma_{{a}_2\cdots {a}_9}\Psi_{{a}_{10}}
\nn \\
{} & & +\f{10880}3\delta^{ b\ {s}_1}_{{a}_1{a}_2}{\Gamma^{{s}_2}}_{{a}_3\cdots {a}_9}\Psi_{{a}_{10}}+\f{2800}3\delta^{ b\ {s}_1}_{{a}_1{a}_2}\Gamma_{{a}_3\cdots {a}_{10}}\Psi^{{s}_2}
\nn \\
{} & & +\f{640}9\delta^{{s}_1}_{{s}_2}\Gamma_{{a}_1\cdots {a}_9  b}\Psi_{{a}_{10}}-\f{1280}{9}\delta^{ b}_{{a}_1}{\Gamma_{{a}_2\cdots {a}_{10}}}^{{s}_1}\Psi^{{s}_2}
\nn \\
{} & & +\f{640}3\delta^{{s}_1}_{ b}{\Gamma^{{s}_2}}_{{a}_1\cdots {a}_9}\Psi_{{a}_{10}}
-1600 \delta^{{s}_1}_{{a}_1}{\Gamma^{{s}_2}}_{{a}_2\cdots {a}_9  b}\Psi_{{a}_{10}}
\nn \\
{}& & +\f{640}{9}\delta_{ b}^{{s}_1}\Gamma_{{a}_1\cdots {a}_{10}}\Psi^{{s}_2}
+\f{3520}9 \delta^{{s}_1}_{{a}_1}\Gamma_{{a}_2\cdots {a}_{10} b}\Psi^{{s}_2}
\nn \\
{} & & 
+40 \delta^{{s}_1}_{{a}_1}{\Gamma^{{s}_2}}_{{a}_2\cdots {a}_{10}} \Psi_{ b}.
\label{J11VectorSpinor1}
\ee

We note that upon a further level decomposition with respect to $\ell_2$, these
expressions for the action of $\mf{k}(\mf{e}_{10})$ on the vector-spinor
representation in the $A_9$ level decomposition are consistent with the
analogous expressions for the $A_8$ decomposition in eqs. (\ref{onpsi10}) and
(\ref{onpsib}). In particular, we see from (\ref{massgen}) that  
\beq
J_{(4)}^{10|10|10\,{\alpha}_1\cdots {\alpha}_{9}}=8 J_{(4,1)}^{\alpha_1\dots \alpha_9},
\eeq
where $\alpha_i=1,\dots,9$, and one can check explicitly that the actions on both
vector-spinor and Dirac-spinor agree.

\end{document}